\documentclass[preprint,12pt]{elsarticle}
\journal{Physica D: Nonlinear Phenomena}

\usepackage{amsmath,amsthm}
\usepackage{amsfonts,amssymb}
\usepackage{mathtools}  
\usepackage{tikz-cd}
\usepackage{pgfplots}
\pgfplotsset{compat=1.18}
\usepackage{hyperref}
\usepackage{subfig}
\usepackage{cleveref}

\theoremstyle{plain}
\newtheorem{lemma}{Lemma}[section]

\theoremstyle{definition}
\newtheorem{definition}[lemma]{Definition}

\newtheorem{example}[lemma]{Example}


\newcounter{parentnumber}

\newcommand{\cost}{\text{cost}}
\newcommand{\dgm}{\text{Dgm}}

\newcommand{\I}{\mathcal{I}}
\DeclareMathOperator{\Ima}{Im}
\DeclareMathOperator{\Ker}{Ker}
\newcommand{\degree}{^{\circ}}
\newcommand{\C}{\mathcal{C}}
\newcommand{\N}{\mathbb{N}}

\newcommand{\R}{\mathbb{R}}

\newcommand{\Z}{\mathbb{Z}}

\usepackage[draft]{changes}
\usepackage{xcolor}
\usepackage{subcaption}

\definecolor{dfcolor}{rgb}{0.85,0.33,0.0}

\begin{document}

\begin{frontmatter}

\title{Topological Data Analysis of Northern Hemisphere SLP Anomalies: Identifying and Tracking the Structural Skeleton of Atmospheric Pressure Systems}

\author[1]{Himanshu Yadav}
\author[2,3]{Gisela D. Charó}
\ead{gisela.charo@cima.fcen.uba.ar}
\author[4,5,6]{Davide Faranda}

\affiliation[1]{organization={Department of Mathematics, University of Florida},
            city={Gainesville},
            postcode={32601}, 
            state={Florida},
            country={USA}}
\affiliation[2]{organization={CONICET – Universidad de Buenos Aires. Centro de Investigaciones
del Mar y la Atmósfera (CIMA)}, postcode={C1428EGA}, city={Ciudad Autónoma de Buenos Aires},country={ Argentina}}

\affiliation[3]{organization={CNRS – IRD – CONICET – UBA. Institut Franco-Argentin d'Études sur le Climat et ses Impacts (IRL 3351 IFAECI)}, postcode={C1428EGA}, city={Ciudad Autónoma de Buenos Aires},country={ Argentina}}

\affiliation[4]{organization={Laboratoire des Sciences du Climat et de l’Environnement, UMR 8212 CEA-CNRS-UVSQ, Université Paris-Saclay \& IPSL},%
            addressline={CEA Saclay l’Orme des Merisiers},
            postcode={91191}, 
            city={Gif-sur-Yvette},
            country={France}}

\affiliation[5]{organization={London Mathematical Laboratory},%
            addressline={8 Margravine Gardens},
            postcode={W6 8RH}, 
            city={London},
            country={UK}}

\affiliation[6]{organization={LMD/IPSL, ENS, Université PSL, École Polytechnique, Institut Polytechnique de Paris, Sorbonne Université, CNRS},%
            city={Paris},
            country={France}}


\begin{abstract}
We propose a novel framework based on Topological Data Analysis (TDA) to identify and track cyclonic and anticyclonic structures in the Northern Hemisphere. Using persistent homology applied to seven decades of daily sea-level pressure anomalies (1948--2023), we represent the atmospheric field as a cubical complex and compute sublevel- and superlevel-set filtrations. This approach allows us to identify 1-dimensional topological features (1-holes) that correspond to coherent pressure systems, which we term 1-cyclones and 1-anticyclones. 

The structural intensity of these features is quantified through their topological depth, while their dynamical evolution is followed using an optimal matching procedure based on the Wasserstein distance between consecutive persistence diagrams. Our results reveal robust seasonal patterns characterized by winter maxima and summer minima in total persistence, frequency, and spatial extent. We show that cyclonic activity is topologically more fragmented and intense, consistent with the seasonal deepening of the Icelandic Low, whereas anticyclones exhibit a heavier long-duration tail associated with persistent blocking episodes. 

Crucially, we demonstrate that TDA metrics can differentiate between distinct dynamical regimes of atmospheric blocking, distinguishing the quasi-stationary, ``frozen'' topology of the 2003 European heatwave from the more volatile and unstable configuration of the 2012 cold spell. Compared to classical geometric tracking algorithms, this framework provides an objective, multiscale, and noise-robust characterization of the atmospheric skeleton, offering a unified mathematical description of the organization and stability of mid-latitude circulation.
\end{abstract}

\begin{highlights}

\item A TDA framework identifies the topological skeleton of atmospheric pressure systems.
\item Wasserstein-based tracking characterizes the dynamical stability of weather systems.
\item Seven decades of Northern Hemisphere data reveal robust seasonal topological signatures.
\item TDA metrics differentiate between quasi-stationary and unstable blocking regimes.
\item Topological depth offers a multiscale alternative to classical storm tracking.

\end{highlights}
\begin{keyword}

Topological data analysis (TDA) \sep
Persistent homology \sep
Atmospheric blocking \sep
Feature tracking \sep
Sea-level pressure anomalies \sep
Northern Hemisphere



\end{keyword}

\end{frontmatter}

\section{Introduction} \label{sec:intro}
Understanding the dynamics of interacting cyclones (low-pressure systems) and anticyclones (high-pressure systems) is crucial for improving our knowledge of North Atlantic climate variability and extreme weather.
Midlatitude cyclones and anticyclones are fundamental features of the extratropical climate system, driving day-to-day weather variability and maintaining the structure of the jet stream through their transports of heat and momentum.
The frequency, intensity, and pathways of these pressure systems have a direct impact on regional climate and extremes, so investigating their behavior over the historical record (1948–2023) is of great scientific and societal importance.

 Traditionally, meteorologists have studied these systems using either Lagrangian tracking of individual pressure vortices or Eulerian statistics of variance and covariance (so-called “storm track” metrics) \cite{Rogers_1997, Hurrell_1995}. While these approaches have yielded important insights, they also have inherent limitations: Lagrangian methods require criteria to identify and follow cyclone centers, which can be challenging in complex flow situations, whereas Eulerian metrics tend to blend the contributions of cyclonic and anticyclonic events \cite{Okajima_2021}.

Early object-based methods relied on the detection of local minima in sea level pressure (SLP), followed by nearest-neighbour tracking in time\cite{murray1991numerical, sinclair1997objective}. Hodges \cite{hodges1994general,hodges1999adaptive} developed a Lagrangian feature-tracking algorithm applied to relative vorticity that allows dynamically consistent cyclone trajectories. Eulerian object-based methods have also been widely used, notably the approach of Wernli and Schwierz \cite{wernli2006surface}, which identifies cyclones as SLP minima enclosed by closed contours, providing both location and spatial extent. More recent developments favor vorticity-based detection, where coherent structures are extracted from thresholded 850 hPa vorticity fields using connected-component labeling, as in Inatsu (2009) \cite{inatsu2009neighbor} and Flaounas et al. (2014) \cite{flaounas2014cyclotrack}. Alternative perspectives have emerged that emphasize the structural and energetic decomposition of atmospheric flow, such as the Okajima and Nakamura \cite{Okajima_2021} framework, which separates cyclonic and anticyclonic contributions. Despite these advances, intercomparison studies such as IMILAST \cite{neu2013imilast} have shown that cyclone statistics remain sensitive to the chosen detection and tracking methodology.

In this work, we introduce a complementary approach based on Topological Data Analysis (TDA), where cyclonic and anticyclonic structures are identified as topological features of evolving scalar fields.
In particular, we apply persistent homology to sea-level pressure anomalies over the Northern Hemisphere, providing a topological characterization of these pressure systems.
Persistent homology provides a quantitative way to characterize the “shape” of data by identifying their topological features, such as connected components, loops, or voids, and tracking their persistence at different threshold levels (Otter et al. 2017 \cite{Otter_2017}). 

A potential source of confusion arises regarding the term ``persistence.'' In meteorology and climate science, persistence typically refers to \textit{temporal duration}. In the context of TDA, however, it refers to \textit{topological persistence} (the robustness of a feature across scales). To avoid ambiguity, we adopt the term \textbf{topological depth} to describe the life span of a feature within a filtration, and the term ``duration'' for its temporal meaning.

Applying persistent homology to climate data is a relatively new endeavor,
and recent studies demonstrate its promise in revealing patterns that are hard to detect with conventional analyzes.
For example, Muszynski et al. (2019 \cite{Muszynski_2019}) used TDA to automatically recognize atmospheric river patterns in the output of the climate model, 
and Tymochko et al. (2020 \cite{Tymochko_2020})
quantified the diurnal cycle of hurricane convection by analyzing satellite images with persistent homology. 
Strommen et al. (2023 \cite{Strommen_2022})
recently showed that persistent homology can identify low-dimensional “weather regimes”
in midlatitude atmospheric flow – highlighting that the presence of multiple persistent topological holes
in the atmospheric state space corresponds to regimes such as the NAO phases.
These applications underscore that TDA is emerging as a powerful tool in geophysical and climate science, capable of capturing qualitative structural characteristics of complex datasets. 

In the context of spatial pressure anomaly fields, “loops” in the pressure field detect closed cycles, allowing the identification of patterns in which a pressure minimum is surrounded by higher values (cyclones) or a maximum is enclosed by lower values (anticyclones). By varying the isobar threshold, persistent homology further allows the detection and tracking of contiguous high-pressure regions (anticyclones) and low-pressure regions (cyclones), thereby quantifying the topological depth of these features.
This approach enables the objective isolation of cyclonic and anticyclonic structures, without the need for arbitrary filtering or tracking parameters. Their significance is quantified through the topological depth of the associated topological features, that is, the range of pressure thresholds over which a given high- or low-pressure region remains a distinct topological feature.
This approach offers several advantages: it reduces sensitivity to arbitrary threshold choices, handles merging and splitting events naturally through topological bifurcations, and provides a noise-robust, multiscale representation of the atmospheric ``skeleton.''
This perspective offers a unified and mathematically grounded description of atmospheric structures, complementing existing Lagrangian and Eulerian approaches.

In this study, we leverage persistent homology to examine over seven decades of Northern Hemisphere SLP anomalies (1948--2023). Our goal is to objectively characterize the recurrent topological structures generated by interacting pressure systems and to determine how the resulting loop structure reflects underlying climate variability. We address whether topological features can serve as robust markers of synoptic phenomena, such as persistent blocking highs or deep cyclonic lows.
This work is organized as follows: Section~\ref{sec:tda} provides the theoretical foundations of persistent homology and details the methodology for tracking topological features. Section~\ref{sec:data} describes the dataset and explains the mapping of topological definitions onto the physical structures of the Northern Hemisphere pressure field. Section~\ref{sec:result} presents our climatological findings, the topological characterization of blocking events, and a comparative analysis between the proposed TDA framework and the Murray--Simmonds algorithm. Finally, Sections~\ref{sec:disc} and~\ref{sec:conc} provide the discussion and concluding remarks, respectively.

\section{Methodology}\label{sec:tda}

\subsection{Topological foundations}
\begin{figure}
    \centering
    \includegraphics[width=0.3\linewidth]{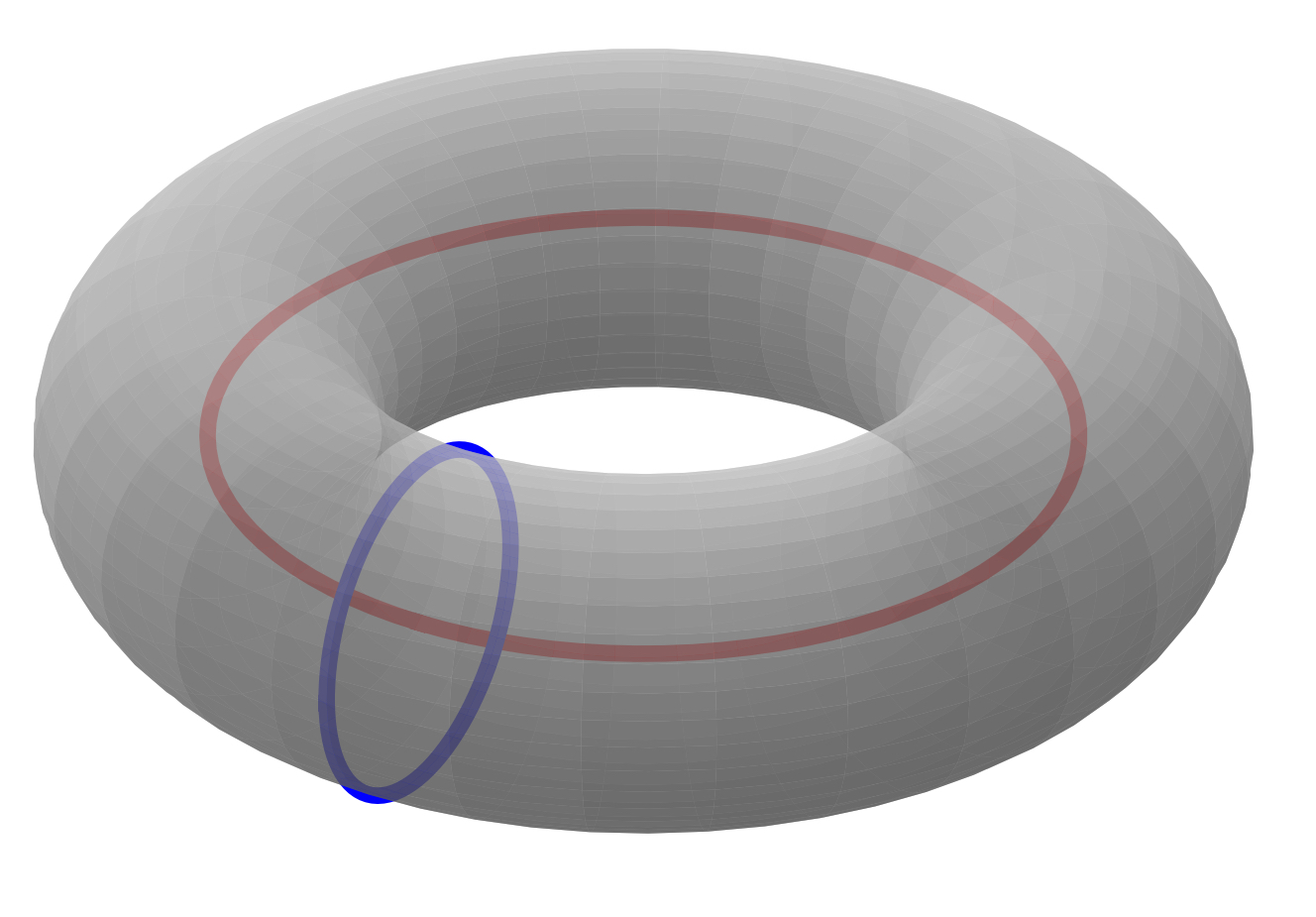}
    \caption{A hollow torus with two non-equivalent 1-holes (respectively blue and red circles).}
    \label{fig:torus}
\end{figure}


TDA focuses on understanding the underlying shape of data by examining how points are intrinsically organized within a topological space $\mathcal{Z}$. This structure in $\mathcal{Z}$ can be characterized across multiple degrees \( d \in \mathbb{N}_0 \).
One of the fundamental concepts in algebraic topology responsible for detecting and quantifying this organization is the notion of homology groups.
Homology groups, a set of scale-invariant topological descriptors,  encode information about the existence, number, and location of holes of different dimensions, known as: n- holes. 
In order to compute the homology groups, it is necessary to approximate the object \( \mathcal{Z} \) by a \emph{cell complex}. A cell complex is a structured collection of simple building blocks, called \emph{cells}, which are glued together in a consistent way to form an approximation of the original space. Each cell corresponds to a basic geometric element of a given dimension: 0-cells are points, 1-cells are line segments, 2-cells are filled polygons (e.g., squares or triangles), and higher-dimensional cells generalize this idea. By assembling these cells, it is possible to construct a combinatorial representation of the space that preserves its essential topological features while being mathematically tractable for homology computations.
The zero-order homology group \( \mathcal{H}_0(K) \) measures the connectivity of the complex, with its rank corresponding to the number of connected components (0-holes); \( \mathcal{H}_1(K) \) identifies non-trivial loops of 1-cells around the complex (1-holes); and \( \mathcal{H}_2(K) \) captures loops of 2-cells (voids or 2-holes). More generally, higher-order homology groups \( \mathcal{H}_d(K) \), with \( d \geq 3 \), correspond to \( d \)-holes \cite{kinsey2012topology}. In the case of a torus, there is one connected component (one 0-hole), two 1-holes and one void (one  2-hole) (see Figure \ref{fig:torus}). For a detailed account of the definitions of homology groups, see \ref{app:hom_groupcomplex}.

\subsection{Cubical representation of images}
\begin{figure}[ht]
    \centering
    \subfloat[Grayscale image]{
    \includegraphics[width=0.45\linewidth]{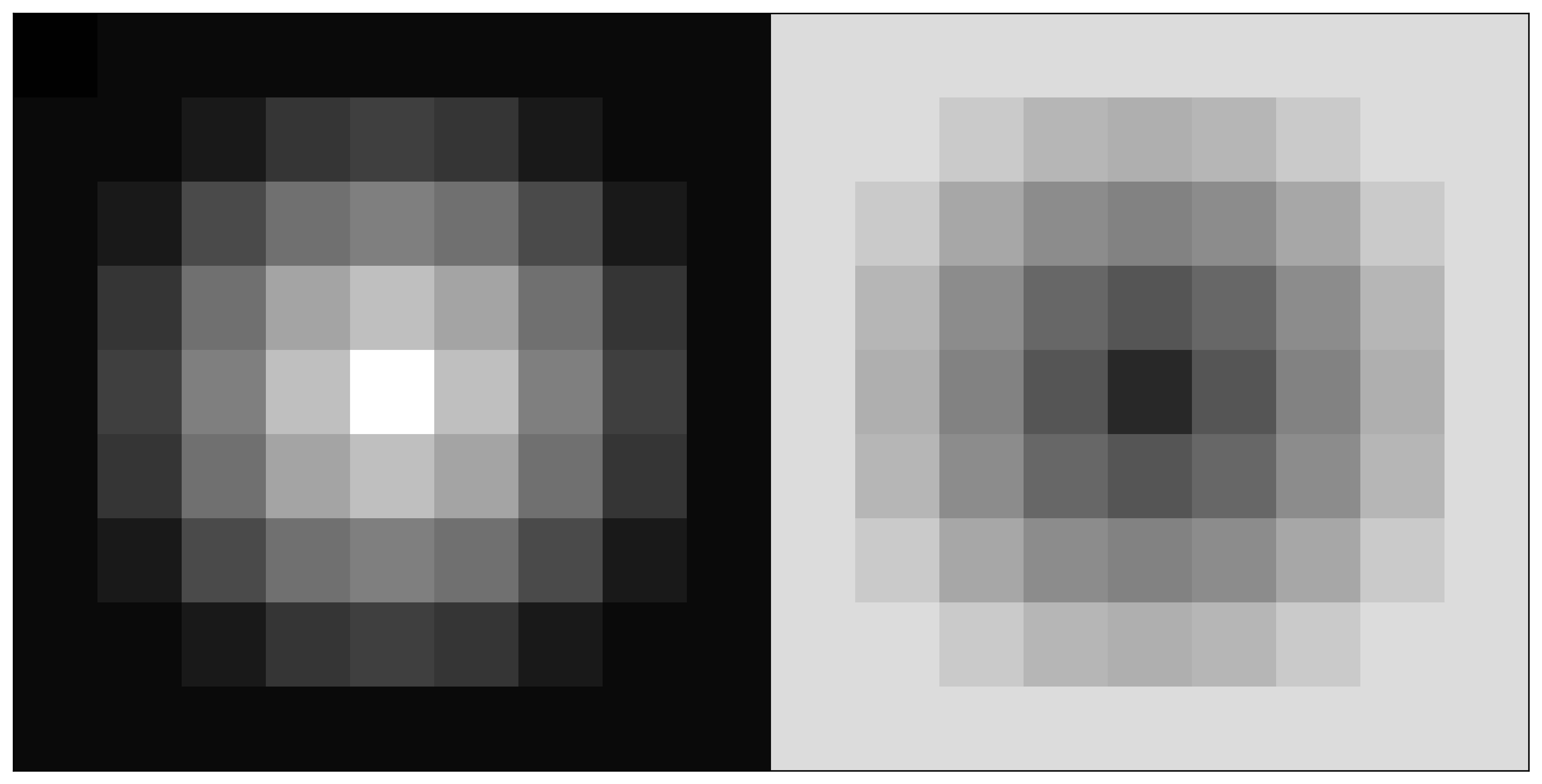}
    \label{subfig:gray}
    }
    \subfloat[Grid representation]{
    \includegraphics[width=0.45\linewidth]{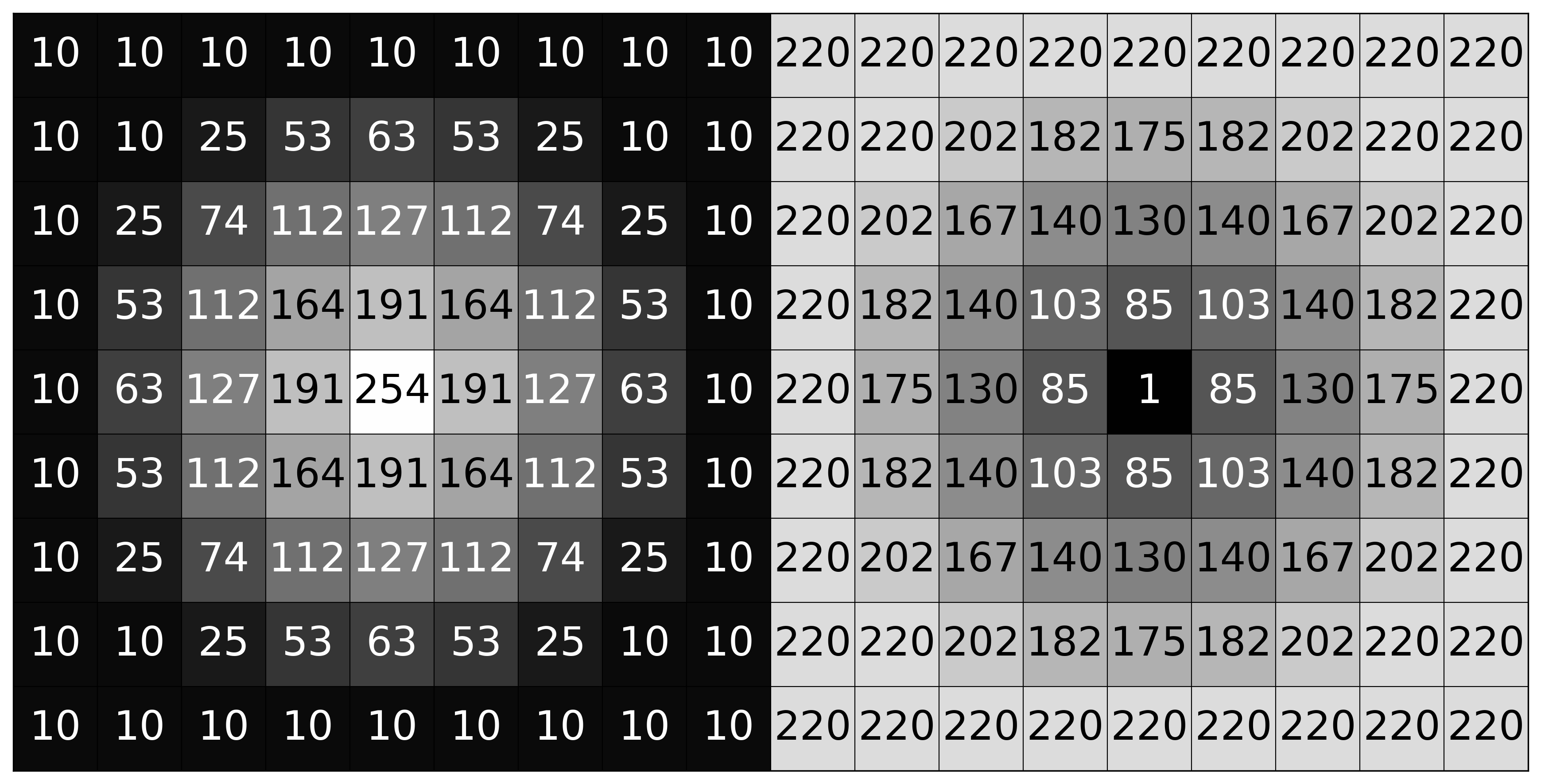}
    \label{subfig:tesl}
    }
    
    \caption{
    A grayscale image and its representation as a grid which is obtained after tesselation by congruent squares.
    Each square has a number associated to it corresponding to the intensity of the light.
    }
    \label{fig:gray_tesl}
\end{figure}

In this study, the object of interest is a grayscale image in \( \mathbb{R}^2 \). A grayscale image is defined as a collection of pixels, with each pixel recording the intensity of light.
A grayscale image can be divided into a grid obtained after tessellation by congruent squares, where each square of the grid corresponds to a pixel of the image. Subsequently, the light intensity of these pixels is associated with their corresponding squares, thus obtaining a gridded representation of the grayscale image, as illustrated in Figure \ref{fig:gray_tesl}.  These square grids are discretized into a cubical complex \( K \),  a specific type of cell complex designed for image data \cite{kaczynski2006computational,Wagner_2011}. A cubical complex is composed of 0-cubes (vertices), 1-cubes (edges), 2-cubes (squares), 3-cubes, and,  in general, l-cubes for $l\ge 4$. A formal definition of cubical complexes is provided in \ref{app:cubical-complexes}.

In the context of grayscale images, constructing a sequence of nested subcomplexes and computing the \( n \)-holes for each subcomplex allows us to track how topological features evolve over a scalar field. 
Filtration is a way to construct this sequence of complexes, and the notion of persistent homology is the correct way to capture homological features across filtration.

\subsection{Filtrations induced by intensity}
The cubical complex defined by the tessellation in Figure \ref{fig:gray_tesl} consists of all unit squares together with their edges and vertices. Each square is associated with a number representing its light intensity. The \emph{intensity function} \( I \) assigns to each $l$-dimensional elementary cube $u$ of a cubical complex $K \subset \mathbb{R}^d$ a value \( I(u) \) satisfying \( 0 \le I(u) \le 255 \). Note that edges and vertices, although part of the cubical complex, are not assigned intensity values. 
We extend $I$ to all faces $\sigma \subseteq K$ (of any dimension) by defining
\[
\I_{\min}(\sigma) \coloneqq 
\min\big\{\, I(\tau_i) : \sigma \text{ is a face of the } \tau_i \text{ $l$-cubes} \,\big\}.
\]
\[
\I_{\max}(\sigma) \coloneqq 
\max\big\{\, I(\tau_i) : \sigma \text{ is a face of the } \tau_i \text{ $l$-cubes} \,\big\}.
\]
For $l$-dimensional cubes, the definition reduces to $\I_{\min}(\tau_i) =\I_{\max}(\tau_i) = I(\tau_i)$.

\begin{definition}
For each threshold \( t \in \mathbb{R} \):
\[
K_{t}^{\text{sub}} \coloneqq 
\{\tau \in K : \I_{\min}(\tau) \le t \}
\quad \text{and} \quad
K^{\text{super}}_{t} \coloneqq 
\{\tau \in K : \I_{\max}(\tau) \ge t \}
\]
define the \emph{sublevel} and \emph{superlevel filtered cubical complexes}\cite{kaczynski2006computational}, respectively.

For increasing values of $t$, $t_1 < t_2 < \cdots < t_m$. The superlevel-set (sublevel-set) filtration shows how the topology of the domain changes as $t$ decreases (increases) (see Figure \ref{fig:syn_pd_filt}).

The corresponding filtrations are the nested sequences
\[
K_{t_1}^{\text{sub}} \subseteq K_{t_2}^{\text{sub}} \subseteq \cdots \subseteq K_{t_m}^{\text{sub}}, 
\quad
K_{t_1}^{\text{super}} \supseteq K_{t_2}^{\text{super}} \supseteq \cdots \supseteq K_{t_m}^{\text{super}} \]
\end{definition}

In this work we focus exclusively on 1-holes, since the features of interest, cyclonic and anticyclonic structures, manifest as closed loops rather than as connected components. For this reason, we develop and present the theoretical framework only at the level of 1-dimensional homology.

\subsection{Persistent homology in dimension 1}
\begin{definition}
For $0 \le i \le j \le n$, let 
$i^{1}_{ij}\colon H_1(K_i)\to H_1(K_j)$ 
be the map induced in first homology. 
The $(i,j)$-persistent first homology group is
\[
H^{ij}_1=\operatorname{Im}(i^{1}_{ij}).
\]
\end{definition}

Persistent homology \cite{kaczynski2006computational} keeps track of the 1-cycles that have not yet become a boundary of any square, i.e., these cycles are 1-holes in the object.
The filtration step at which the 1-hole first appears is called its \textit{birth}, and the step at which it disappears is called its \textit{death}.

\begin{definition}
A persistence diagram $\operatorname{Dgm}$ for the 1- holes is a collection of  points $\{(b_j,d_j)\}_j$,  $0 \le j \le n$ where each point corresponds to a $1$-hole that appears at parameter value $b_j$ (birth) and disappears at parameter value $d_j$ (death) in the filtration.
\end{definition}
\begin{definition}
The \emph{topological persistence} of the $j$-th $1$-hole
is defined as 
\[
\mathrm{pers_{top}}(j) = |d_j - b_j|,
\]
The topological persistence quantifies how long this particular 1-hole
survives throughout the filtration.
\end{definition}

\begin{definition}
Given a persistence diagram $\operatorname{Dgm}$, the \emph{p-order total persistence} ~\cite{cohen2010lipschitz} $\mathrm{TP}_p$  ($p>0$) is defined as
\[
\mathrm{TP}_p(\operatorname{Dgm})
= \sum_{j} \mathrm{pers_{top}}(j) ^p
= \sum_{j} |d_j - b_j|^p.
\]
\end{definition}
The first-order total persistence, $\mathrm{TP}_1$ provides a scalar measure of the overall topological activity 
in dimension $1$, and can be interpreted as a measure of the overall topological organization present in the field.  In contrast, the second-order total persistence, $\mathrm{TP}_2$, assigns greater weight to highly persistent features and is therefore more sensitive to dominant large-scale structures. While $\mathrm{TP}_1$ reflects the total amount of topological signal distributed across all cycles, $\mathrm{TP}_2$ emphasizes the contribution of the most robust features. The combined analysis of $\mathrm{TP}_1$ and $\mathrm{TP}_2$ provides information on whether the topological organization is distributed among many moderately persistent structures or concentrated in a few highly persistent ones.

Persistence diagrams are stable under small perturbations or random noise in the input data \cite{cohen2005stability}. 
For a sublevel-set filtration, the points lie above the \textit{diagonal} $\Delta$ ($x=y$), while for a superlevel-set filtration, they appear below it. 
The distance of a point from the diagonal quantifies the \emph{persistence} of the corresponding n-hole. 
Points farthest from the diagonal correspond to the most persistent features, although multiple features may share this distinction if they exhibit the same maximal persistence.

\begin{figure}

    \subfloat[Superlevel-set filtration]{
    \centering
    \begin{tikzpicture}
        \draw[<-, thick] (0,-1) -- (0,5.5);
        
        \node[left] at (0,-0.5) {0};
        \node[left] at (0,0.8) {10};
        \node[left] at (0,2.5) {100};
        \node[left] at (0,4) {200};
        \node[left] at (0,5.5) {253};

        \draw[thick] (-0.1,-0.5) -- (0.1,-0.5);  
        \draw[thick] (-0.1,0.8) -- (0.1,0.8);  
        \draw[thick] (-0.1,2.5) -- (0.1,2.5); 
        \draw[thick] (-0.1,4) -- (0.1,4);  
        \draw[thick] (-0.1,5.5) -- (0.1,5.5);     
        \node[rotate=90] at (-1,2.5) {Intensity};
        \node at (2.75,2.5) {\includegraphics[width = 0.17\linewidth]{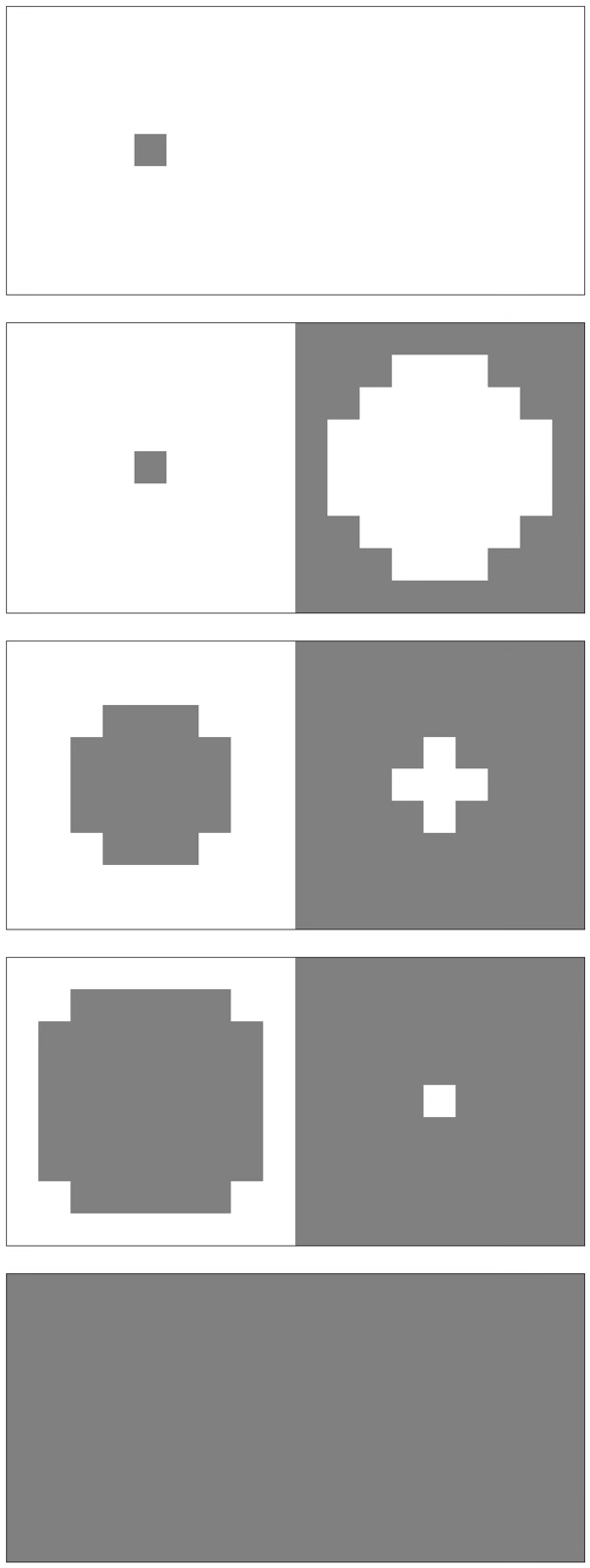}};

    \end{tikzpicture}
    \label{subfig:syn_filt_sup}
    }
    \subfloat[Persistence Diagram]{
    {\includegraphics[width = 0.5\linewidth]{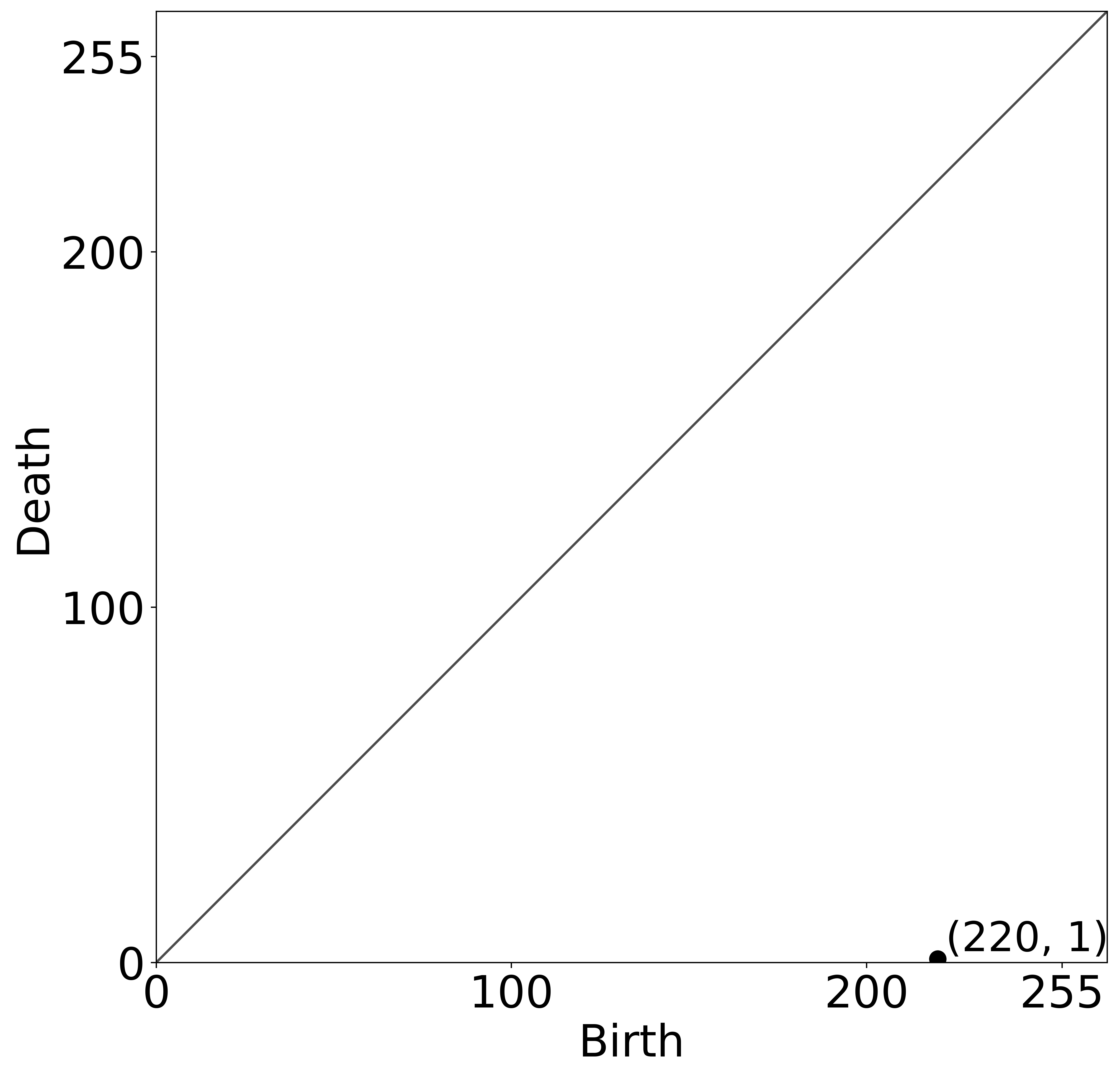}}
    \label{subfig:syn_pd_filt_sup}
    }
    
    \subfloat[Sublevel-set filtration]{
    \centering
    \begin{tikzpicture}
        \draw[->, thick] (0,-0.5) -- (0,6);
        
        \node[left] at (0,-0.5) {2};
        \node[left] at (0,0.8) {10};
        \node[left] at (0,2.5) {100};
        \node[left] at (0,4) {200};
        \node[left] at (0,5.5) {255};

        \draw[thick] (-0.1,-0.5) -- (0.1,-0.5);  
        \draw[thick] (-0.1,0.8) -- (0.1,0.8);  
        \draw[thick] (-0.1,2.5) -- (0.1,2.5); 
        \draw[thick] (-0.1,4) -- (0.1,4);  
        \draw[thick] (-0.1,5.5) -- (0.1,5.5);     
        \node[rotate=90] at (-1,2.5) {Intensity};
        
        \node at (2.75,2.5) {\includegraphics[width = 0.17\linewidth]{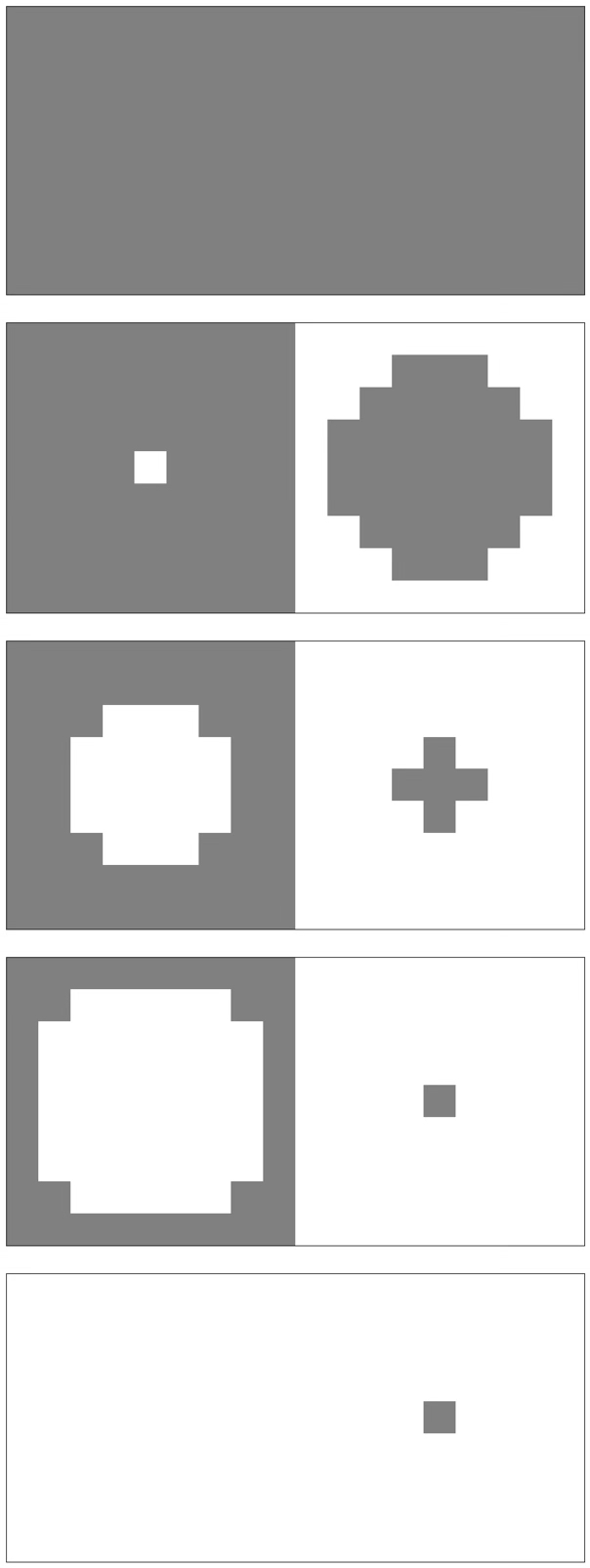}};

    \end{tikzpicture}
    \label{subfig:syn_filt_sub}
    }
    \subfloat[Persistence Diagram]{
    {\includegraphics[width = 0.5\linewidth]{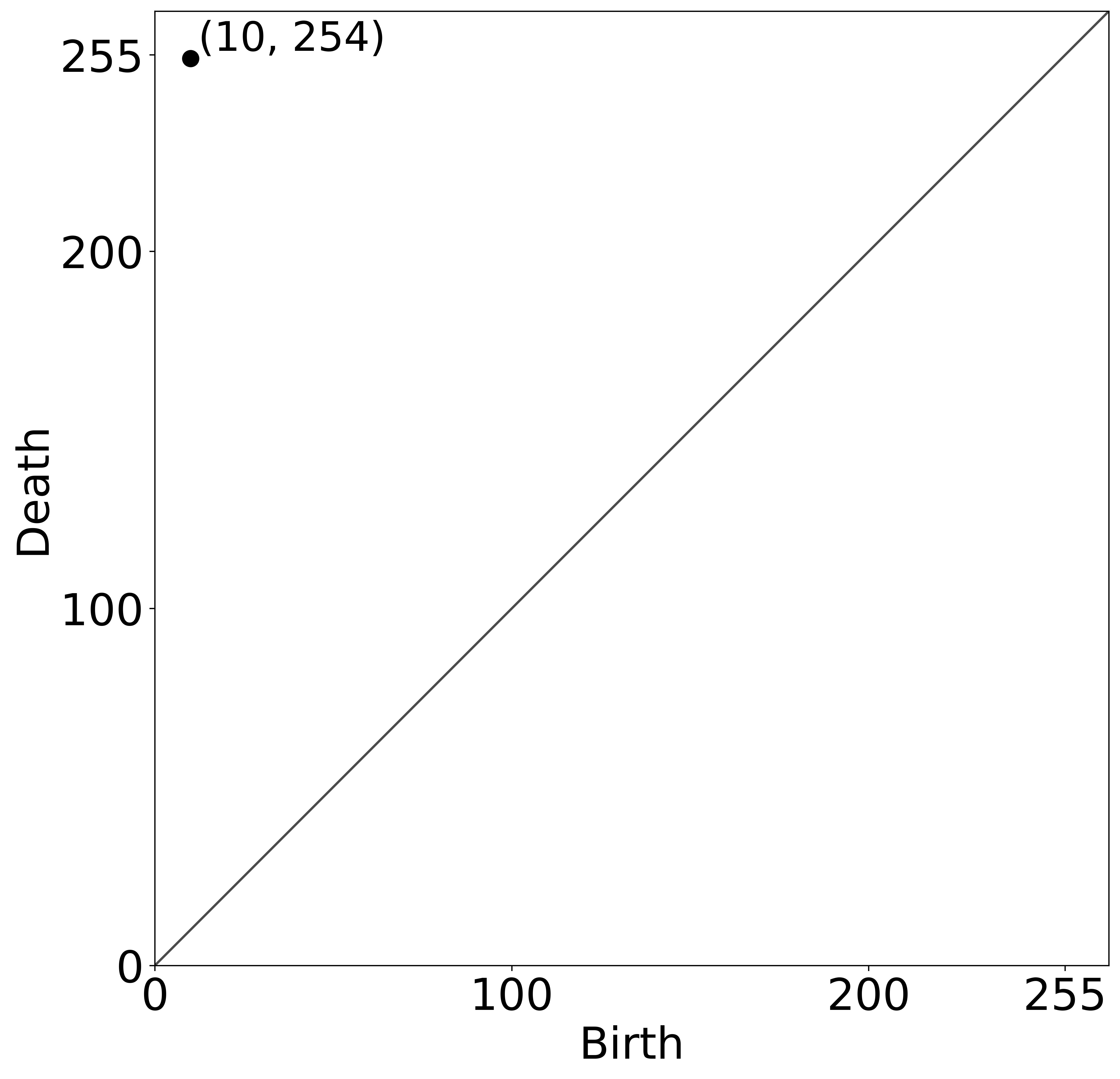}}
    \label{subfig:syn_pd_filt_sub}
    }
    
    \caption{Filtered cubical complexes (shown in gray) and their corresponding persistence diagrams. In all cases, birth and death values are expressed in terms of intensity. The top row corresponds to the superlevel-set filtration, whereas the bottom row corresponds to the sublevel-set filtration.}
    \label{fig:syn_pd_filt}
\end{figure}

Figure~\ref{fig:syn_pd_filt} illustrates the two filtrations considered in this work, namely the superlevel-set and sublevel-set filtrations, applied to the grayscale image shown in Figure~\ref{fig:gray_tesl}. The corresponding persistence diagrams are displayed alongside the filtered cubical complexes.
The superlevel-set cubical filtration captures the 1-hole on the right, represented by the point $(220,1)$ in (b). 
In the sublevel-set cubical filtration (c), the 1-hole on the left is captured and represented by the point $(10,254)$ in (d).

Of particular significance is the interpretation of the death value in each persistence pair. In a cubical filtration, the death of a 1-hole is associated with the addition of a unique square to the complex. For example, in the superlevel-set filtration and for the persistence pair $(220,1)$, there is a square with intensity $1$ whose inclusion causes the corresponding $1$-hole to disappear. Similarly, in the sublevel-set filtration and for the pair $(10,254)$, there is a square with intensity $254$ responsible for the death of that $1$-hole. We refer to these cells as \emph{death squares}.

While each 1-hole has a unique death square, this square does not necessarily coincide with the absolute intensity extremum of the corresponding region. Rather, it identifies the location at which the topological feature closes and therefore contains information about the local structure of the intensity field in the vicinity of the 1-hole.

When applied to atmospheric pressure fields, death squares acquire a natural physical interpretation. In superlevel-set filtrations, they tend to be associated with local low-pressure centers, whereas in sublevel-set filtrations they tend to be associated with local high-pressure centers. Consequently, death squares provide physically meaningful reference points that can be exploited for cyclone and anticyclone tracking.

\subsection{Temporal tracking of topological features}\label{sec:topological_tracking}
In order to compare topological features across a time series of datasets, we consider a sequence of 1-persistence diagrams. Each 1-persistence diagram may contain a lot of non-equivalent 1-holes, and understanding their temporal evolution requires a method to associate 1-holes from one diagram to the next. In this section, we employ optimal matching to track topological features across consecutive diagrams, enabling the analysis of their dynamics along the timeline. This approach naturally leads to the use of the Wasserstein distance, which quantifies the minimal cost of matching points between diagrams and provides a rigorous metric for comparing their topological structures.

To compare two persistence diagrams quantitatively, we need to define a distance between them.
A key step in defining such a distance is specifying a method to match points between the diagrams. However, persistence diagrams may have different numbers of points. To address this, we augment each diagram with points in $\Delta$, corresponding to features with birth equal to death. Since such features are topologically insignificant, this augmentation does not alter the diagram.

\begin{definition}
Let $f_1$ and $f_2$ be two filtrations, and let $\dgm(f_1)$ and $\dgm(f_2)$ denote their corresponding persistence diagrams. A \emph{matching} M between them is a set
\[
M \subseteq (\dgm(f_1) \cup \Delta) \times (\dgm(f_2) \cup \Delta),
\]
 The matching $M$ satisfies the property that each point $a \in \dgm(f_1)$ appears in exactly one pair in $M$, either matched with a point in $\dgm(f_2)$ or with a point on $\Delta$, and similarly for each $b \in \dgm(f_2)$. Points on the diagonal $\Delta$ may appear in multiple pairs.
\end{definition}

Once a matching has been established, a corresponding cost can be assigned. Intuitively, this cost should increase with the distance between the matched points: pairs of points that are closer to each other incur a lower cost, whereas pairs that are farther apart incur a higher cost.

\begin{definition}
The cost of a matching $M$ \cite{cohen2010lipschitz} is defined as
\[
\mathrm{cost}(M)
=
\sum_{(p,q)\in M}
\|p-q\|_1
=
\sum_{(p,q)\in M}
\left(
|b_p-b_q|
+
|d_p-d_q|
\right).
\]
\end{definition}
Using this notion of cost, we can define a distance between persistence diagrams by seeking the optimal matching that minimizes it.

\begin{definition}
Let $\dgm(f_1)$ and $\dgm(f_2)$ be two persistence diagrams.
The \emph{1-Wasserstein distance} \cite{cohen2010lipschitz} between them is defined as $$W_{1}(\dgm(f_1),\dgm(f_2))\coloneqq\inf_{M}\cost(M)$$
where the infimum is taken over all matchings $M$ between $\dgm(f_1)$ and $\dgm(f_2)$.
\end{definition}

In this work, we consider two types of persistence diagrams corresponding to super and sublevel-set filtrations. Matching and Wasserstein distances are computed only between diagrams of the same type.

Given a sequence of 1-persistence diagrams, each diagram may contain different 1-holes. We employ optimal matching to analyze the temporal evolution of these 1-holes.
For each pair of consecutive diagrams, we compute the optimal matching and use it to follow the corresponding 1-holes along the timeline. 
To ensure physically meaningful correspondences, we apply a distance threshold to the matched pairs:
if the death squares of two matched 1-holes are spatially separated by more than a prescribed distance,
the match is rejected and the trajectory is not continued. 
This filtering step ensures that only 1-holes whose death squares remain in close spatial proximity are linked across consecutive diagrams, 
preventing spurious long-range associations that would not correspond to the continuous evolution of a feature.

\begin{figure}
    \centering
    \subfloat[Sequence of grayscale images]{
        \includegraphics[width = 0.8\linewidth]{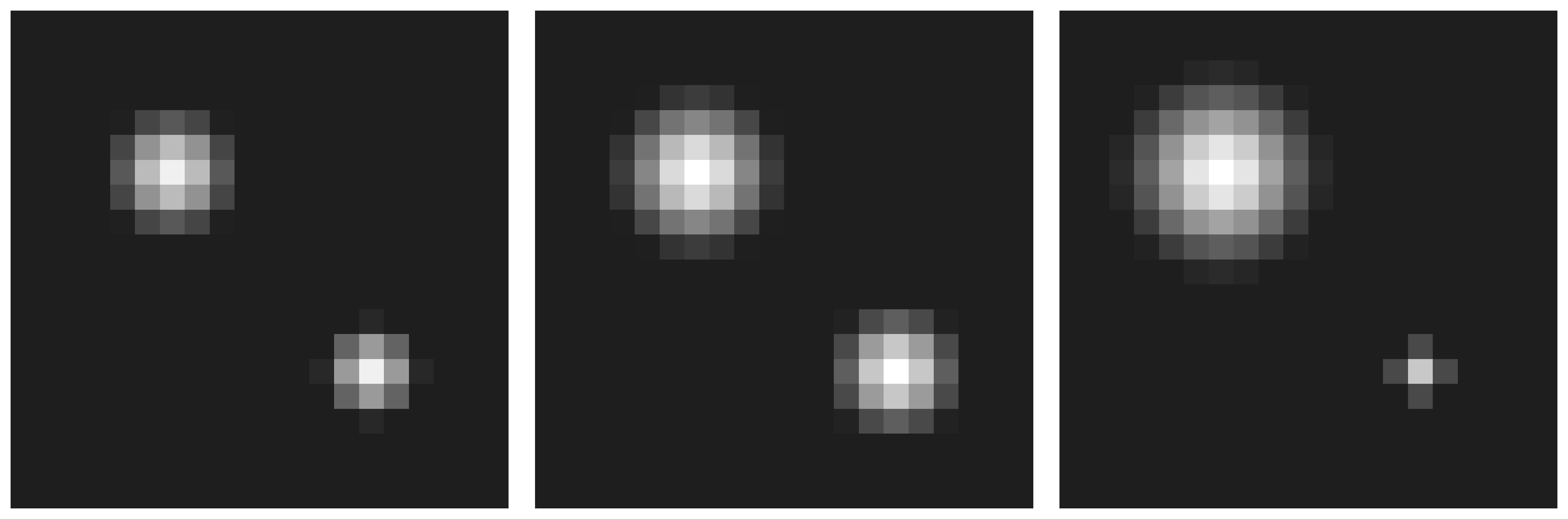} }
    
    \subfloat[Sequence of persistence diagrams]{
    \centering
    \begin{tikzpicture}[scale=0.5]
    \begin{axis}[
        xlabel=birth,
        ylabel=death,
        axis lines=middle,
        xmin=0, xmax=255,
        ymin=0, ymax=255,
        grid=none,
        axis equal,
        ticks=both,
        legend pos=south east,
        xlabel style={at={(axis cs:180,-100)}, font=\Huge},
        ylabel style={at={(axis cs:-110,150)}, font=\Huge, rotate=90},
        xticklabel style={font=\Huge, rotate=90},
        yticklabel style={font=\Huge}
    ]
    \addplot[domain=0:255, black] {x};
    \addplot[only marks, mark=*, orange, mark size=4pt] coordinates {
        (100,190)
    };
    \addplot[only marks, mark=*, green, mark size=4pt] coordinates {
        (70,140)
    };
    \end{axis}
    \end{tikzpicture}
    \begin{tikzpicture}[scale=0.5]
    \begin{axis}[
        xlabel=birth,
        ylabel=death,
        axis lines=middle,
        xmin=0, xmax=255,
        ymin=0, ymax=255,
        grid=none,
        axis equal,
        ticks=both,
        legend pos=south east,
        xlabel style={at={(axis cs:180,-100)}, font=\Huge},
        ylabel style={at={(axis cs:-110,150)}, font=\Huge, rotate=90},
        xticklabel style={font=\Huge, rotate=90},
        yticklabel style={font=\Huge}
    ]
    \addplot[domain=0:255, black] {x};
    \addplot[only marks, mark=*, orange, mark size=4pt] coordinates {
        (100,230)
    };
    \addplot[only marks, mark=*, green, mark size=4pt] coordinates {
        (70,150)
    };
    \end{axis}
    \end{tikzpicture}
    \begin{tikzpicture}[scale=0.5]
    \begin{axis}[
        xlabel=birth,
        ylabel=death,
        axis lines=middle,
        xmin=0, xmax=255,
        ymin=0, ymax=255,
        grid=none,
        axis equal,
        ticks=both,
        legend pos=south east,
        xlabel style={at={(axis cs:180,-100)}, font=\Huge},
        ylabel style={at={(axis cs:-110,150)}, font=\Huge, rotate=90},
        xticklabel style={font=\Huge, rotate=90},
        yticklabel style={font=\Huge}
    ]
    \addplot[domain=0:255, black] {x};
    \addplot[only marks, mark=*, orange, mark size=4pt] coordinates {
        (90,240)
    };
    \addplot[only marks, mark=*, green, mark size=4pt] coordinates {
        (50,120)
    };
    \end{axis}
    \end{tikzpicture}
    }

    \subfloat[Sequence of grayscale images with representative 1-cycles]{
        \includegraphics[width = 0.8\linewidth]{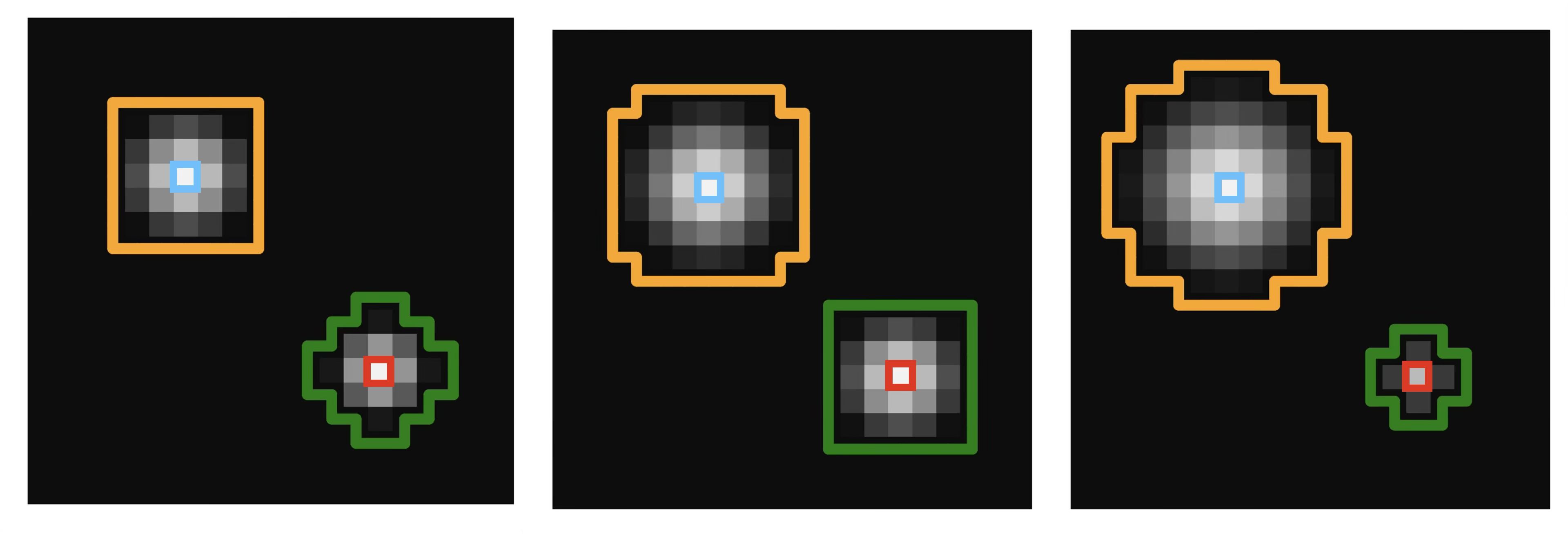}
    }
    \caption{
(a) Sequence of grayscale images.
(b) Sequence of persistence diagrams, with 1-holes tracked across the timeline using optimal matching. Birth and death values correspond to intensity levels.
(c) Sequence of grayscale images showing the regions associated with the tracked 1-holes, outlined by orange and green contours. The death squares corresponding to each 1-hole are also indicated by blue and red markers.}   
    \label{fig:syn_pd_tracking}
\end{figure}
Figure~\ref{fig:syn_pd_tracking} shows a sequence of three grayscale images, each containing three holes of varying size. The corresponding persistence diagrams reveal two points, one with greater topological persistence than the other. Using optimal matching, we track these features across the sequence of diagrams and plot the 1-holes in orange and green contours.
We also plot death squares for each 1-holes, we will later use it to filter out some matching.

\section{Data Description and Topological Implementation} \label{sec:data}

We collected daily Sea-Level Pressure (SLP) data from the NCEP-NCAR Reanalysis 1 dataset provided by the National Centers for Environmental Prediction (NCEP) and the National Center for Atmospheric Research (NCAR) for the Northern Hemisphere ($0\degree N-90\degree N$, all longitudes), spanning 1 January 1948 to 31 December 2023.
To isolate the transient weather systems from the seasonal cycle, which is the dominant source of non-stationarity in the SLP field, we compute SLP anomalies. For each grid point, the anomaly is defined as the deviation of the daily value from the 76-year mean corresponding to that specific calendar day. This procedure is primarily designed to remove the seasonal climatology rather than long-term trends: unlike temperature or moisture variables, Northern Hemisphere SLP does not exhibit a significant secular trend over the 1948--2023 period that would confound the identification of topological features. The standard daily-anomaly calculation is therefore sufficient to provide a stable reference state for isolating synoptic-scale variability, without introducing additional detrending steps that could create artifacts in the persistence diagrams.

\begin{figure}
    \centering
    \includegraphics[width=0.6\linewidth]{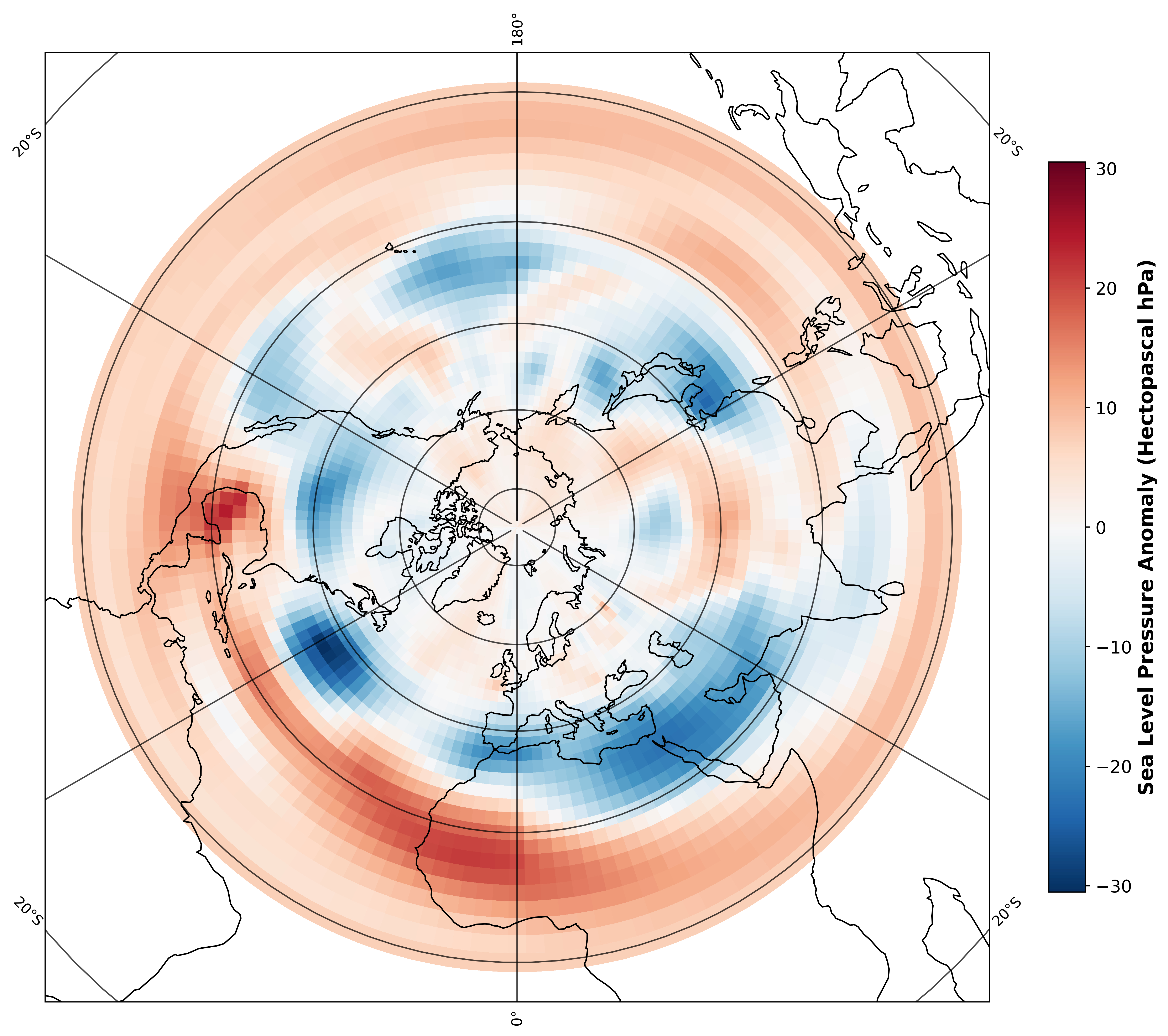}
    \caption{Sea level pressure (SLP) anomalies over the Northern Hemisphere on 31 December 2022. Blue shading indicates cyclonic regions, and red shading indicates anticyclonic regions.}
    \label{fig:global_projection}
\end{figure}

We analyze SLP anomalies to study cyclonic (negative) and anticyclonic (positive) regions. Cyclonic regions embedded within anticyclonic regions reflect dynamics governing storm formation and intensity in the Northern Hemisphere (see Figure ~\ref{fig:global_projection})\cite{holton2013introduction}. 
The resulting anomaly fields are treated as grayscale images, where intensity values correspond to pressure deviations in hectopascals (hPa). These fields are mapped onto a square grid to construct the cubical complexes required for persistent homology computations.

\subsection{Identification of 1-Cyclones and 1-Anticyclones}
The core of our methodology lies in the dual use of filtrations to capture the full spectrum of atmospheric pressure systems:

\begin{itemize}
    \item \textbf{1-Anticyclones:} Identified as persistent 1-holes through a sublevel-set cubical filtration. These represent coherent regions of positive pressure anomalies.
    \item \textbf{1-Cyclones:} Identified as persistent 1-holes through a superlevel-set cubical filtration. These represent coherent regions of negative pressure anomalies.
\end{itemize}

\begin{figure}
    \centering
    \begin{minipage}{0.5\linewidth}
    \subfloat[Superlevel-set filtration]{
        \centering
        \begin{tikzpicture}
            \node[inner sep=0pt] (figures) {
                \begin{minipage}{0.8\linewidth}
                    \centering
                    \begin{minipage}{\linewidth}
                        \centering
                        \scriptsize 10 hPa\\
                        \includegraphics[width=0.3\linewidth]{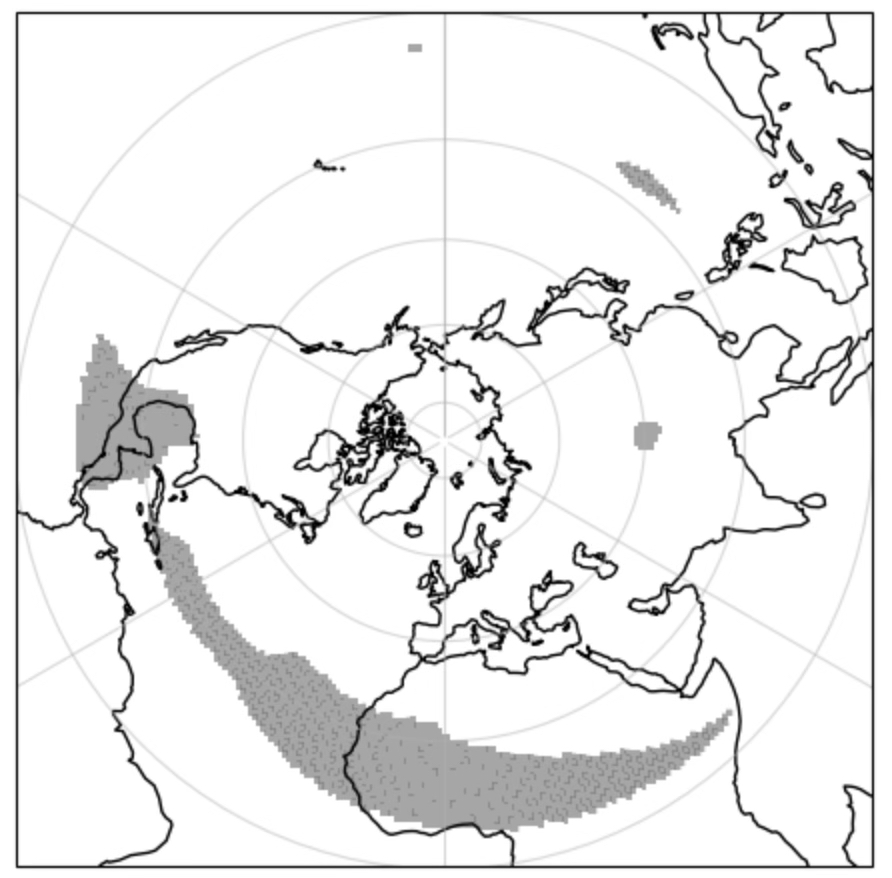}
                    \end{minipage}\\[2.5mm]
                    \begin{minipage}{\linewidth}
                        \centering
                        \scriptsize 0 hPa\\
                        \includegraphics[width=0.3\linewidth]{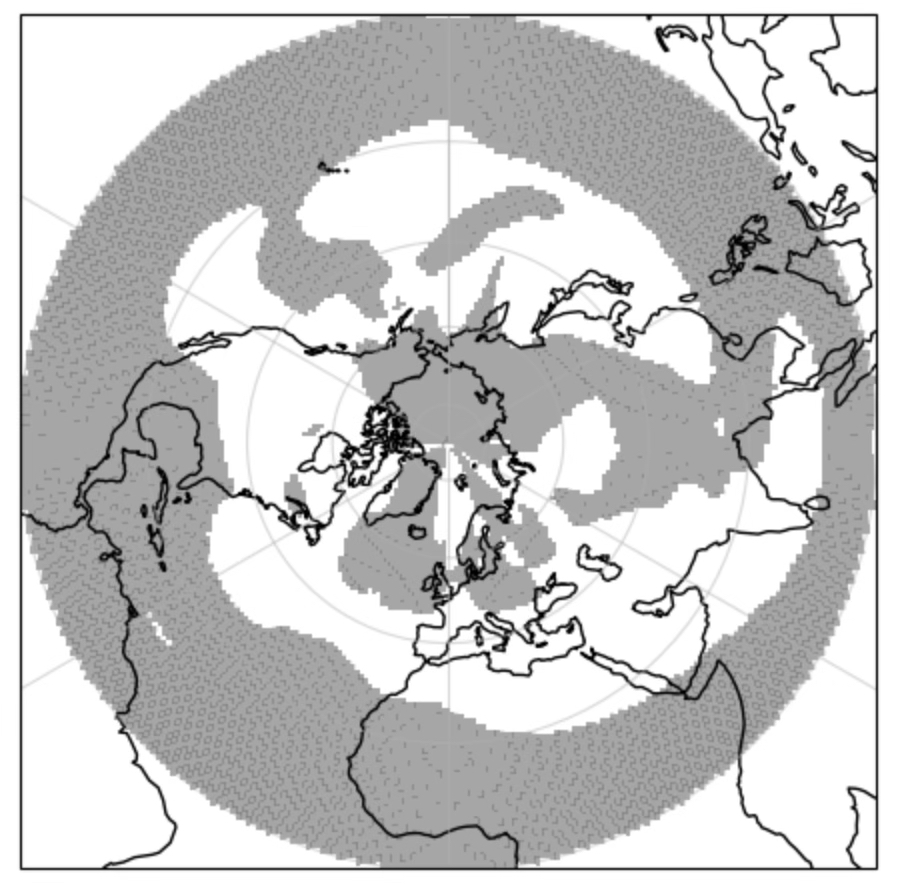}
                    \end{minipage}\\[2.5mm]
                    \begin{minipage}{\linewidth}
                        \centering
                        \scriptsize -10 hPa\\
                        \includegraphics[width=0.3\linewidth]{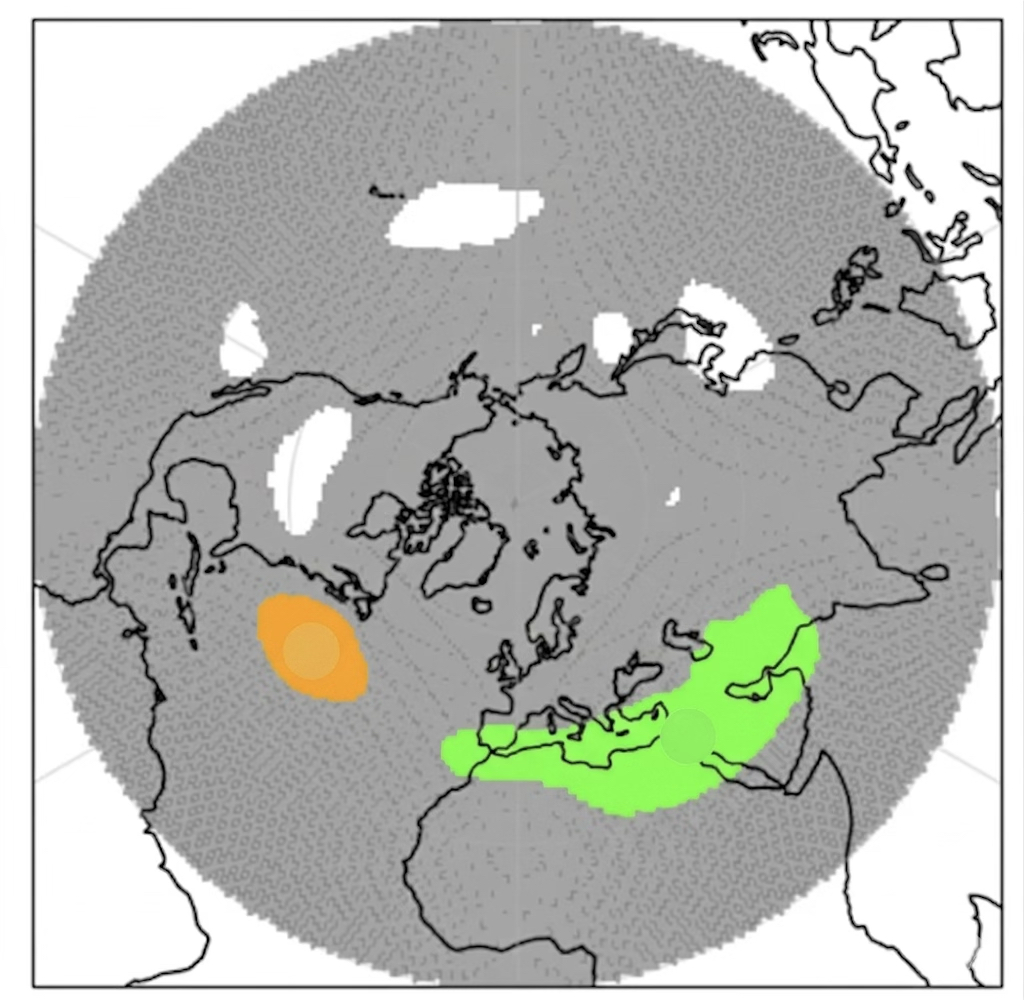}
                    \end{minipage}\\[2.5mm]
                    \begin{minipage}{\linewidth}
                        \centering
                        \scriptsize -20 hPa\\
                        \includegraphics[width=0.3\linewidth]{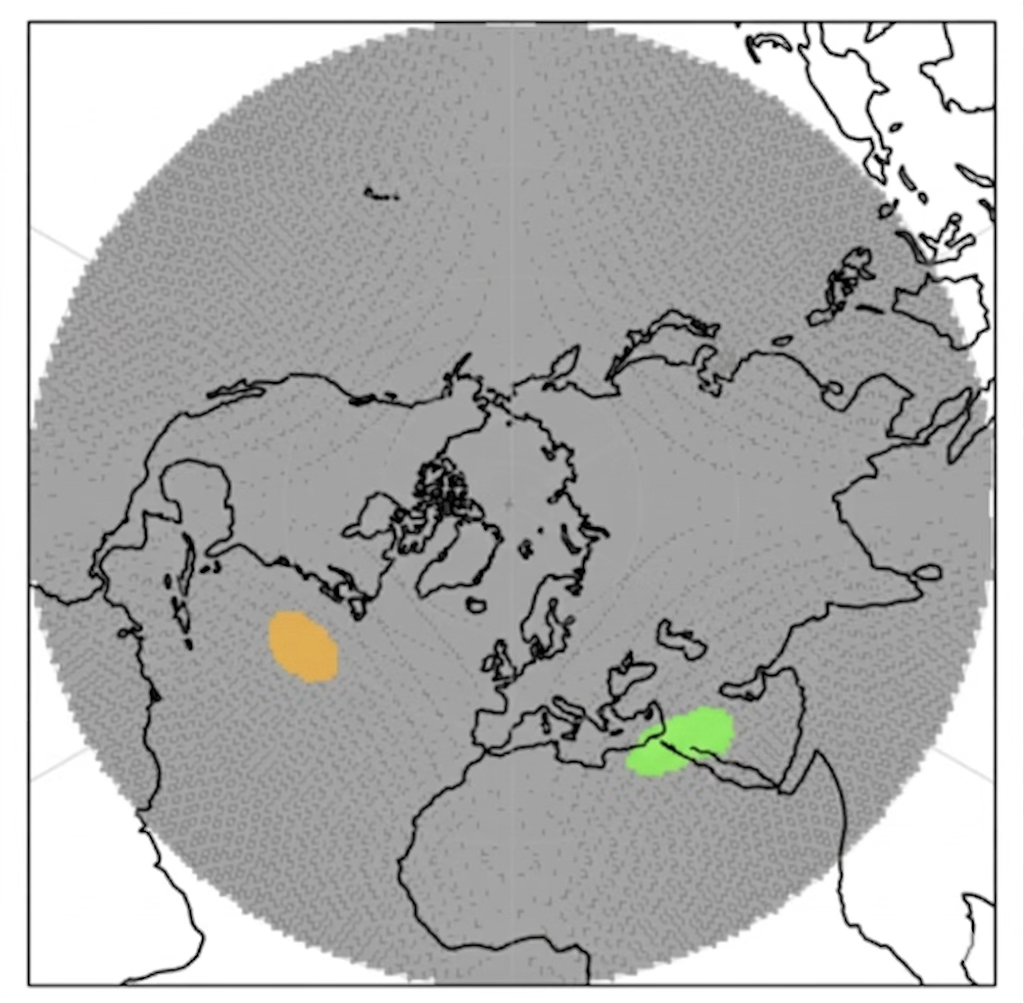}
                    \end{minipage}\\[2mm]
                \end{minipage}
            };
            \draw[->, thick] ([xshift=18mm]figures.north west) -- ([xshift=18mm]figures.south west);
        \end{tikzpicture}
        \label{subfig:filt_dec}
    }
    \end{minipage}
    \begin{minipage}{0.45\linewidth}
        \subfloat[Persistence diagram for the superlevel-set filtration.]{
            \centering
            \includegraphics[width=0.9\linewidth]{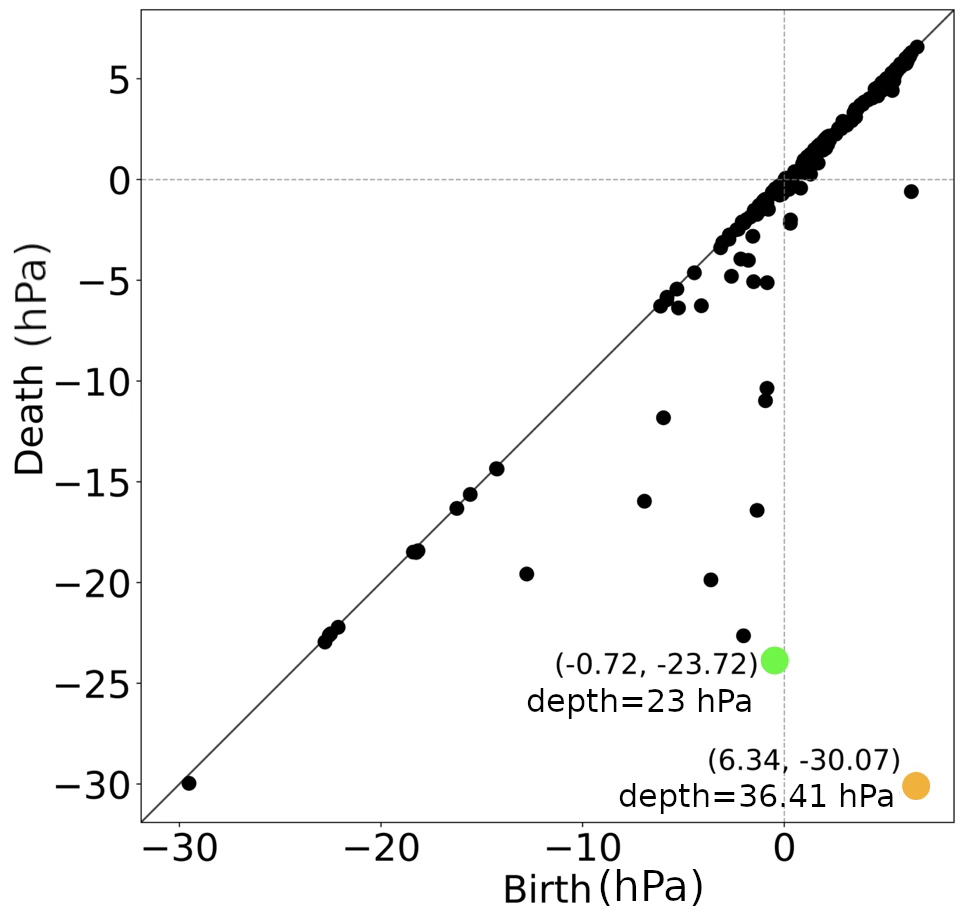}
            \label{subfig:pd_filt_dec}
        }
    \end{minipage}

    \begin{minipage}{0.5\linewidth}
    \subfloat[Sublevel-set filtration]{
        \centering
        \begin{tikzpicture}
            \node[inner sep=0pt] (figures) {
                \begin{minipage}{0.8\linewidth}
                    \centering
                    \begin{minipage}{\linewidth}
                        \centering
                        \scriptsize 20 hPa\\
                        \includegraphics[width=0.3\linewidth]{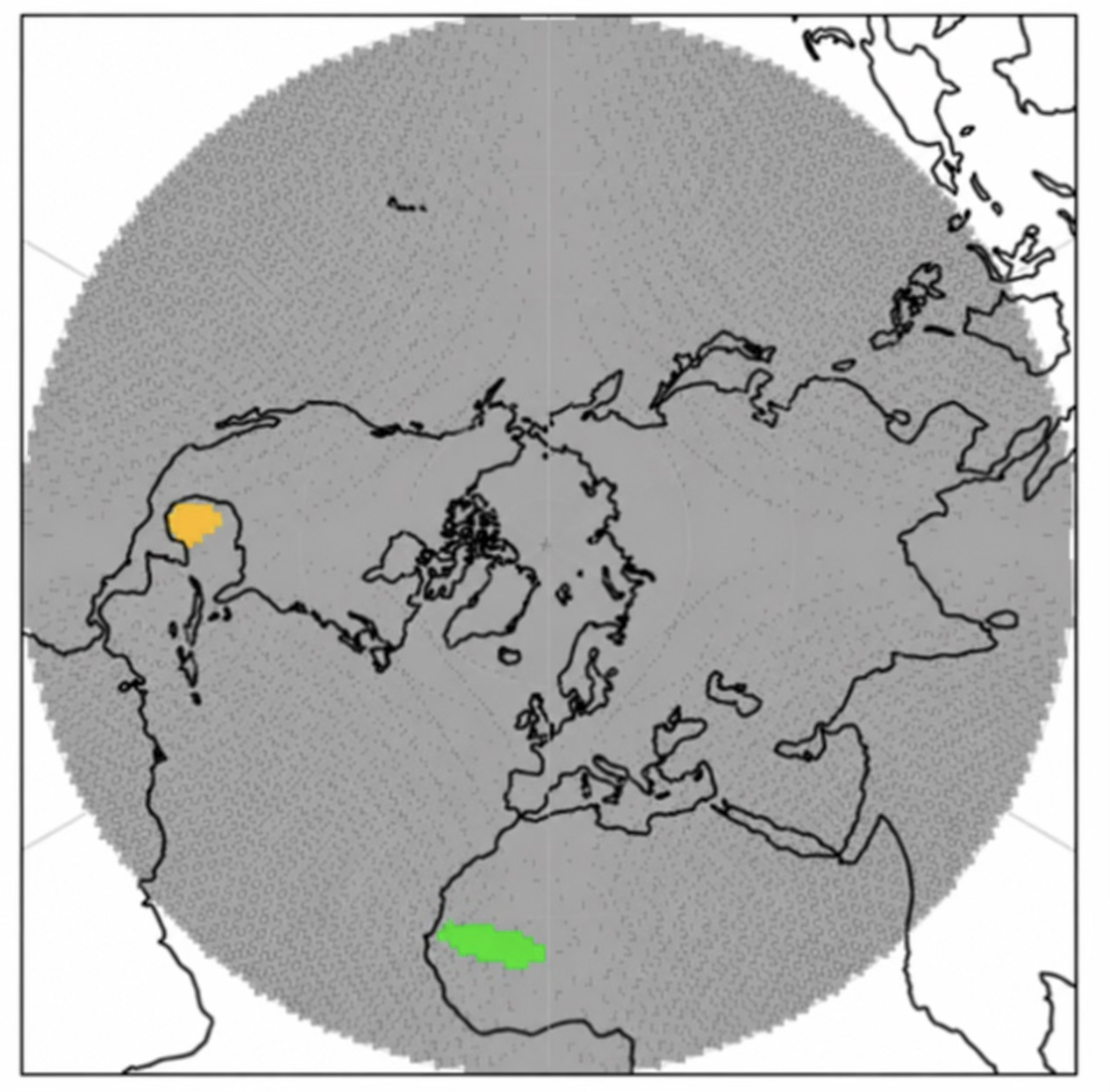}
                    \end{minipage}\\[2mm]
                    \begin{minipage}{\linewidth}
                        \centering
                        \scriptsize 10 hPa\\
                        \includegraphics[width=0.3\linewidth]{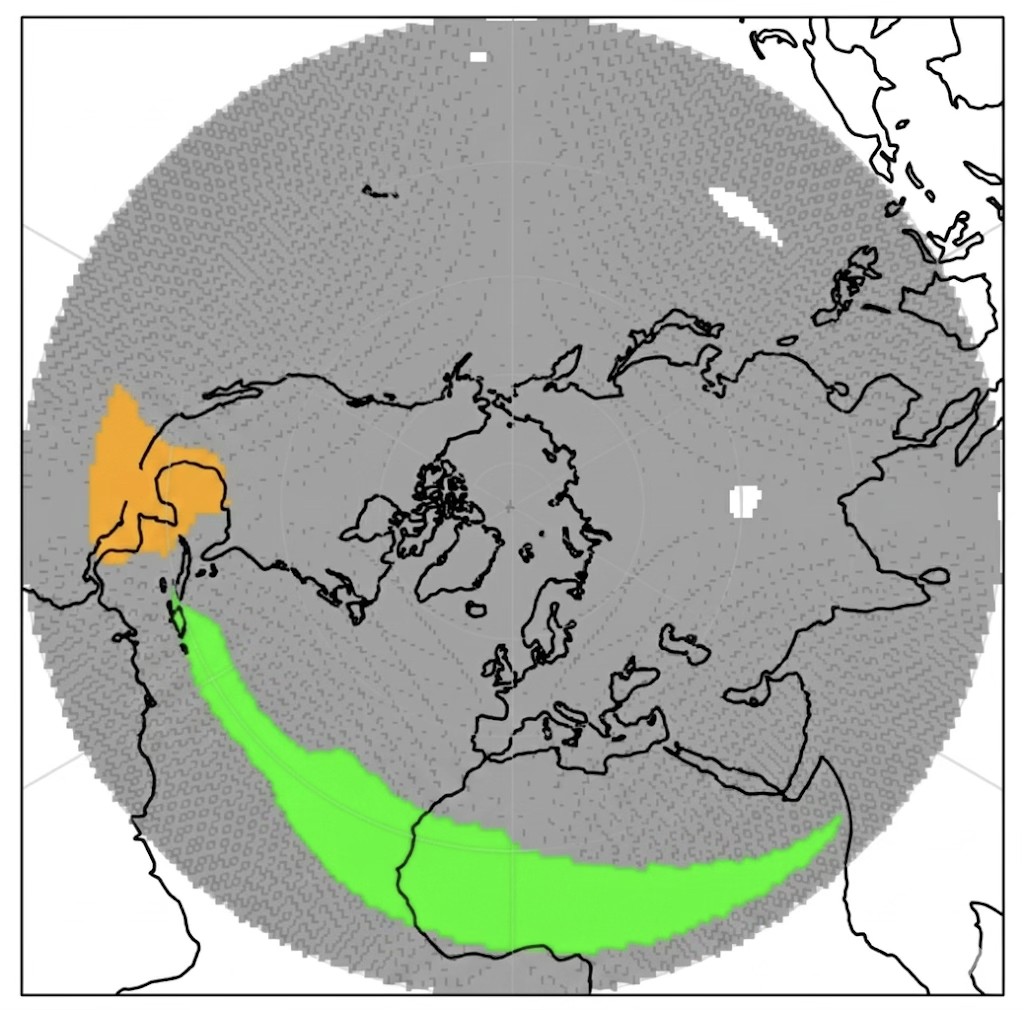}
                    \end{minipage}\\[2mm]
                    \begin{minipage}{\linewidth}
                        \centering
                        \scriptsize 0 hPa\\
                        \includegraphics[width=0.3\linewidth]{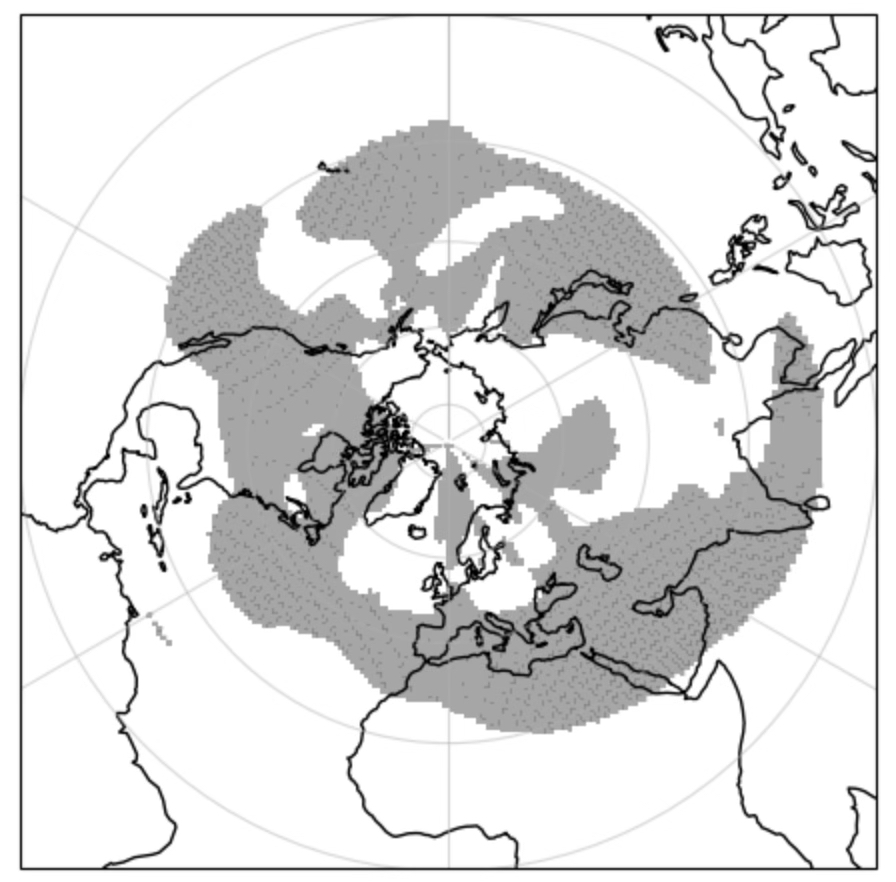}
                    \end{minipage}\\[2mm]
                    \begin{minipage}{\linewidth}
                        \centering
                        \scriptsize -10 hPa\\
                        \includegraphics[width=0.3\linewidth]{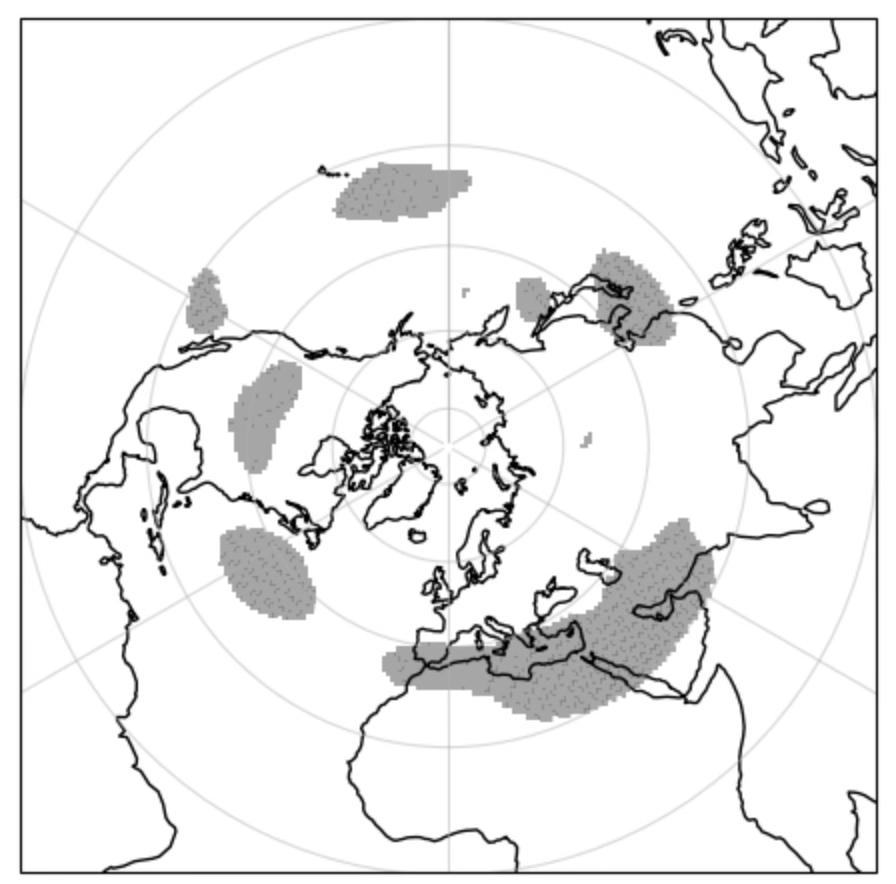}
                    \end{minipage}\\[2mm]
                \end{minipage}
            };
            \draw[<-, thick] ([xshift=18mm]figures.north west) -- ([xshift=18mm]figures.south west);
        \end{tikzpicture}
        \label{subfig:filt_inc}
    }
    \end{minipage}
    \begin{minipage}{0.45\linewidth}
        \subfloat[Persistence diagram for the sublevel-set filtration.]{
            \centering
            \includegraphics[width=0.9\linewidth]{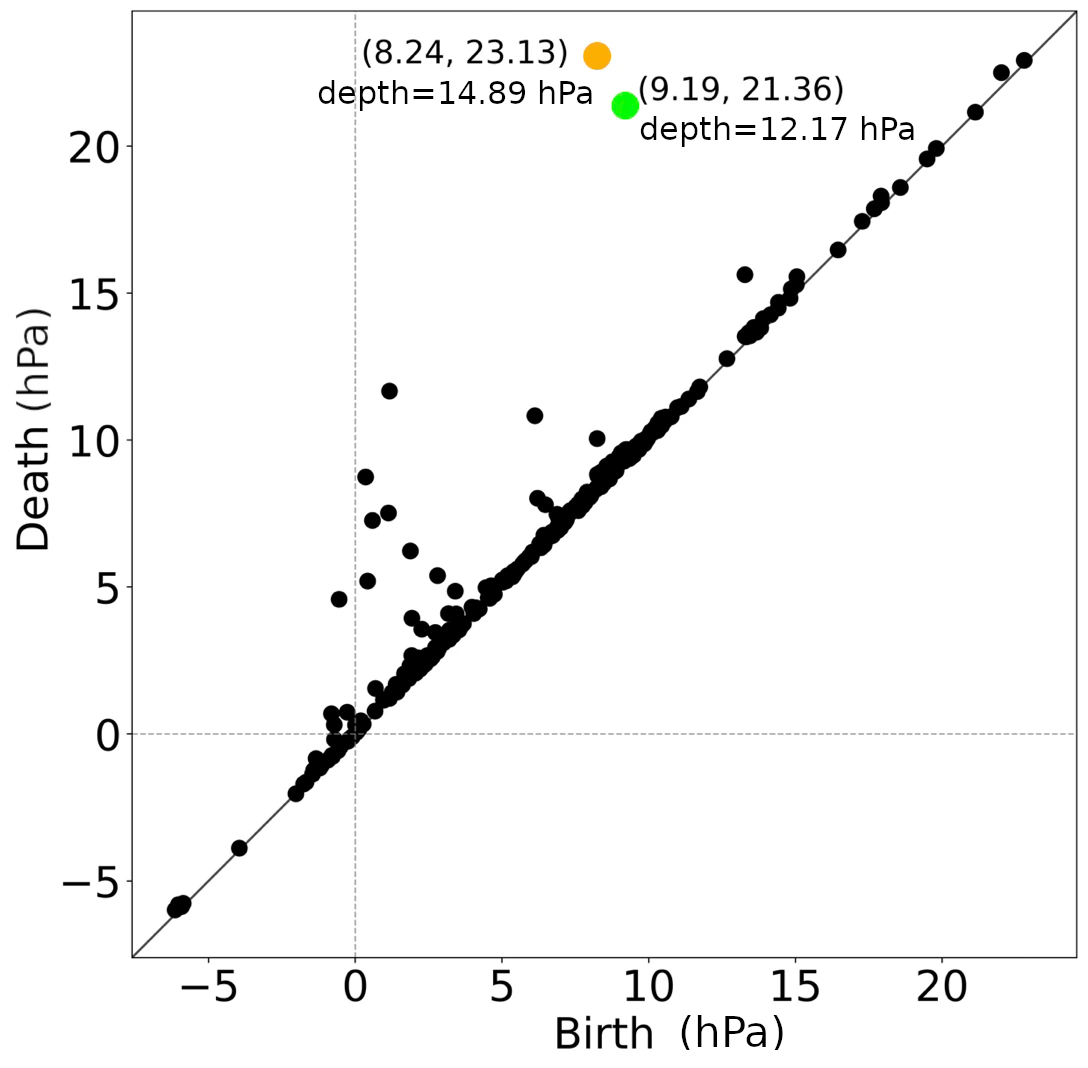}
            \label{subfig:pd_filt_inc}
        }
    \end{minipage}

\caption{Filtered cubical complexes (gray) and corresponding persistence diagrams. The 1-cyclones (a,b) and 1-anticyclones (c,d) with the largest topological depth are highlighted in green and orange and linked to their corresponding points in the persistence diagrams. Birth and death values are expressed in hPa.}
    \label{fig:pd_filt}
\end{figure}

As established in the Introduction \ref{sec:intro}, we denote the lifespan of these features as \textbf{topological depth} (depth=$|d-b|$).  Figure~\ref{fig:pd_filt} provides a concrete example of this mapping for 31 December 2022. 
Consider, for example, the two 1-cyclones highlighted in green and orange in Figure~\ref{fig:pd_filt}(a). These structures correspond to the green and orange points in the persistence diagram shown in Figure~\ref{fig:pd_filt}(b). The topological depth of a 1-cyclone is defined as the difference between its birth and death filtration values, corresponding to the pressure at which the feature first appears and the minimum pressure within the associated low-pressure region, respectively. The topological depths of the green and orange 1-cyclones are $23\,\mathrm{hPa}$ and $36.41\,\mathrm{hPa}$, respectively.

Similarly, the two 1-anticyclones highlighted in green and orange in Figure~\ref{fig:pd_filt}(c) correspond to the green and orange points in the persistence diagram shown in Figure~\ref{fig:pd_filt}(d). The topological depth of a 1-anticyclone is defined as the difference between its birth and death filtration values, corresponding to the pressure at which the feature first appears and the maximum pressure within the associated high-pressure region, respectively. The topological depths of the green and orange 1-anticyclones are $12.17\,\mathrm{hPa}$ and $14.89\,\mathrm{hPa}$, respectively. 

The global organization of these fields is further summarized by the $p$-order total persistence ($\mathrm{TP}_{p}$), allowing us to quantify the integrated intensity of all cyclonic ($\mathrm{TP_{p}D}_{C}$) and anticyclonic ($\mathrm{TP_{p}D}_{A}$) features present on a given day. In the example of Figure \ref{fig:pd_filt} we find $\mathrm{TP_{1}D}_{C}=232$ hPa, $\mathrm{TP_{1}D}_{A}=158$ hPa, $\mathrm{TP_{2}D}_{C} = 3249$ hPa, and $\mathrm{TP_{2}D}_{A} = 779$ hPa.

\subsection{Tracking and Trajectory Characteristics}
The temporal evolution of these features is followed using the tracking procedure described in Section~\ref{sec:topological_tracking}.

\begin{figure}\centering
    \subfloat[]{
        \includegraphics[width = 0.8\linewidth]{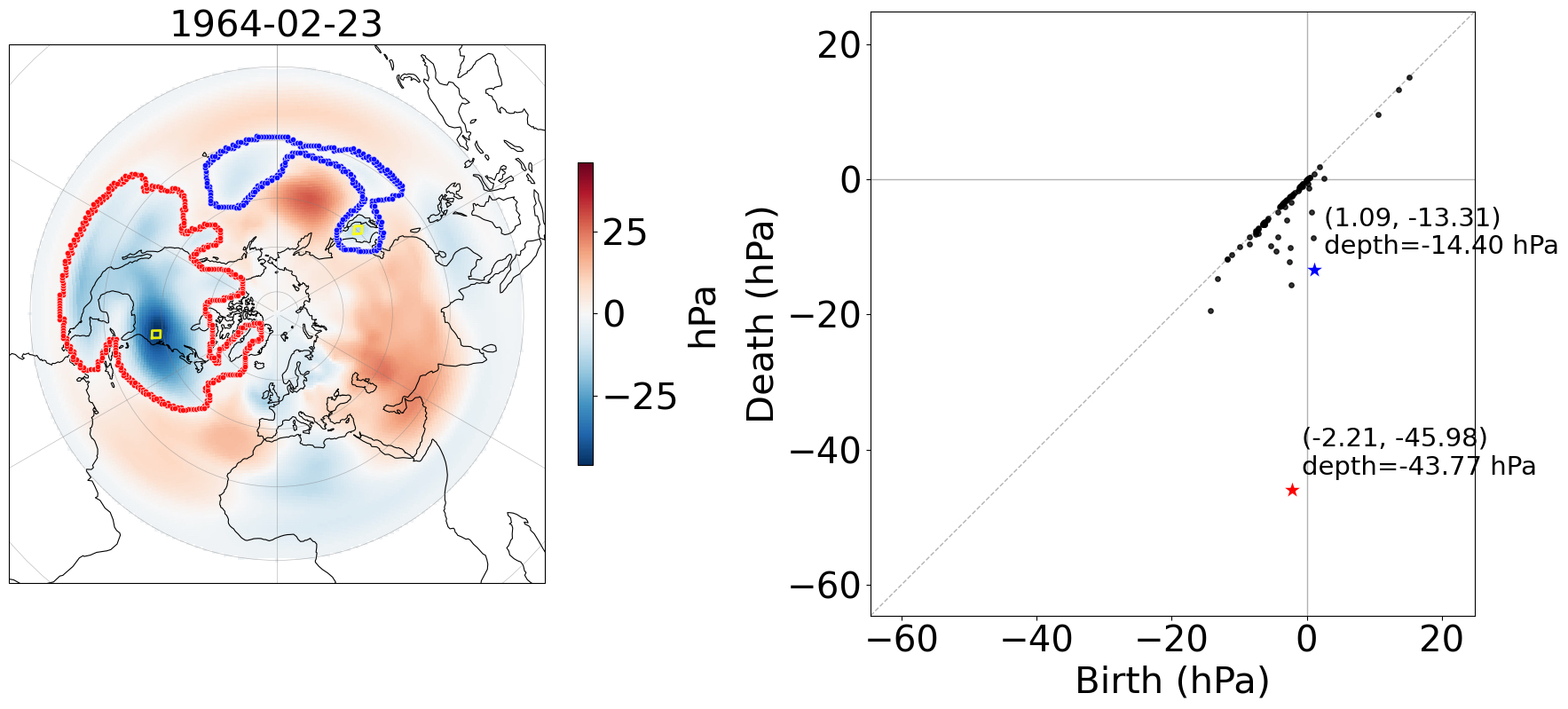}
    }  

    \subfloat[]{
        \includegraphics[width = 0.8\linewidth]{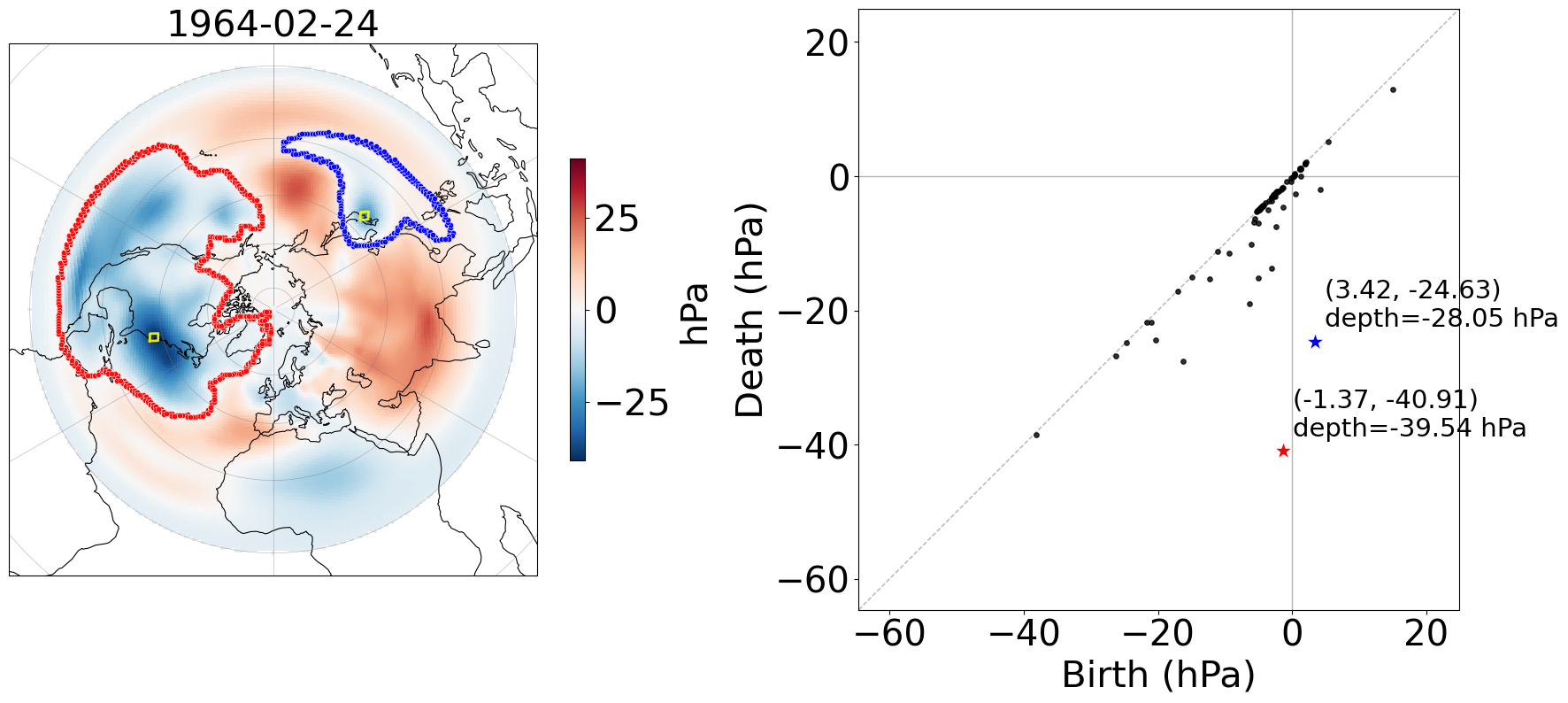}
    }  

    \subfloat[]{
        \includegraphics[width = 0.8\linewidth]{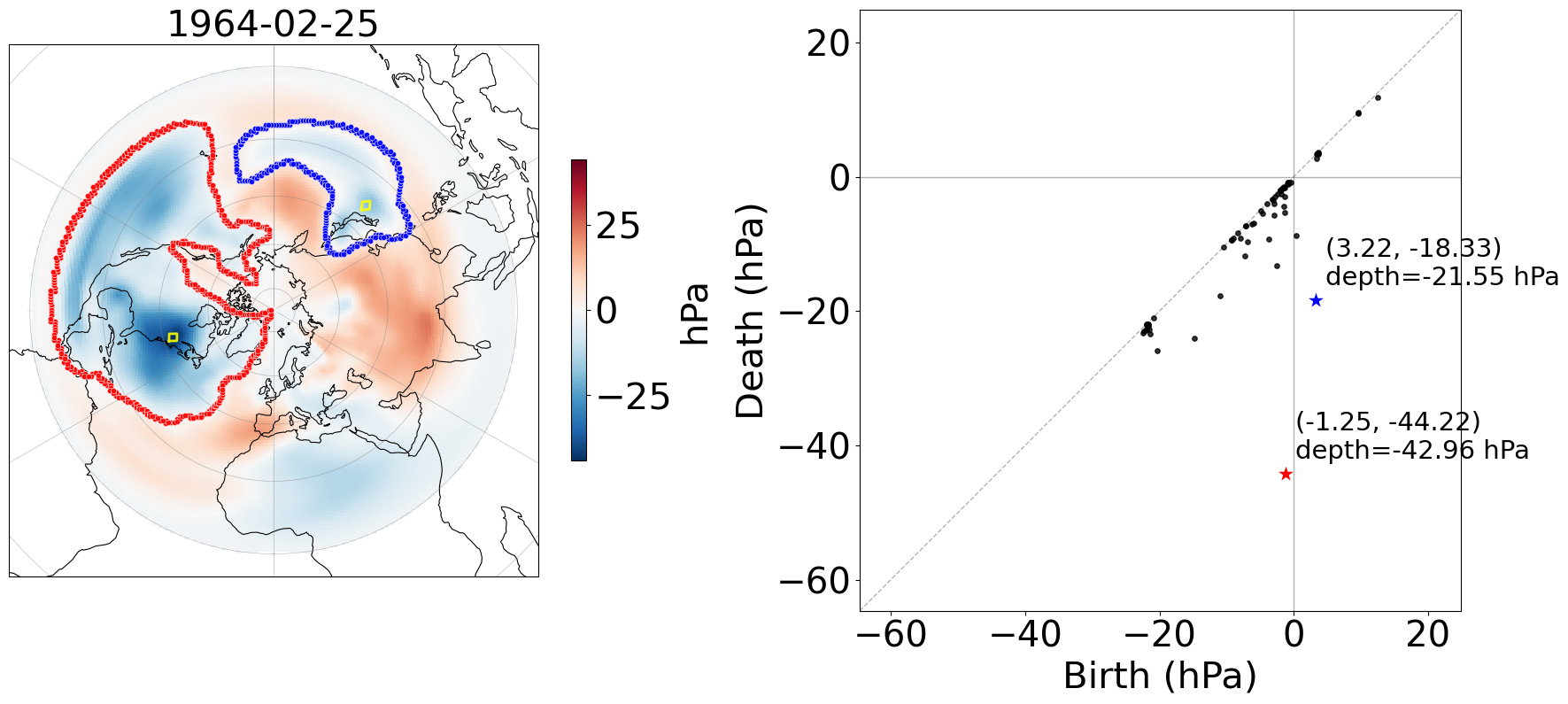}
    }  
    
    \caption{
    Tracking of 1-cyclones using the superlevel-set cubical filtration for the 
    period February 23--25, 1964. 
    Each row shows the SLP anomaly field (left) 
    and the corresponding persistence diagram (right) for a given day.
    In the persistence 
    diagrams, the red and blue stars represent two 1-cyclones that are tracked 
    across the three consecutive days.
    The corresponding spatial extent of each 
    tracked 1-cyclone is outlined by red and blue contours in the SLP anomaly 
    fields. The death square of each tracked 1-cyclone is marked by a yellow 
    square.}
    \label{fig:climate_pd_tracking}
\end{figure}

Figure~\ref{fig:climate_pd_tracking} illustrates a characteristic 3-day trajectory of a 1-cyclone for February 23--25, 1964. The red point marked in the diagrams represents a feature tracked across these three consecutive days, with birth and death values corresponding to filtration levels expressed in hectopascals (hPa). It is important to note that these coordinates do not represent the central pressure anomaly of the cyclone itself, but rather the threshold levels at which the associated topological loop appears and disappears during the filtration. Consequently, points located close to the diagonal correspond to features with very small topological depth and are interpreted as weak topological structures, whereas robust cyclonic systems are associated with points farther from the diagonal.

\begin{figure}\centering
    \subfloat[]{
        \includegraphics[width = 0.75\linewidth]{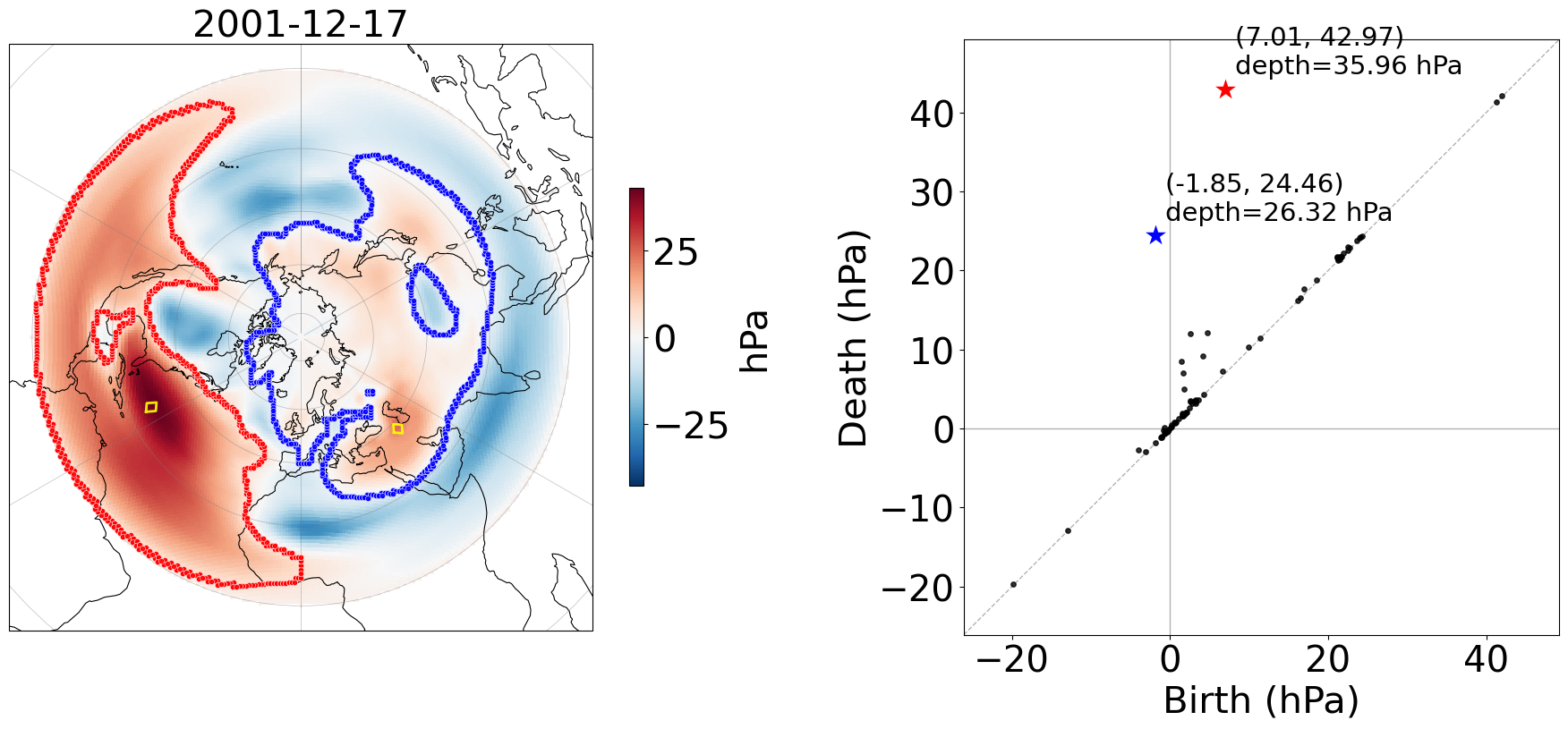}
    }  
    
    \subfloat[]{
        \includegraphics[width = 0.75\linewidth]{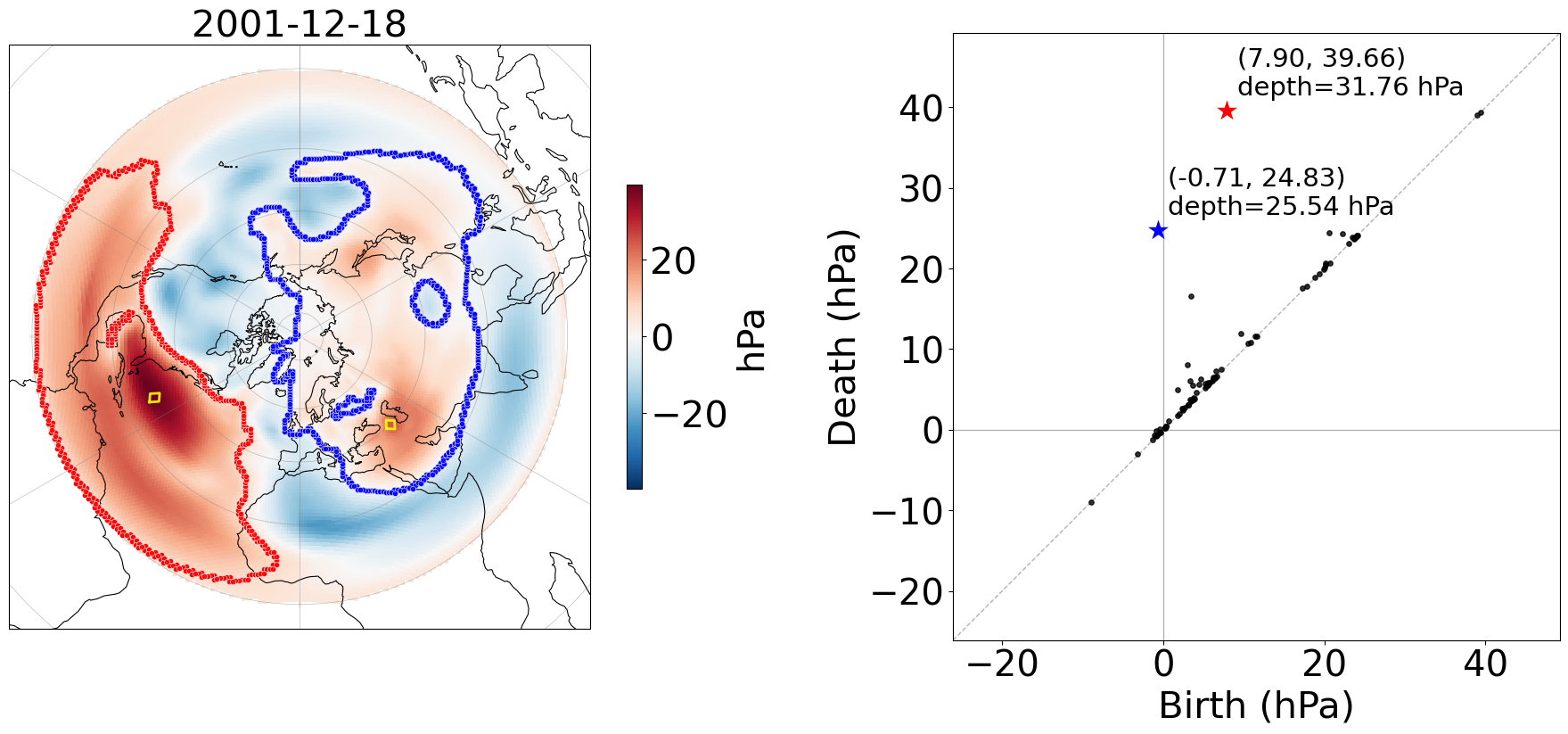}
    }  

    \subfloat[]{
        \includegraphics[width = 0.75\linewidth]{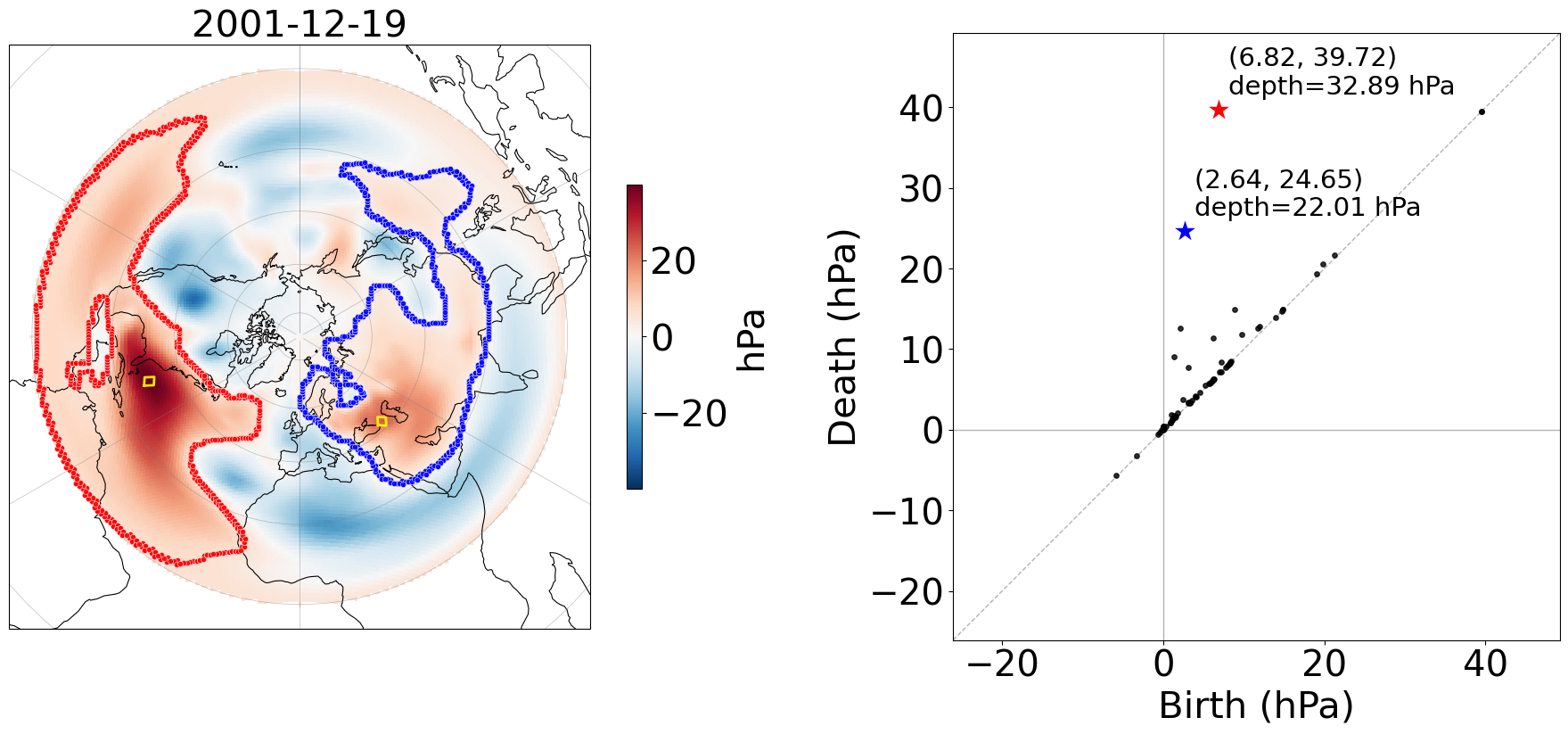}
    }  

    \caption{
    Tracking of 1-anticyclones using the sublevel-set cubical filtration for the
    period December 17--19, 2001.
    Each row shows the SLP anomaly field (left) 
    and the corresponding persistence diagram (right) for a given day.
    In the persistence 
    diagrams, the red and blue stars represent two 1-anticyclones that are tracked 
    across the three consecutive days.
    The corresponding spatial extent of each 
    tracked 1-anticyclone is outlined by red and blue contours in the SLP anomaly 
    fields. The death square of each tracked 1-anticyclone is marked by a yellow 
    square.
    }
    \label{fig:climate_pd_tracking_sub}
\end{figure}

The corresponding SLP anomalies sequence is shown alongside the  tracked 1- cyclone. A more extreme example of structural stability is shown in Figure~\ref{fig:climate_pd_tracking_sub}, which displays the onset of the longest 1-anticyclone trajectory in our record, a system that maintained its topological identity for 17 consecutive days (17 December 2001 to 2 January 2002).

\subsection{Temporal aggregation of trajectory-based metrics}
To derive climatological statistics from these trajectories, we implement a consistent temporal aggregation scheme. All metrics are conditioned on the initiation time: a trajectory is assigned to a specific calendar month based on its start date.

For each trajectory, we first compute its mean characteristic values (topological depth, spatial area, and intensity) by averaging over all daily ``snapshots'' in which the feature was tracked. Here, a ``snapshot'' refers to the representation of a cyclone or anticyclone on a single day. These individual means are then aggregated to compute the following monthly statistics:
\begin{enumerate}
    \item \textit{Trajectory frequency:} The average number of newly initiated trajectories per month.
    \item \textit{Mean lifetime:} The average duration (in days) of trajectories starting in a given month.
    \item \textit{Mean topological depth:} The average topological depth and spatial extent of features initiated in that month.
    \item \textit{Mean area:} The average geodetic area of the convex hull spanned by the 1-cycle identified per day. 
\end{enumerate}

This aggregation procedure ensures that our analysis reflects the life cycle characteristics of weather systems as they emerge and evolve across different seasons.
\subsection{Topological Descriptors for Blocking Characterization}
To characterize the topological evolution of atmospheric blocking, we analyze a set of complementary descriptors that capture both the intensity and the global organization of the anomaly field. The \textit{maximum topological depth} ($P_{\max}$) measures the prominence of the dominant topological structure, providing an estimate of the depth of the most significant cyclonic or anticyclonic feature. The global topological activity is quantified through the \textit{first-order total persistence} ($\mathrm{TP}_1$), while the \textit{second-order total persistence} ($\mathrm{TP}_2$) is used to emphasize the contribution of the most robust, large-scale structures.
The abundance of stable features is measured by the \textit{number of structures} whose topological depth exceeds 10~hPa, providing a measure of the abundance of robust cyclonic and anticyclonic structures in the circulation pattern. Furthermore, we track the \textbf{spatial area} of the feature exhibiting the highest topological depth to quantify the geographical footprint and temporal expansion of the dominant blocking structure.
Finally, we assess the dynamical stability of the circulation through the \textit{Wasserstein distance} ($d_W(D_t, D_{t+1})$) between consecutive persistence diagrams. Small values of $d_W$ indicate a quasi-stationary and slowly evolving circulation pattern, whereas large values signal rapid topological reorganization associated with the formation or decay of coherent structures. These observables are computed as temporal sequences, enabling a quantitative comparison of the robustness and stability of distinct blocking regimes, such as the July--August 2003 and February 2012 episodes.

\section{Results}\label{sec:result}

\subsection{Climatological Overview of Total Persistence}

We begin by analyzing the long-term properties of the $p$-order total persistence, $\mathrm{TP}_p$, for both sublevel- and superlevel-set filtrations at the monthly scale, considering $p=1$ and $p=2$. We denote the monthly values obtained from the sublevel-set filtration (capturing 1-anticyclones) by $\mathrm{TP_{p}M}_{A}$ and those obtained from the superlevel-set filtration (capturing 1-cyclones) by $\mathrm{TP_{p}M}_{C}$. 

\begin{figure}[ht]
\centering
\includegraphics[width=0.9\linewidth]{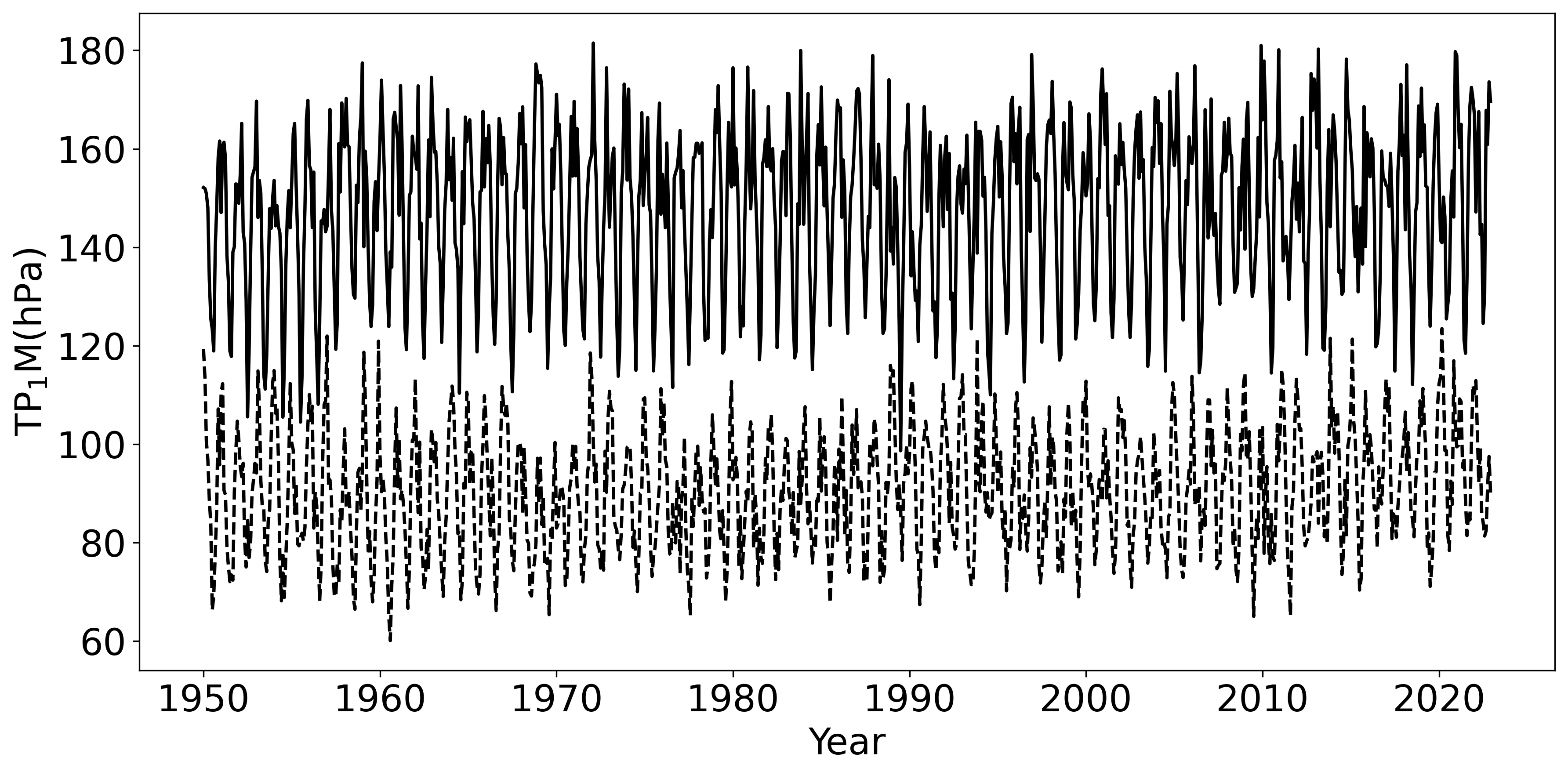}


\includegraphics[width=0.9\linewidth]{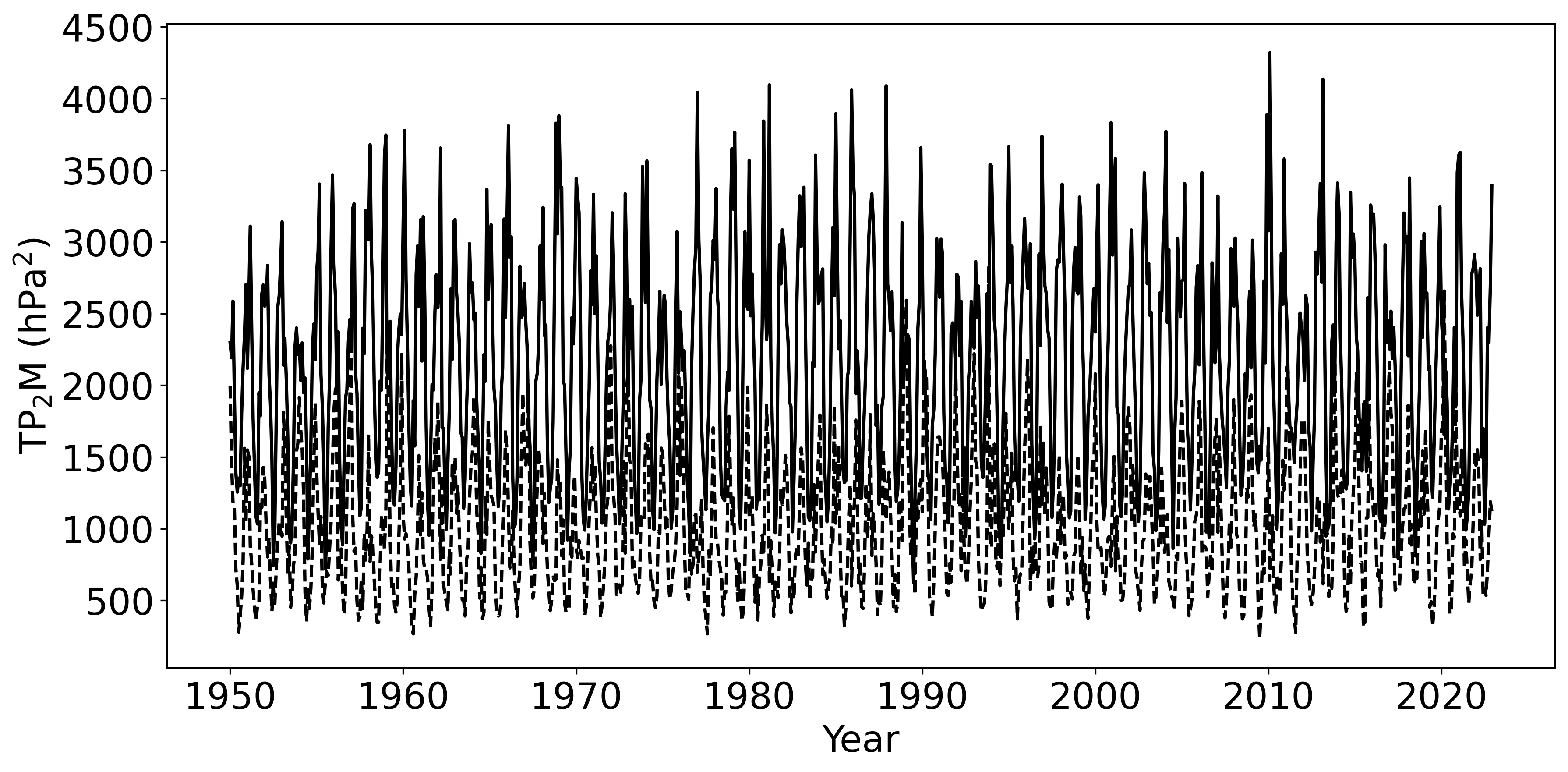}
\caption{
Monthly total persistence of Northern Hemisphere SLP anomalies during 1948--2023. Panel (a) shows the first-order total persistence, $\mathrm{TP_{1}M}$, while panel (b) shows the second-order total persistence, $\mathrm{TP_{2}M}$. Solid lines correspond to 1-cyclones identified from the superlevel-set filtration and dashed lines correspond to 1-anticyclones identified from the sublevel-set filtration. Both metrics exhibit a pronounced seasonal cycle, with larger values for cyclonic structures throughout the year and enhanced wintertime activity.
}.
\label{fig:TP_monthly}
\end{figure}

Figure~\ref{fig:TP_monthly} shows the monthly evolution of these metrics over the period 1948--2023. A pronounced seasonal cycle is evident in both metrics, indicating strong seasonal variability in the topological organization of the atmospheric circulation. Notably, the values associated with 1-cyclones (solid lines) are systematically larger than those associated with 1-anticyclones (dashed lines). This contrast is particularly pronounced for $\mathrm{TP}_2$, which emphasizes features with higher topological depth. The larger values of $\mathrm{TP}_1$ indicate that cyclonic structures contribute more strongly to the overall topological organization of the SLP anomaly field. Likewise, the larger values of $\mathrm{TP}_2$ suggest that high-depth cyclonic structures contribute more strongly to the total topological persistence than their anticyclonic counterparts. Overall, these results indicate that cyclonic structures dominate the topological organization of Northern Hemisphere SLP anomalies throughout the annual cycle. Both metrics exhibit winter maxima and summer minima, consistent with enhanced cyclonic activity and increased atmospheric variability during the cold season.

\begin{figure}[ht]
\centering
\includegraphics[width=0.9\linewidth]{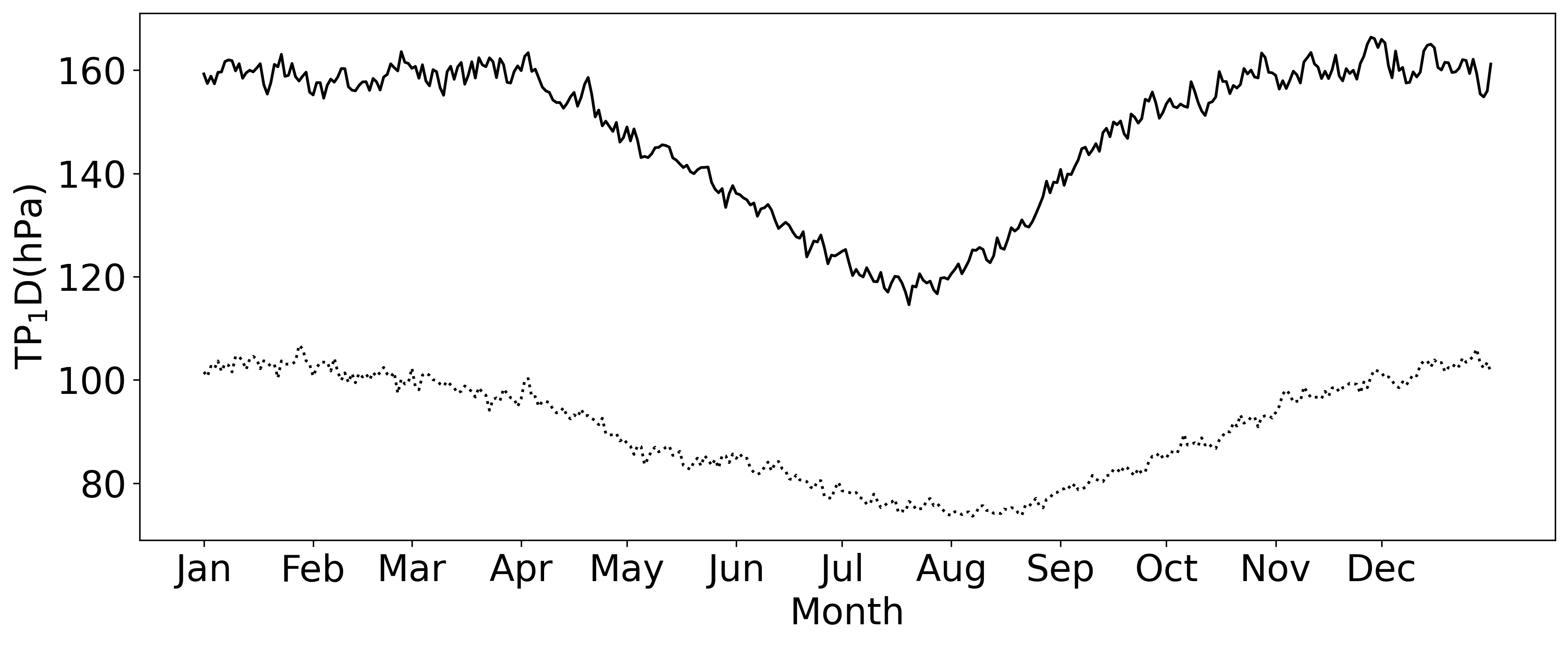}


\includegraphics[width=0.9\linewidth]{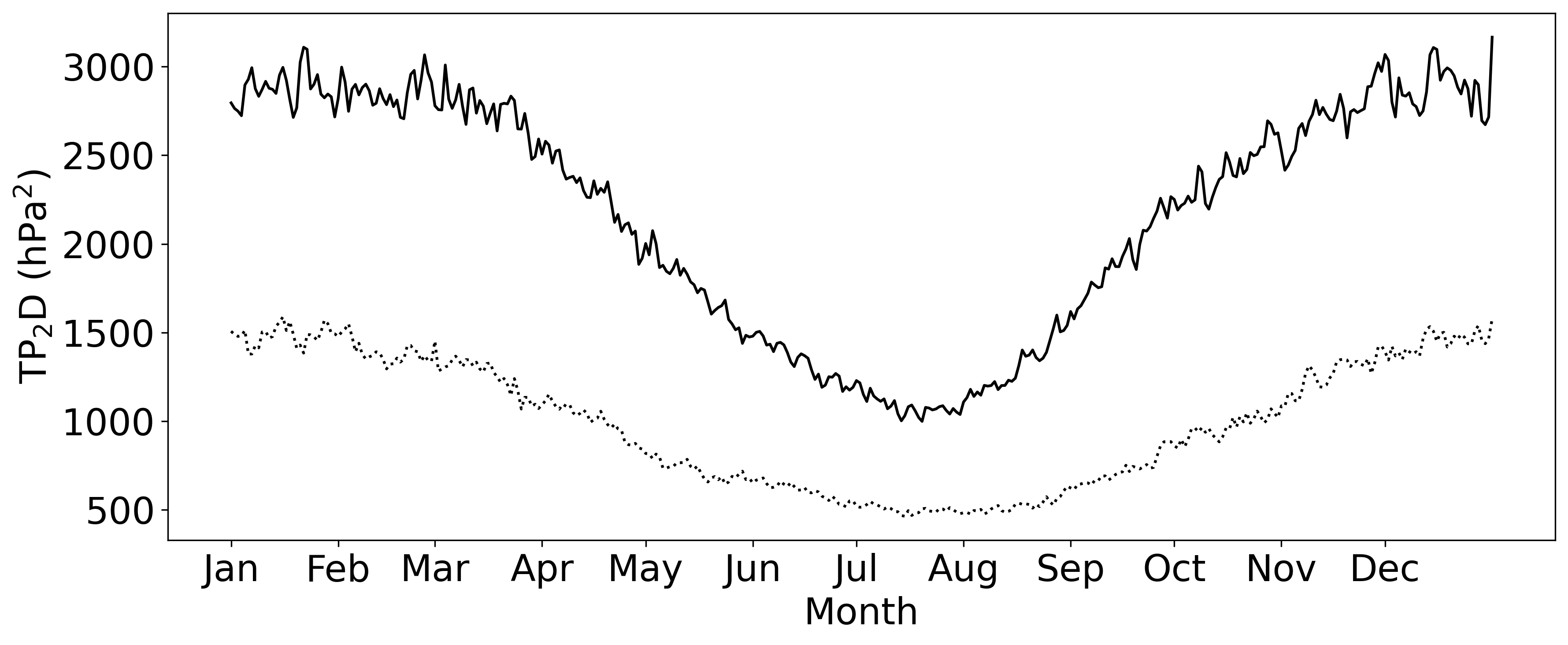}
\caption{
Climatological daily evolution of the first- and second-order total persistence of Northern Hemisphere SLP anomalies during 1948--2023. Panel (a) shows the daily climatology of $\mathrm{TP_{1}D}_{C}$ and $\mathrm{TP_{1}D}_{A}$, while panel (b) shows the corresponding climatology of $\mathrm{TP_{2}D}_{C}$ and $\mathrm{TP_{2}D}_{A}$. Here, $\mathrm{TP_{p}D}_{C}$ and $\mathrm{TP_{p}D}_{A}$ denote the daily total persistence computed from the superlevel-set (1-cyclones) and sublevel-set (1-anticyclones) filtrations, respectively. Solid lines correspond to 1-cyclones and dashed lines to 1-anticyclones. Both persistence measures exhibit a pronounced annual cycle, with maxima during boreal winter and minima during boreal summer. Cyclonic structures consistently contribute larger total persistence values than anticyclonic structures throughout the year, particularly for $\mathrm{TP}_2$, which emphasizes features with large topological depth.
}
\label{fig:TP_daily}
\end{figure}
We denote the daily $p$-order total persistence values for the sublevel-set filtration (capturing 1-anticyclones) as $\mathrm{TP_{p}D}_{A}$, and the daily $p$-order total persistence values for the superlevel-set filtration (capturing 1-cyclones) as $\mathrm{TP_{p}D}_{C}$. Figure~\ref{fig:TP_daily} shows the climatological daily evolution of these metrics, where the daily values were obtained by averaging each calendar day over the full period 1948--2023. In both panels, the values associated with 1-cyclones (solid lines) are systematically larger than those associated with 1-anticyclones (dashed lines). A pronounced annual cycle is evident in both metrics, with maximum values occurring during boreal winter and minimum values during boreal summer. In particular, $\mathrm{TP_{1}D}_{C}$ exhibits a strong seasonal modulation, with winter maxima reaching approximately 160~hPa and summer minima near 115~hPa, while $\mathrm{TP_{1}D}_{A}$ varies between approximately 74 and 102~hPa throughout the year.  This contrast is even more pronounced for $\mathrm{TP}_{2}$, which assigns greater weight to features with large topological depth, indicating that high-depth cyclonic structures contribute more strongly to the total persistence than their anticyclonic counterparts. Moreover, the larger seasonal amplitude observed in $\mathrm{TP}_{2}$ compared to $\mathrm{TP}_{1}$ suggests that seasonal variability is particularly pronounced among the most topologically depth structures. The seasonal cycle is remarkably coherent across both filtration types and persistence orders, with minima consistently occurring during July--August and maxima during December--February. This behavior is consistent with the well-known strengthening of North Atlantic storm-track activity and atmospheric variability during the cold season. The substantial difference in magnitude between the cyclone and anticyclone curves throughout the year indicates that cyclonic systems contribute more strongly to the topological organization of the SLP anomaly field, generating a larger amount of topological structure across pressure thresholds than anticyclonic systems.

\begin{figure}\centering
    \includegraphics[width=0.8\linewidth]{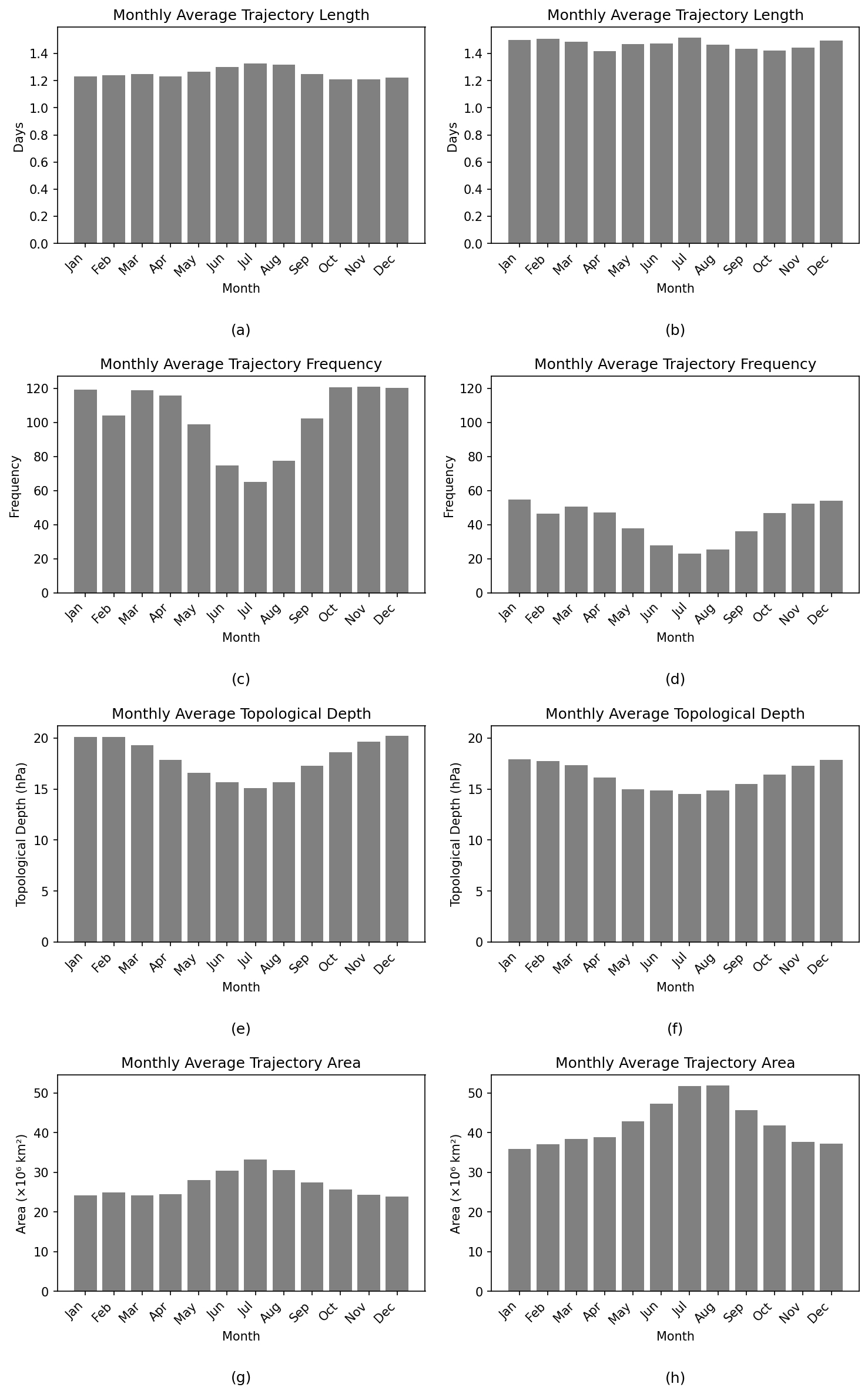}
    \caption{
    Monthly summaries of 1-cyclone trajectories (left) and 1-anticyclone
    trajectories (right). }
    \label{fig:traj_sup_sub_level}
\end{figure}

\subsection{Lagrangian Statistics of 1-Cyclones and 1-Anticyclones}

To analyze the temporal evolution of these structures, we employ optimal matching between consecutive persistence diagrams. Prior to matching, we retain only features whose topological depth exceeds 10~hPa to remove weak and likely spurious structures. This 10~hPa threshold corresponds to the typical amplitude separating synoptically relevant pressure systems from background noise in the SLP anomaly field, and was found to yield robust, reproducible structures across the Northern Hemisphere independently of the local background pressure gradient. For each pair of consecutive diagrams, we compute the optimal matching and use it to track the corresponding features through time.

To ensure physically meaningful correspondences, we apply an additional spatial filtering criterion: if the death squares of two matched features are separated by more than 1{,}000~km, the match is rejected and the corresponding trajectory is terminated. This filtering step ensures that only features whose death squares remain in close spatial proximity are linked across consecutive days, preventing spurious long-range associations that are unlikely to represent the continuous evolution of a single cyclonic or anticyclonic structure. After applying both filters, 22.68\% of matched pairs were retained for 1-cyclones and 38.3\% for 1-anticyclones.

\subsubsection{1-cyclone trajectories}

We identified a total of 94{,}169 1-cyclone trajectories over the period 1948--2023. The duration distribution is strongly right-skewed: 81.7\% of trajectories lasted a single day, 18.0\% lasted 2--4 days, 0.4\% lasted 5--9 days, and only two trajectories lasted 10 days or longer, with the longest reaching 11 days. The mean trajectory length is 1.25 days (std $= 0.61$ days). The mean topological depth across all trajectories is 18.32~hPa (std $= 8.10$~hPa), with a maximum value of 87.49~hPa. Spatially, trajectories travel on average 132.42~km over their lifetime (99.73~km/day), while the most mobile trajectory covers 6{,}511~km over 11 days. The average distance between matched trajectory pairs is 535.90~km.

Figure~\ref{fig:traj_sup_sub_level} (left column) summarizes the monthly distributions of length, frequency, topological depth, and area of the trajectory. 
The monthly trajectory frequency exhibits a broad maximum during the extended cold season (October to March), reaching approximately 120 trajectories per month, while a pronounced minimum is observed during June--July with approximately 65 trajectories per month. Although secondary fluctuations are visible within the winter months, the dominant signal is the strong seasonal contrast driven by the intensification of the storm track during boreal winter.
The mean topological depth per trajectory follows a marked seasonal cycle, reaching approximately 20~hPa during winter and decreasing to approximately 15~hPa during summer. In contrast, the mean trajectory length remains nearly constant throughout the year, varying only weakly with season. The trajectory area also exhibits a relatively weak seasonal cycle, with slightly larger values during summer than during winter.

\subsubsection{1-anticyclone trajectories}

We identified a total of 38{,}221 1-anticyclone trajectories over the period 1948--2023. As in the cyclone case, durations are concentrated at short timescales: 72.3\% of trajectories lasted a single day, 25.7\% lasted 2--4 days, 1.9\% lasted 5--9 days, and 46 trajectories (0.1\%) lasted between 10 and 17 days, with the longest reaching 17 days. The mean trajectory length is 1.47 days (std $= 1.01$ days). The mean topological depth is 16.59~hPa (std $= 7.20$~hPa), with a maximum value of 72.81~hPa. Spatially, trajectories travel on average 225.58~km over their lifetime (138.92~km/day), while the most mobile trajectory covers 6{,}590~km over 17 days. The average distance between matched trajectory pairs is 482.02~km.

Figure~\ref{fig:traj_sup_sub_level} (right column) summarizes the monthly distributions of trajectory length, frequency, topological depth, and area for 1-anticyclones.
Monthly trajectory frequency for 1-anticyclones shows a robust peak during the winter months (November to January), reaching approximately 55--60 trajectories per month, and decreases significantly to 23--26 trajectories during the summer.
The mean topological depth per trajectory peaks at approximately 18~hPa during winter and decreases to approximately 14~hPa during summer. As for cyclones, the mean trajectory length remains nearly constant throughout the year. Anticyclone trajectories occupy substantially larger areas than cyclone trajectories and exhibit a clear seasonal cycle, reaching their maximum extent during boreal summer.

\subsection{Synoptic Case Studies: Atmospheric Blocking}\label{sec:blocking}
\begin{figure}[htbp]
    \centering
    \includegraphics[width=0.85\linewidth]{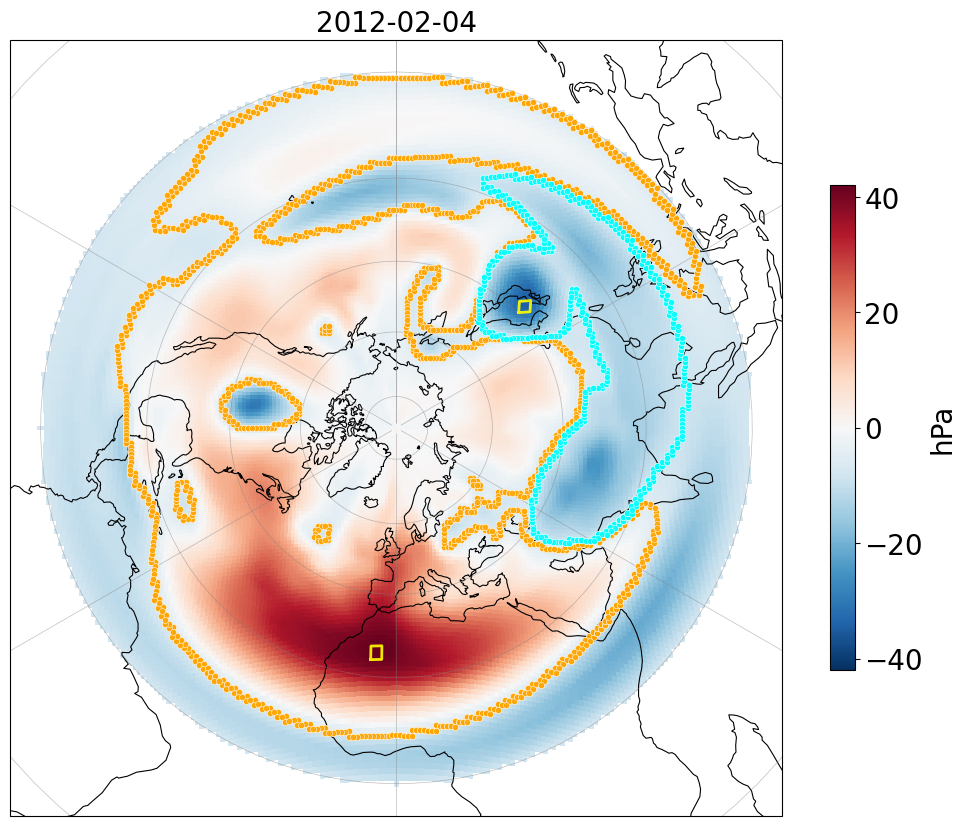}
\caption{
SLP anomaly field over the Northern Hemisphere on 4 February 2012, corresponding to the peak of a blocking event. Red (blue) shading indicates positive (negative) anomalies in hPa. The orange contour outlines the most prominent 1-anticyclone identified from the sublevel-set filtration, enclosing the strong positive anomaly ($\gtrsim 40$~hPa) centered over Southern Europe and the Mediterranean. The cyan contour outlines the most prominent 1-cyclone identified from the superlevel-set filtration, enclosing the negative anomaly over Scandinavia and the Barents Sea. Yellow squares indicate the corresponding death squares of the two features. The temporal evolution of this event is shown in Figure~\ref{fig:blocking_2012}.
}
    \label{fig:blocking_peak}
\end{figure}

We analyze the February 2012 event and the July--August 2003 European blocking episode, allowing us to assess the robustness and generality of the topological signatures across distinct synoptic regimes and dynamical configurations. 

We first analyze the February 2012 blocking event, which was associated with exceptional snowfall over Rome and large parts of Italy. Figure~\ref{fig:blocking_peak} shows the SLP anomaly field over the Northern Hemisphere on 4 February 2012, corresponding to the mature phase of this blocking event. The orange contour identifies the most prominent 1-anticyclone, enclosing a strong positive anomaly ($\gtrsim 40$~hPa) over Southern Europe and the Mediterranean. The cyan contour indicates the most prominent 1-cyclone, associated with a negative anomaly over Scandinavia and the Barents Sea.

\begin{figure}[htbp]
    \centering
    \subfloat[31 Jan 2012]{
        \includegraphics[width=0.45\linewidth]{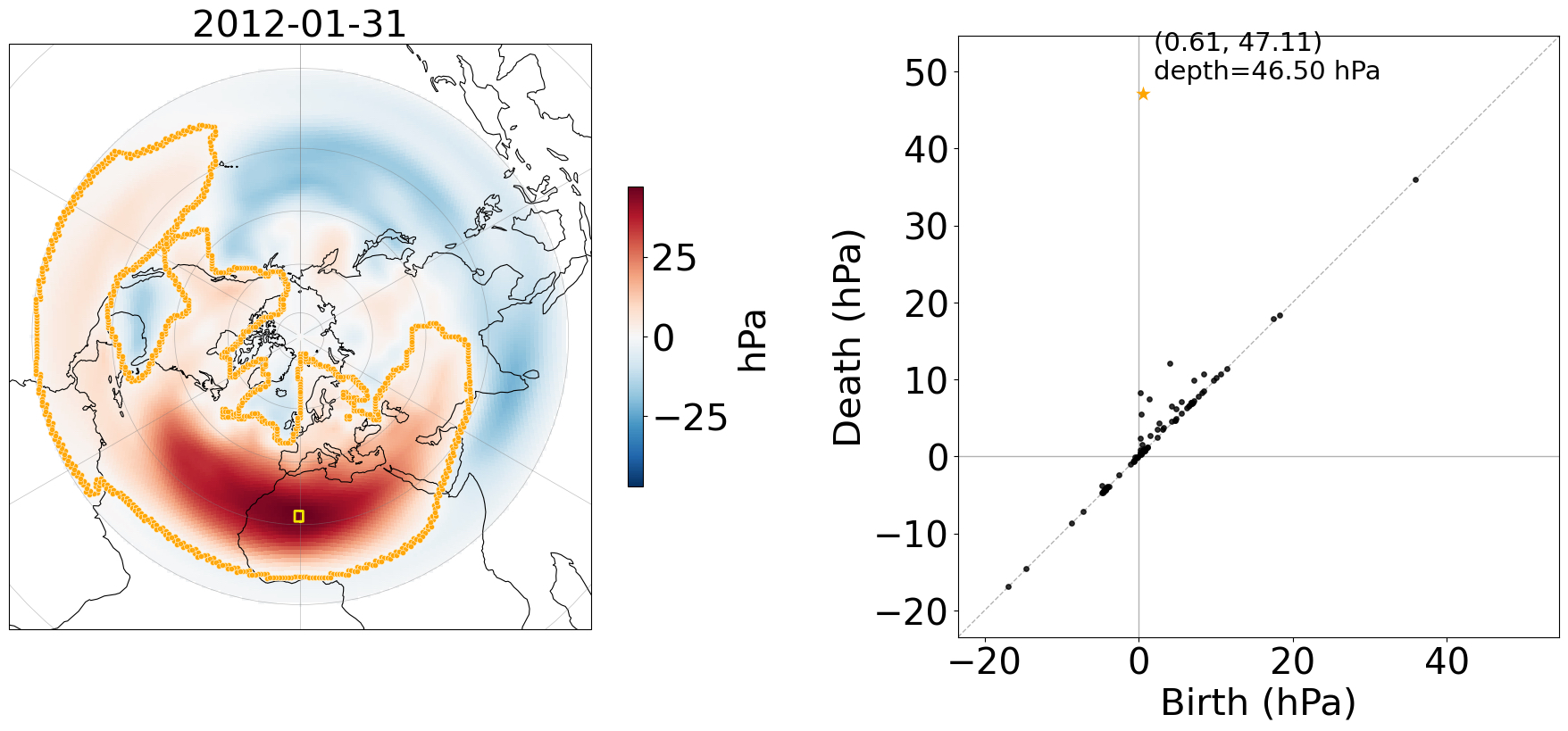}
        \label{subfig:block_0131}
    }
    \subfloat[1 Feb 2012]{
        \includegraphics[width=0.45\linewidth]{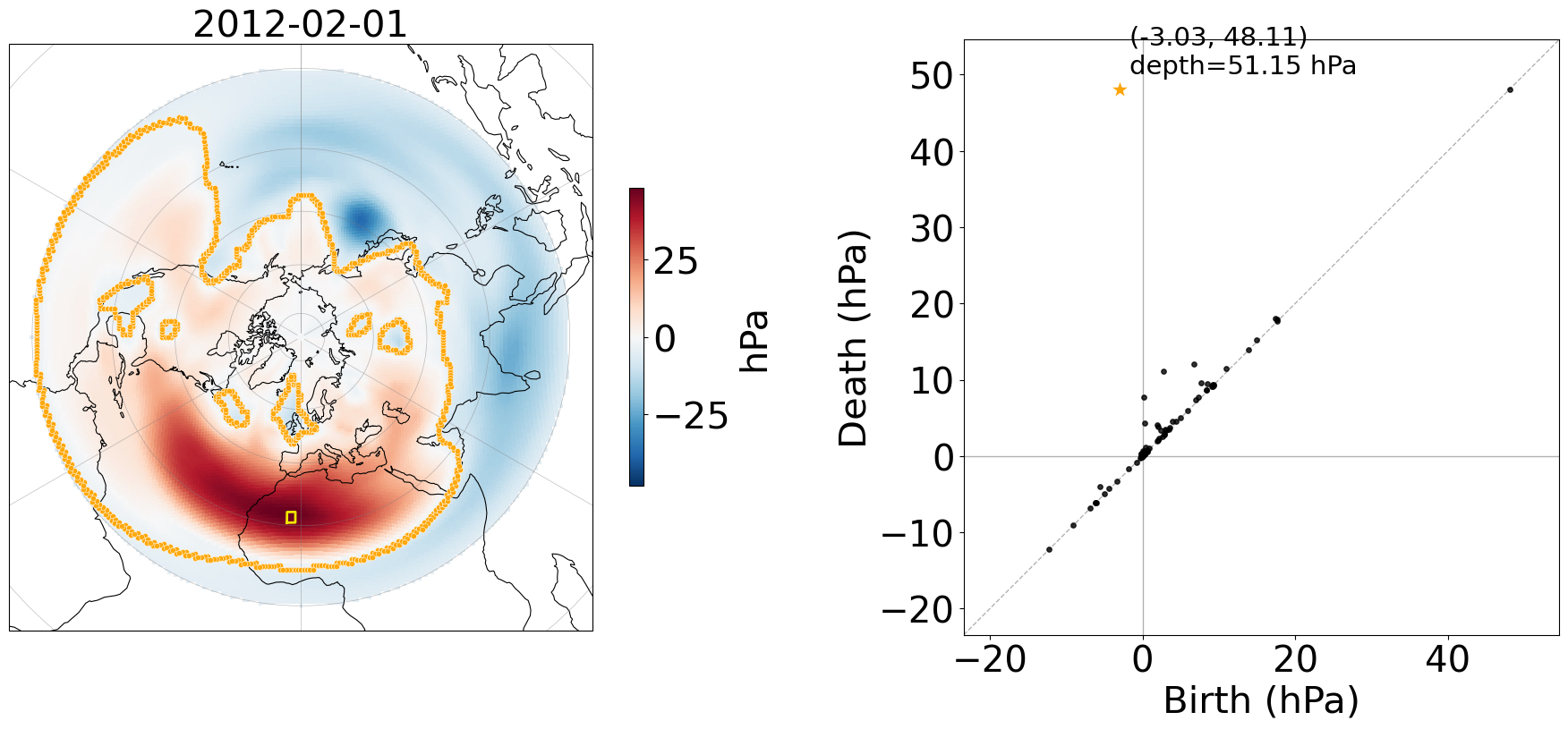}
        \label{subfig:block_0201}
    }\\
    \subfloat[2 Feb 2012]{
        \includegraphics[width=0.45\linewidth]{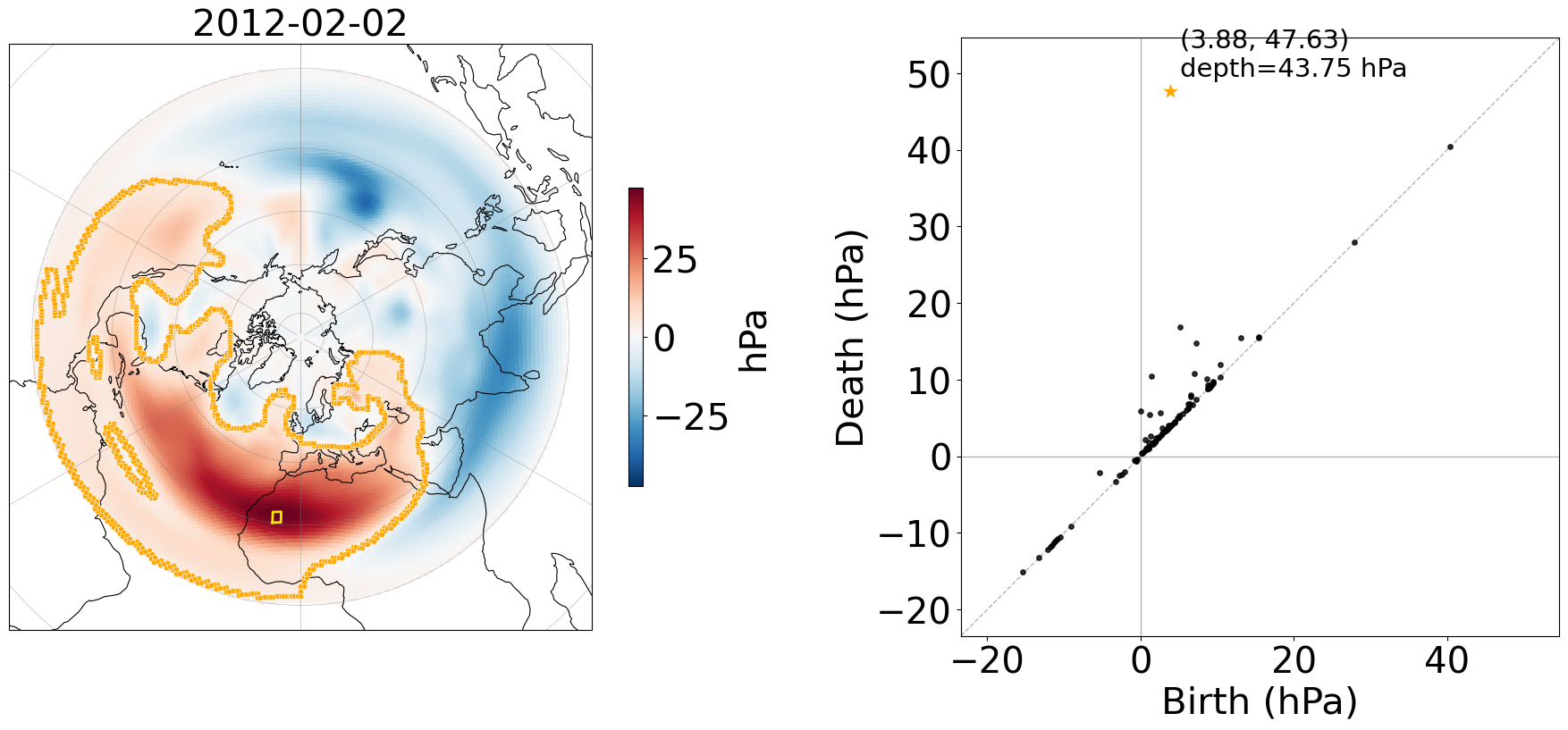}
        \label{subfig:block_0202}
    }
    \subfloat[3 Feb 2012]{
        \includegraphics[width=0.45\linewidth]{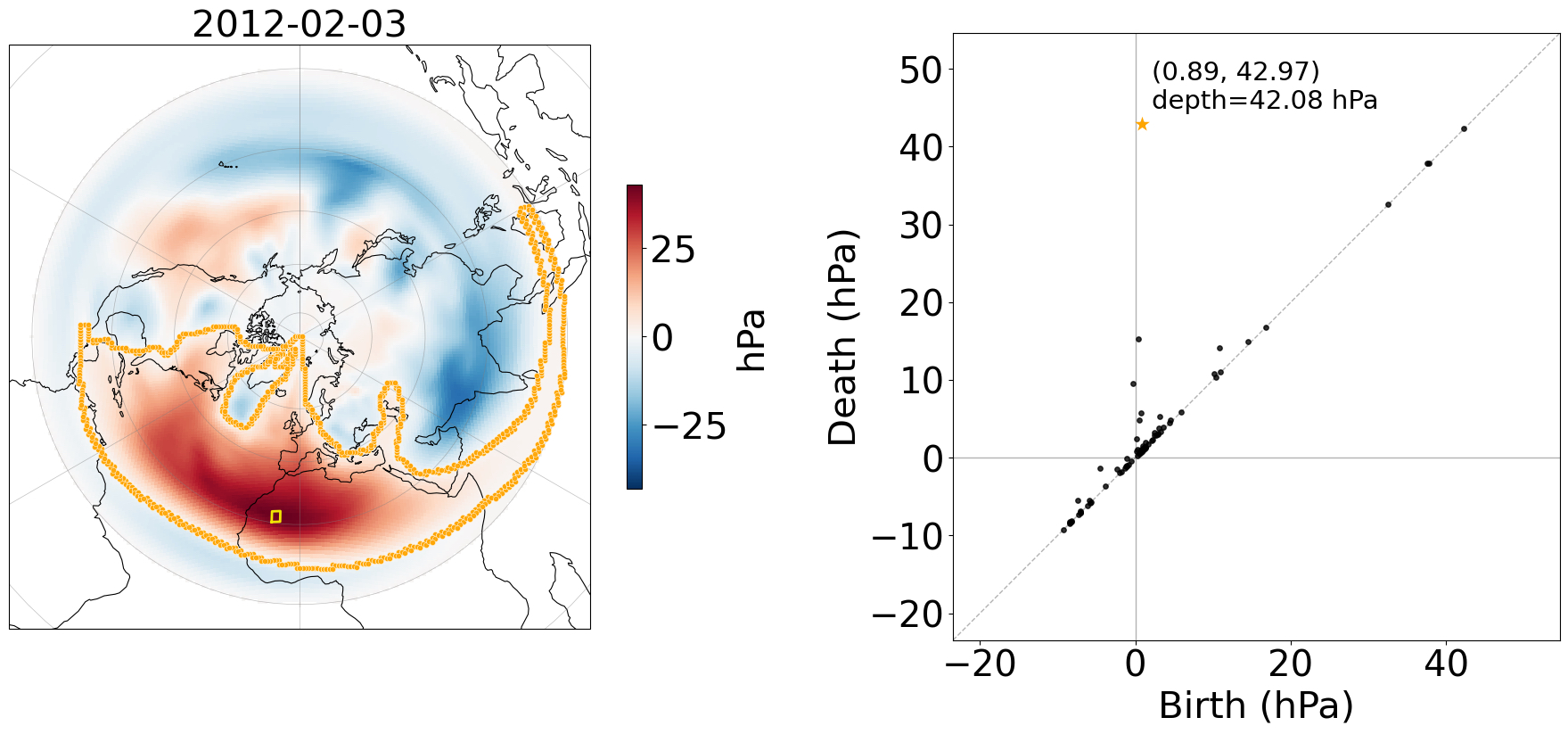}
        \label{subfig:block_0203}
    }\\
    \subfloat[4 Feb 2012 (peak)]{
        \includegraphics[width=0.45\linewidth]{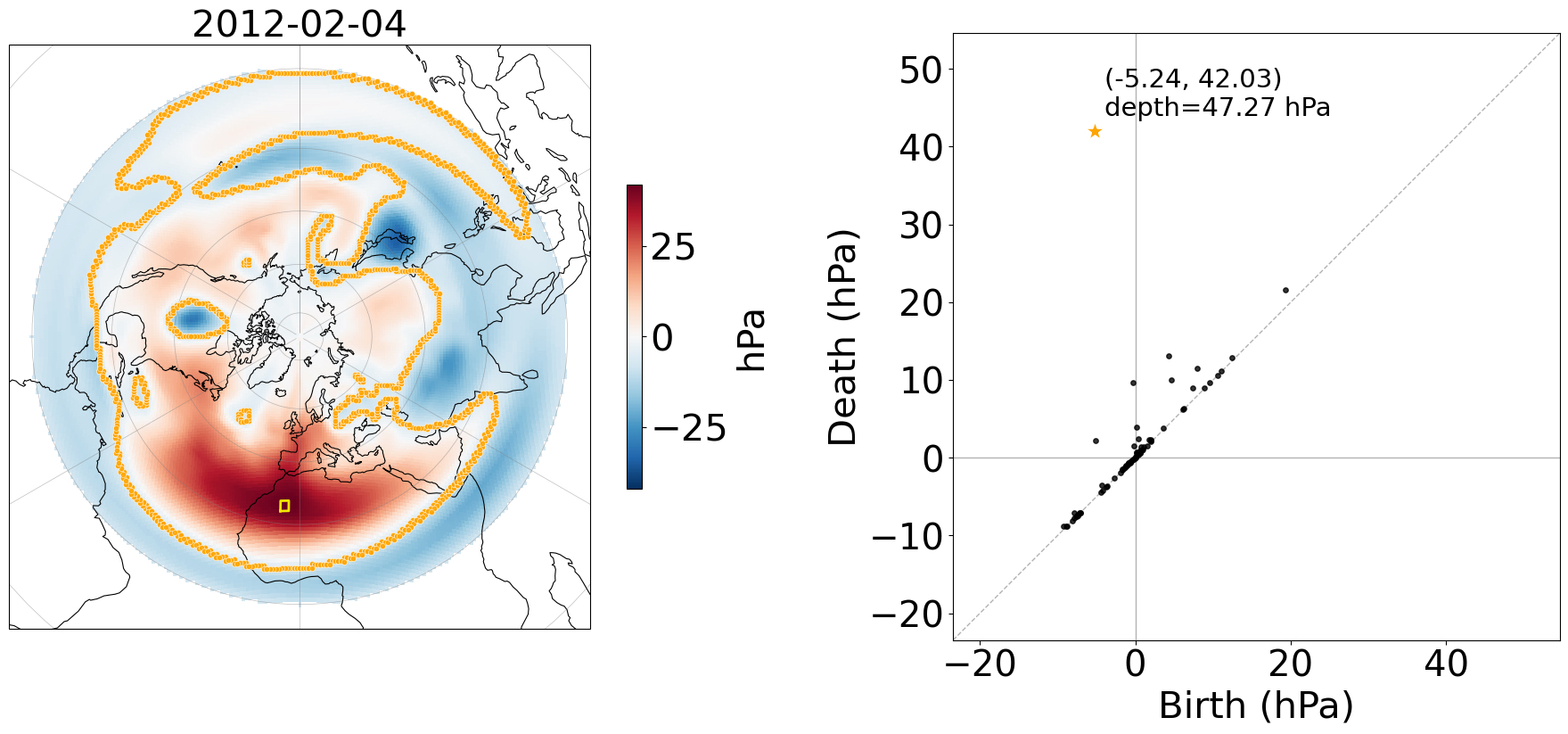}
        \label{subfig:block_0204}
    }
    \subfloat[5 Feb 2012]{
        \includegraphics[width=0.45\linewidth]{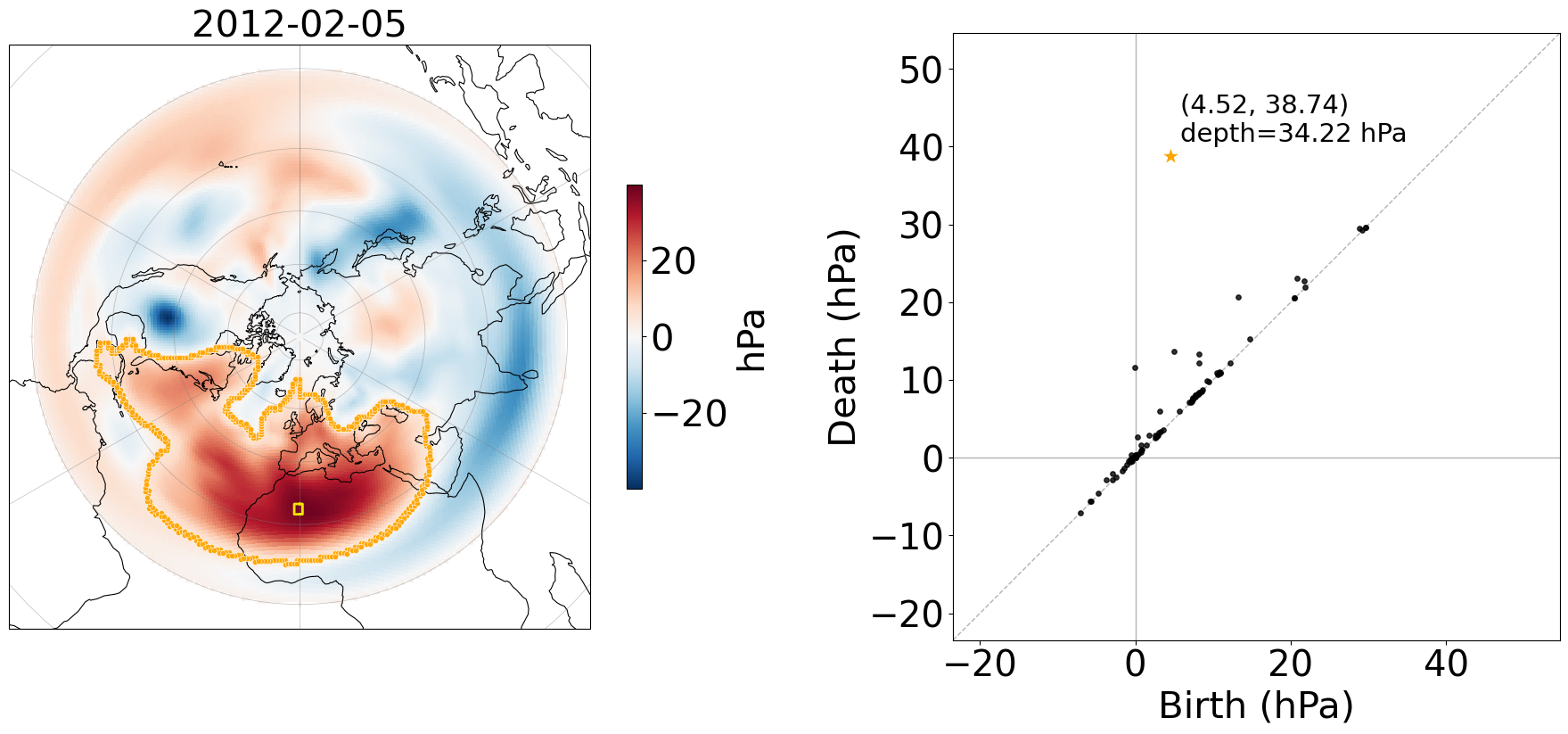}
        \label{subfig:block_0205}
    }
    \caption{Day-by-day evolution of the February 2012 blocking event via the 
    sublevel-set filtration (31 January -- 5 February 2012). Each panel shows 
    the SLP anomaly map (left) and the corresponding persistence diagram (right). 
    The orange star in each persistence 
    diagram identifies the tracked feature.
    The orange contour encloses the tracked 1-anticyclone; the yellow 
    square marks its death square. The feature intensifies from 31 
    January, reaches peak persistence on 4 February (panel e; see also 
    Figure~\ref{fig:blocking_peak}), and weakens by 5 February.}
    \label{fig:blocking_2012}
\end{figure}

Figure~\ref{fig:blocking_2012} shows the temporal evolution of the blocking event, tracked from 31 January to 5 February 2012. The tracked 1-anticyclone progressively intensifies during the onset phase, reaches maximum topological depth on 4 February, and weakens thereafter. The corresponding persistence diagrams reveal the emergence of a dominant 1-anticyclone that becomes increasingly separated from the rest of the topological features, indicating the formation of a coherent and robust anticyclonic structure associated with the mature stage of the blocking event.

To assess the robustness of these topological signatures, we compare the February 2012 event with the July--August 2003 blocking event. This event was associated with the exceptional European heatwave of summer 2003, sustained by a persistent and quasi-stationary anticyclonic circulation that remained over Europe for several weeks. 

\begin{figure}[htbp]
    \centering

    \subfloat[July 2003]{
        \includegraphics[width=0.45\linewidth]{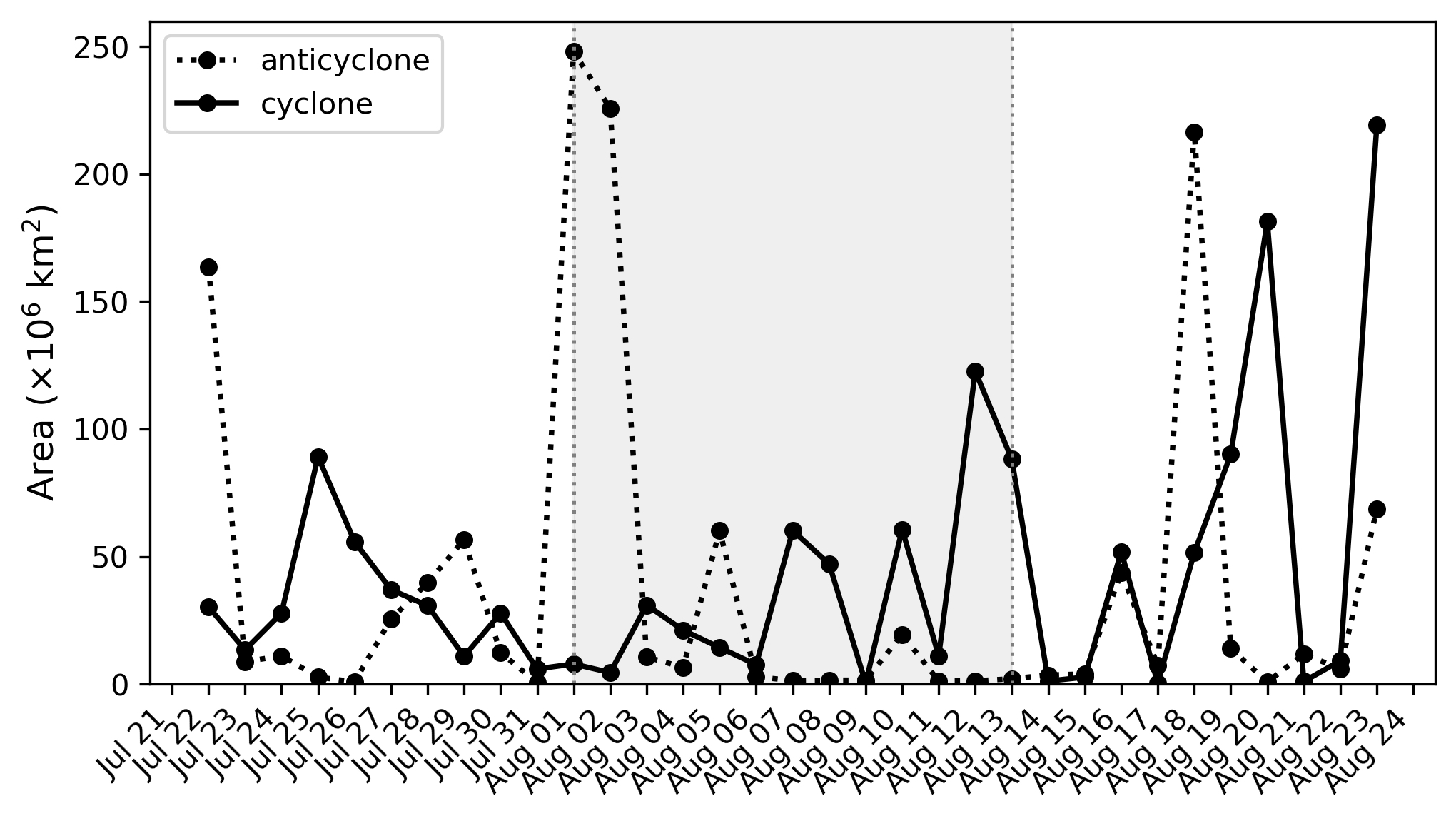}
    }
    \subfloat[February 2012]{
        \includegraphics[width=0.45\linewidth]{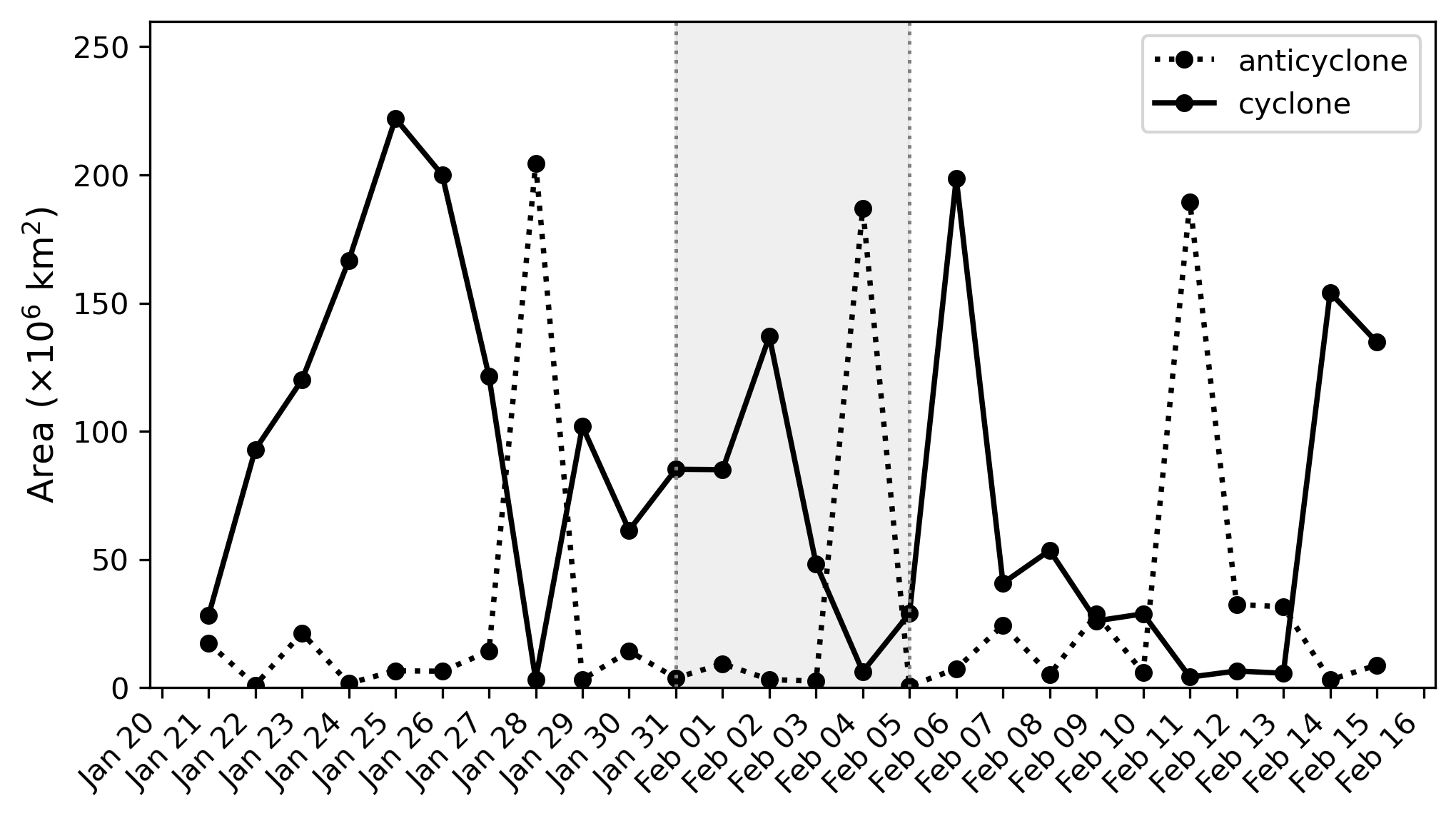}
    }

    \vspace{0.2cm}

    \subfloat[July 2003]{
        \includegraphics[width=0.45\linewidth]{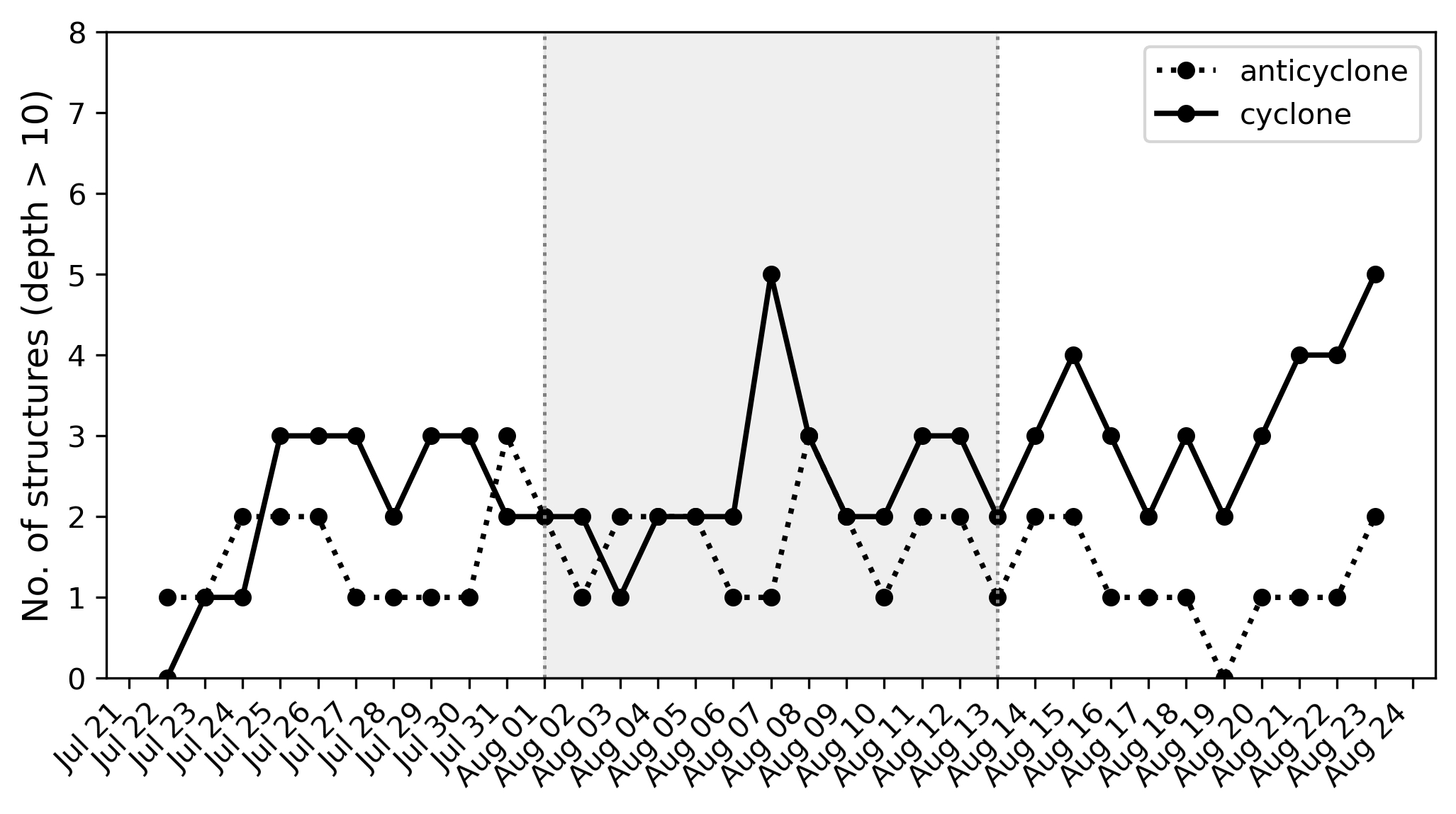}
    }
    \subfloat[February 2012]{
        \includegraphics[width=0.45\linewidth]{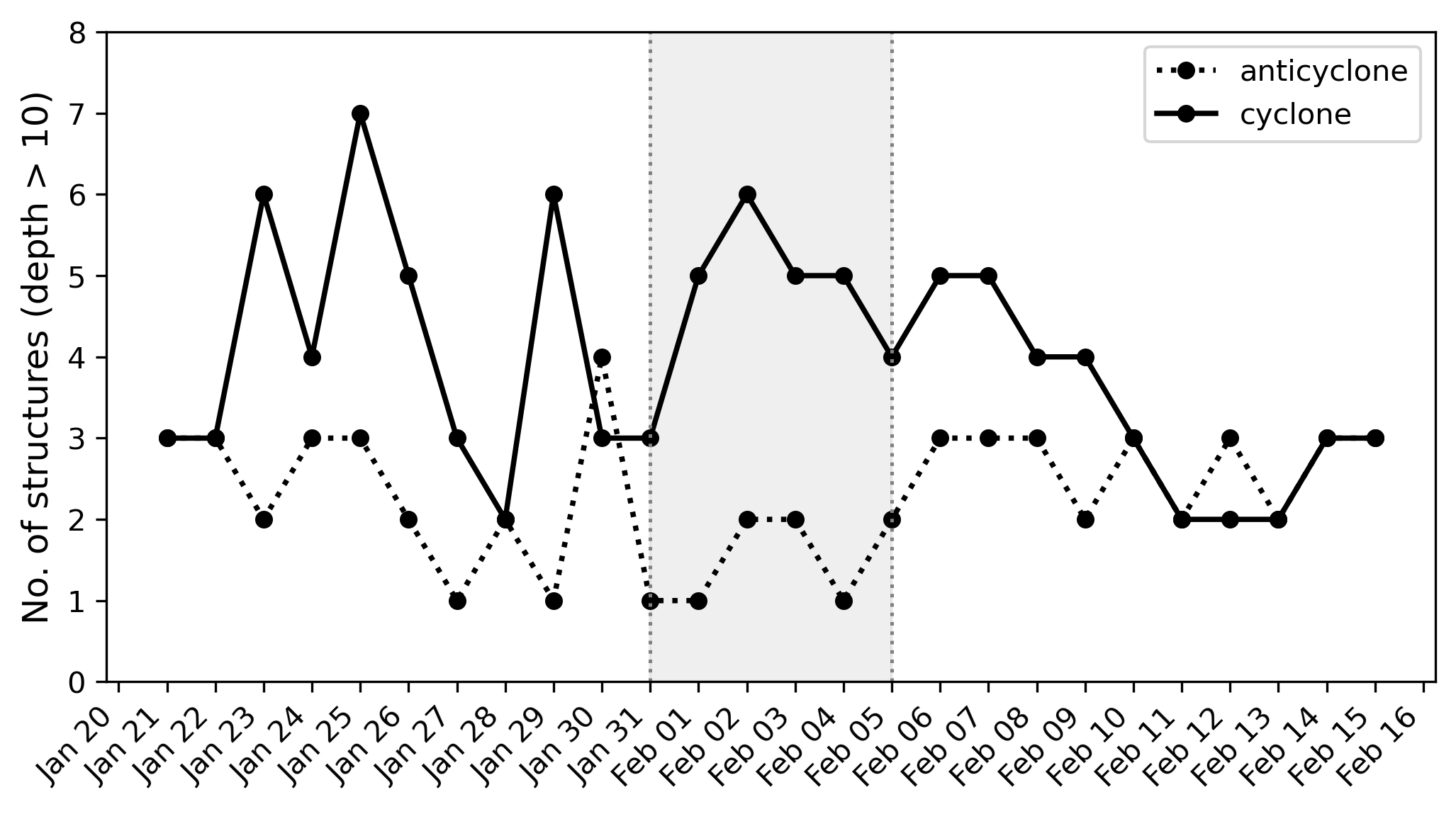}
    }

    \vspace{0.2cm}

    \subfloat[July 2003]{
        \includegraphics[width=0.45\linewidth]{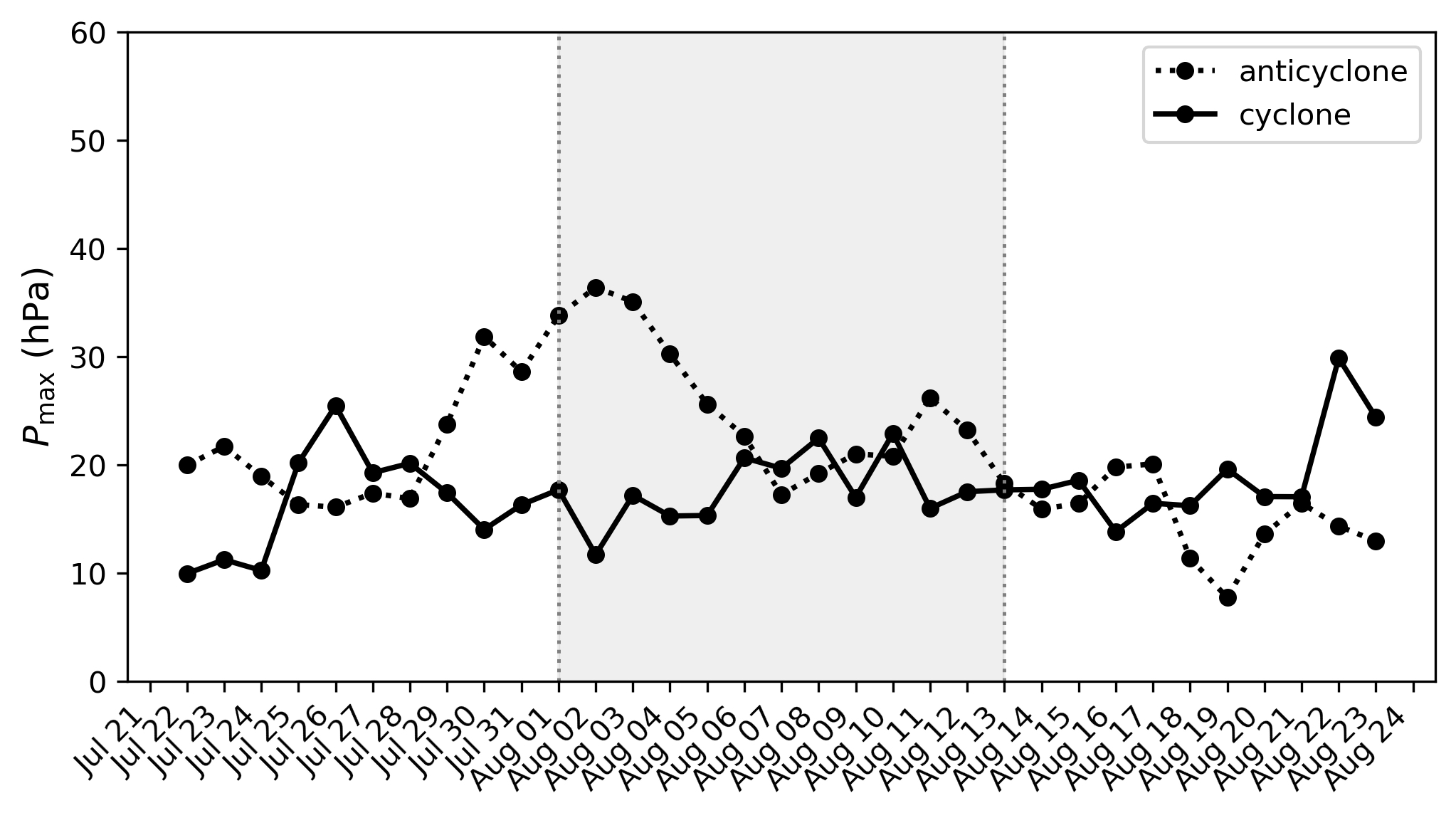}
    }
    \subfloat[February 2012]{
        \includegraphics[width=0.45\linewidth]{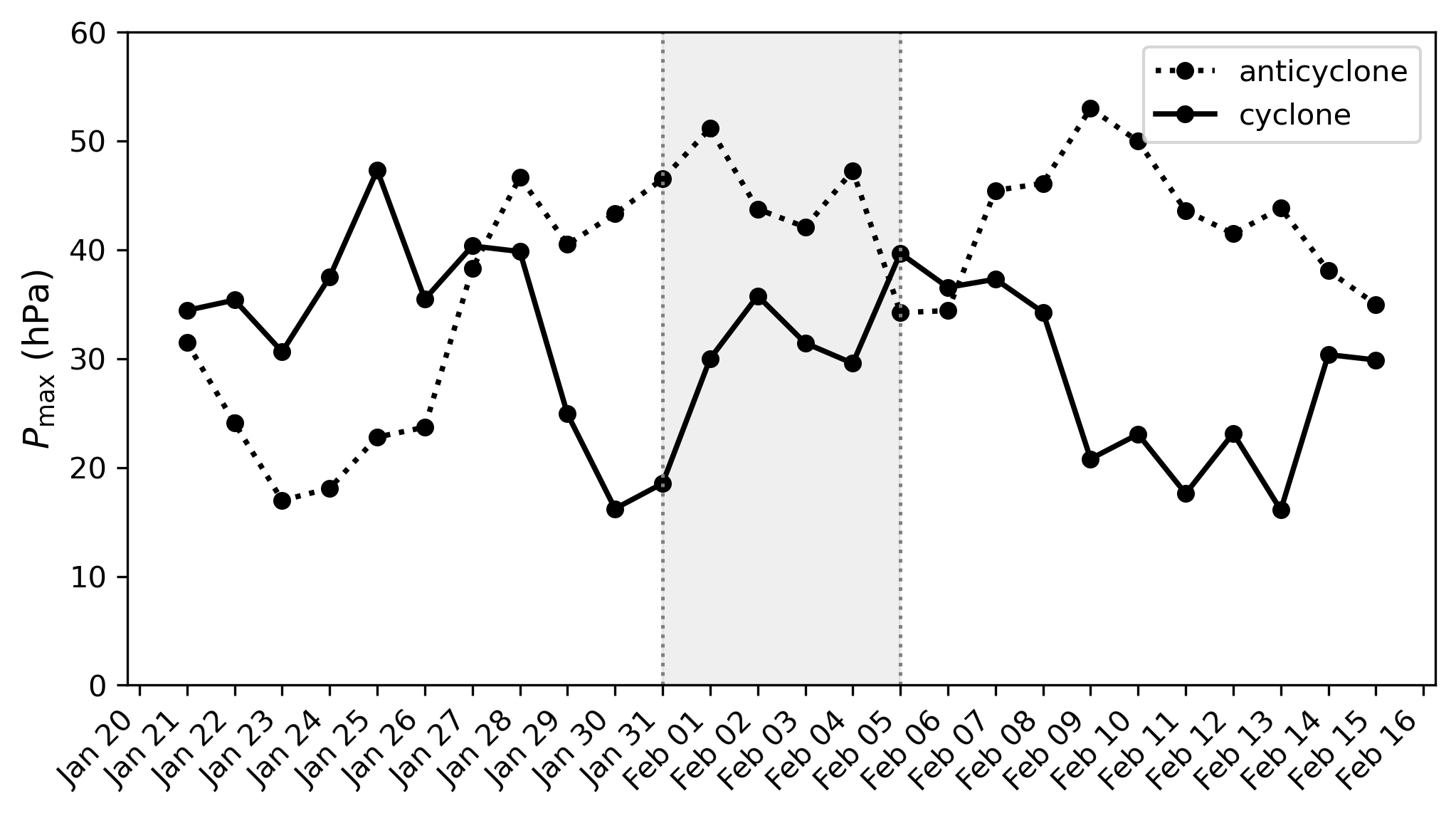}
    }
    \caption{
    Comparison of topological descriptors during the July--August 2003 and February 2012 blocking events. From top to bottom: area associated with the most persistent feature, number of persistent cycles with persistence greater than 10 hPa, and maximum topological depth $P_{\max}$. The left column corresponds to the July--August 2003 blocking event, while the right column corresponds to the February 2012 blocking event.
    }
    \label{fig:blocking_comparison}
\end{figure}

Figure~\ref{fig:blocking_comparison} illustrates the physical evolution of the circulation features. The area associated with the 1-anticyclone that has the highest topological depth (top row) exhibits a sharp increase during the blocking phase (shaded area), significantly surpassing the area of cyclonic features. Notably, there is a clear difference in the temporal timing of these spatial peaks: in the July--August 2003 event, the maximum area is reached at the very beginning of the blocking phase, suggesting a rapid initial expansion. In contrast, the February 2012 event shows the peak area towards the end of the mature phase (around 4 February), indicating a more gradual intensification in terms of spatial extent. Regarding the number of persistent features (middle row), 1-cyclones are consistently more numerous than 1-anticyclones, suggesting that while a single anticyclone dominates the map, cyclonic activity remains fragmented into multiple smaller vortices. 

The maximum topological depth $P_{\max}$ (bottom row) further characterizes the intensity of these structures. In the February 2012 event, the anticyclonic $P_{\max}$ is remarkably high and stable, averaging around 40~hPa and peaking near 50~hPa, which indicates an exceptionally deep and robust high-pressure system. Conversely, in the July--August 2003 event, the $P_{\max}$ is comparatively lower, reaching a maximum of approximately 35~hPa and showing a steady decay as the blocking progresses. This suggests that while the 2003 event was more persistent in time, the 2012 event reached a higher level of topological prominence and maintained its intensity more effectively during its peak.

\begin{figure}[htbp]
    \centering

    \subfloat[July--August 2003]{
        \includegraphics[width=0.45\linewidth]{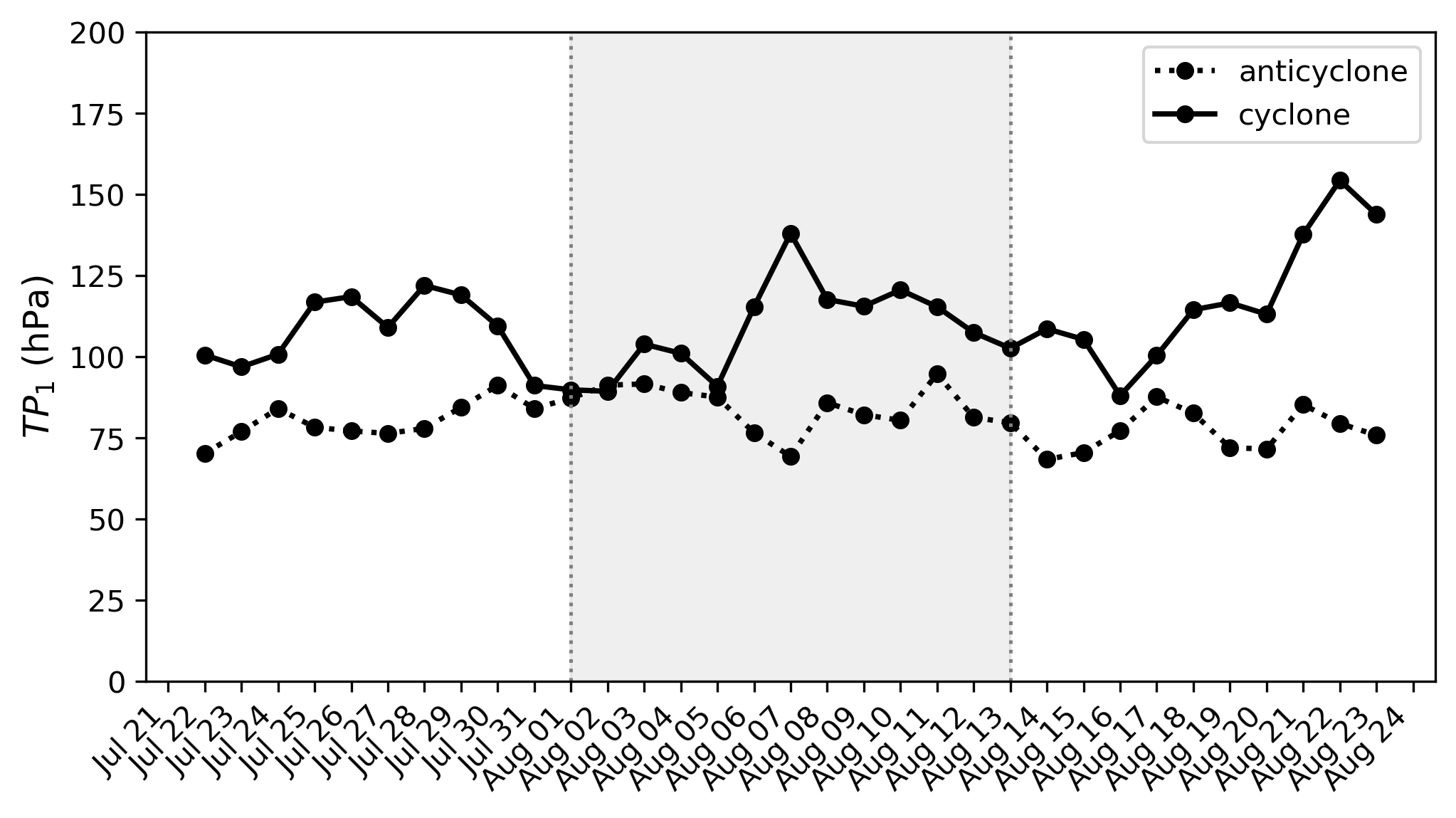}
    }
    \subfloat[February 2012]{
        \includegraphics[width=0.45\linewidth]{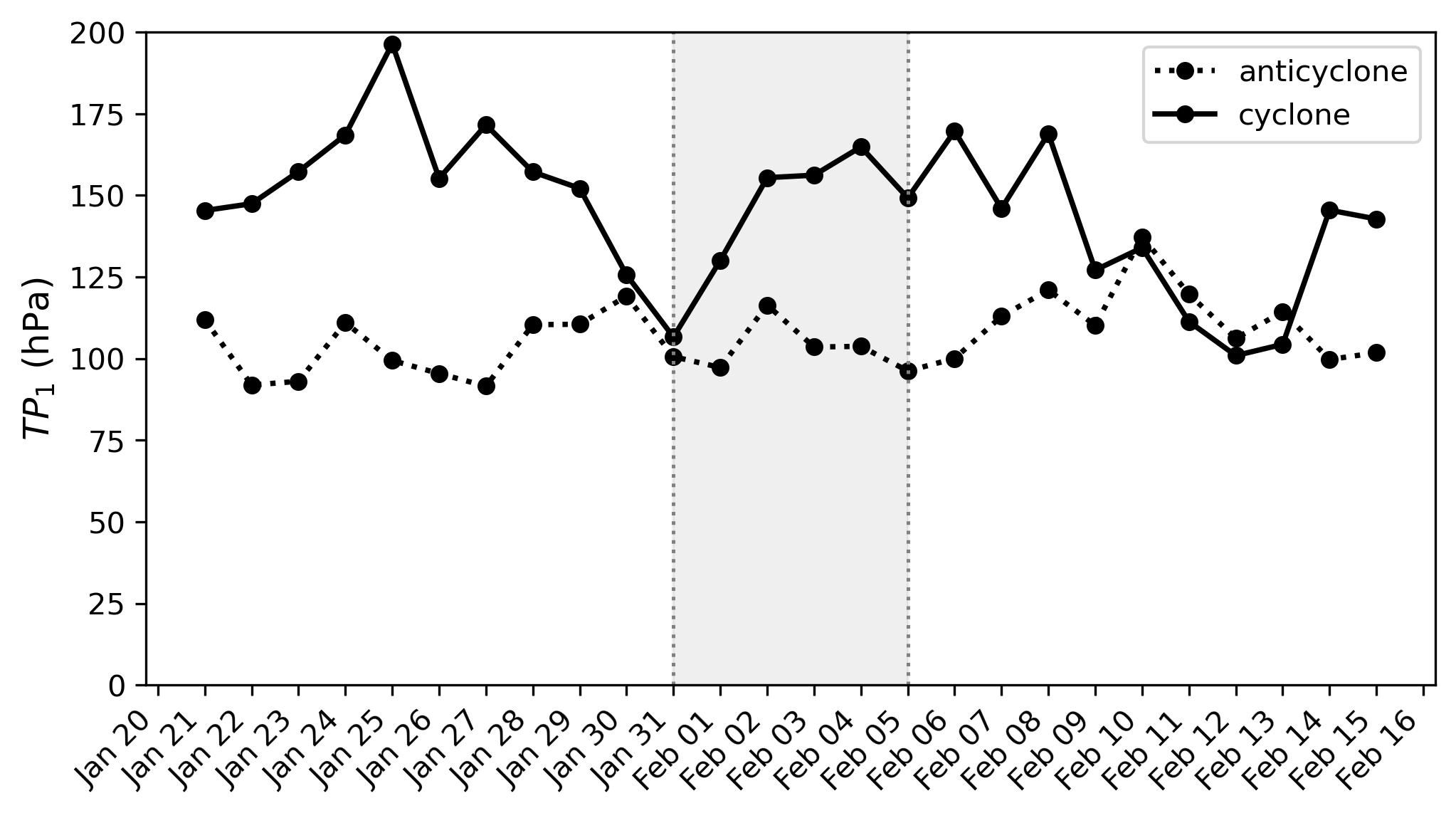}
    }

    \vspace{0.2cm}

    \subfloat[July--August 2003]{
        \includegraphics[width=0.45\linewidth]{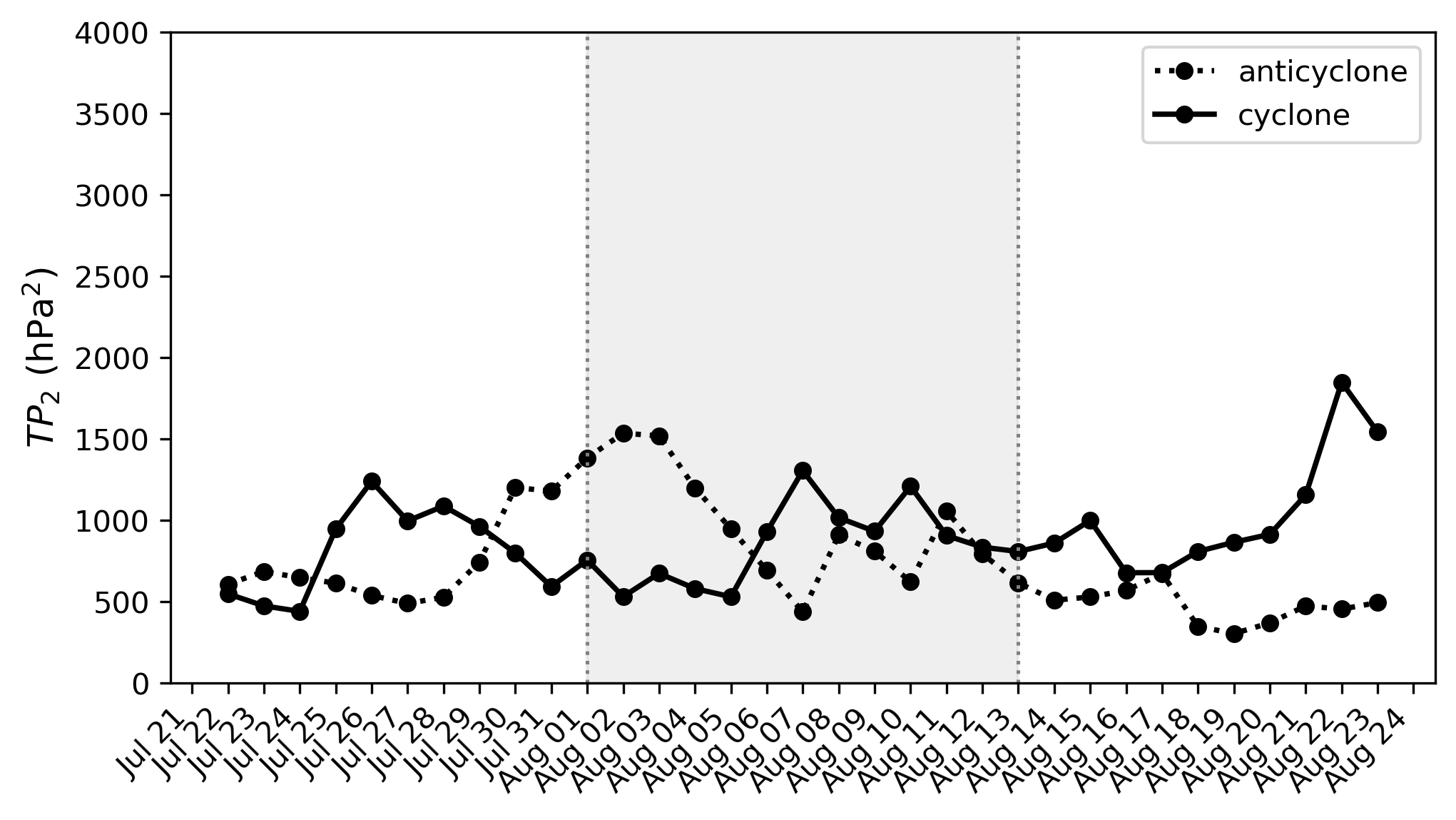}
    }
    \subfloat[February 2012]{
        \includegraphics[width=0.45\linewidth]{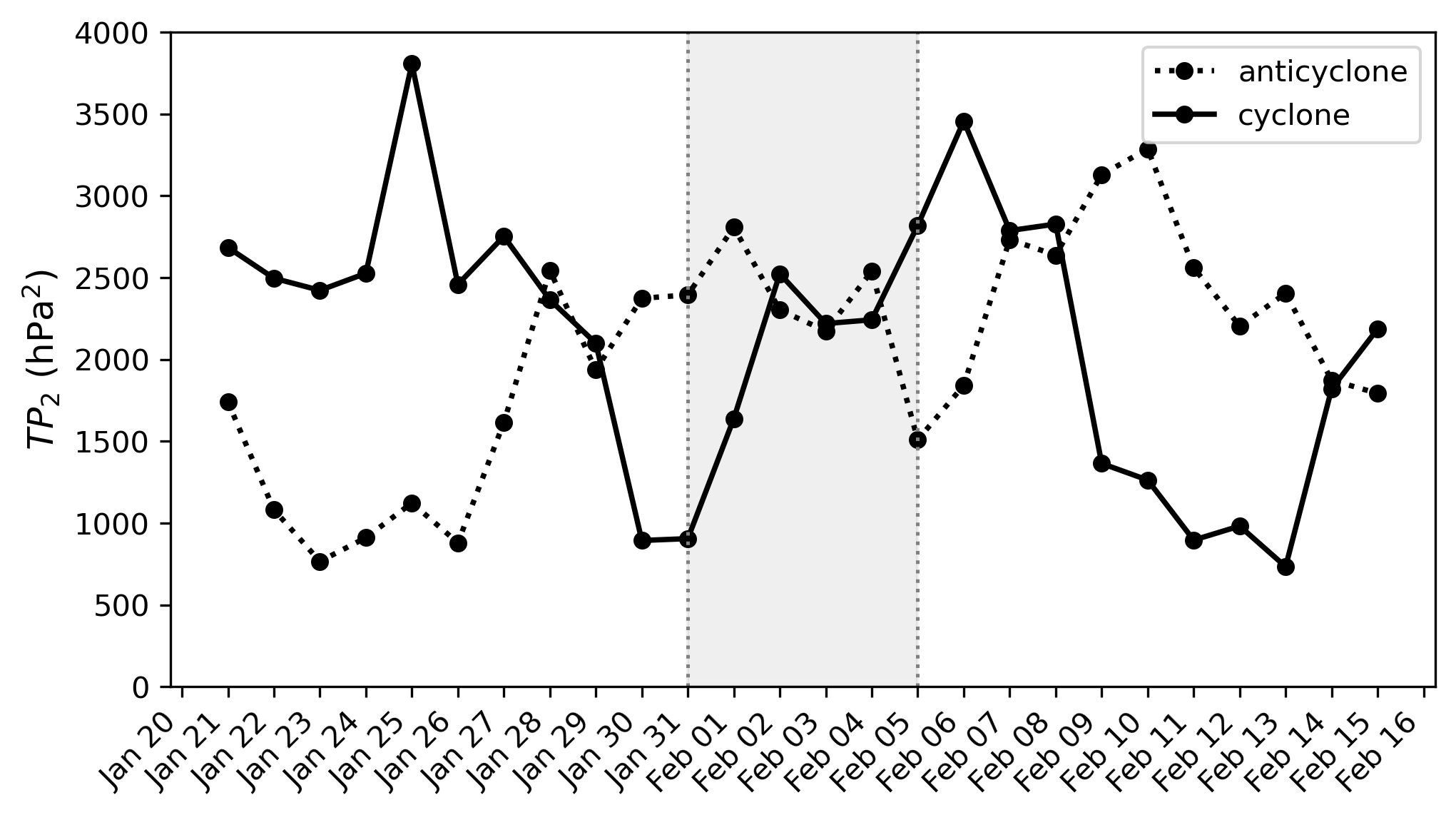}
    }

    \vspace{0.2cm}

    \subfloat[July--August 2003]{
        \includegraphics[width=0.45\linewidth]{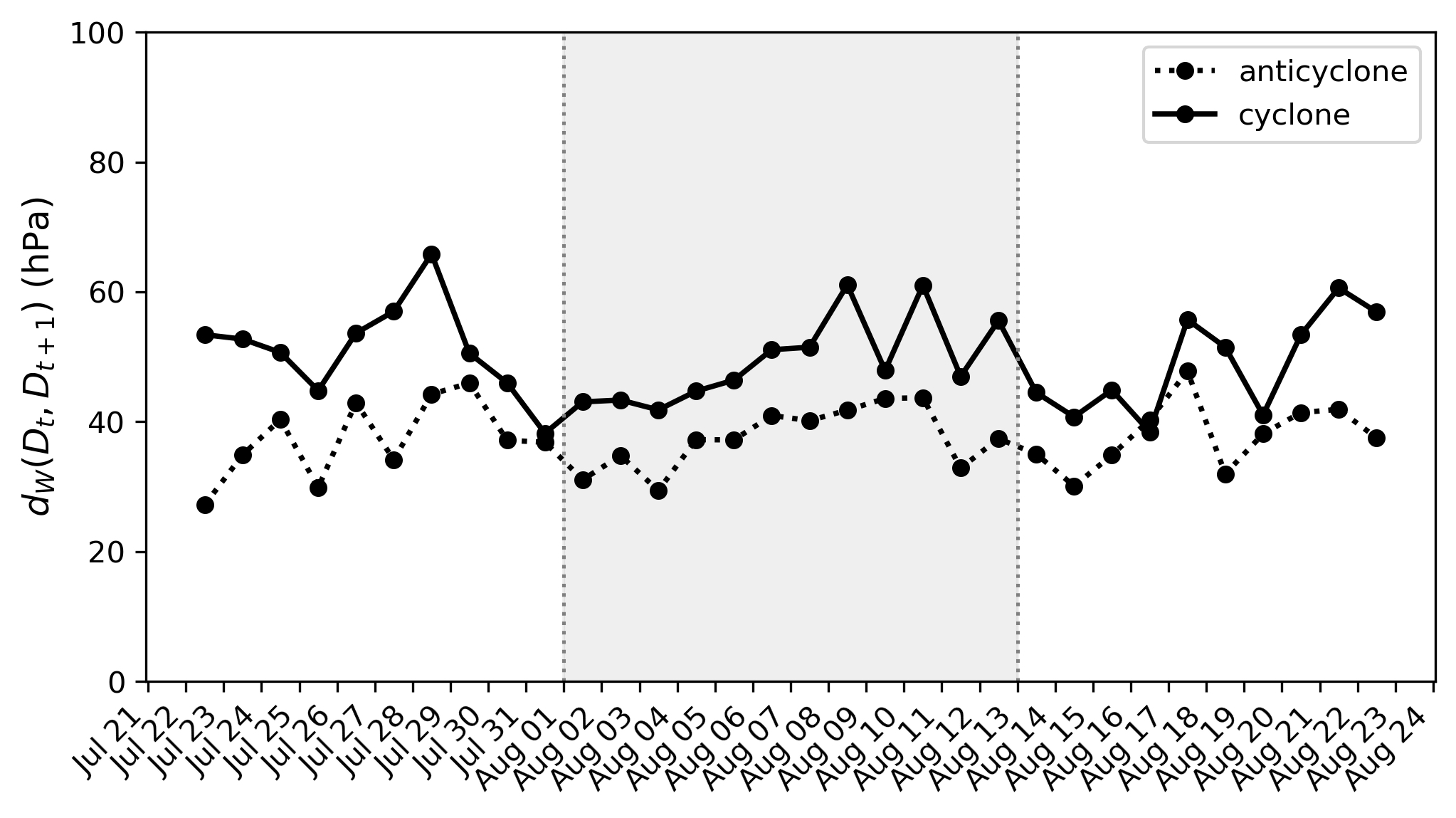}
    }
    \subfloat[February 2012]{
        \includegraphics[width=0.45\linewidth]{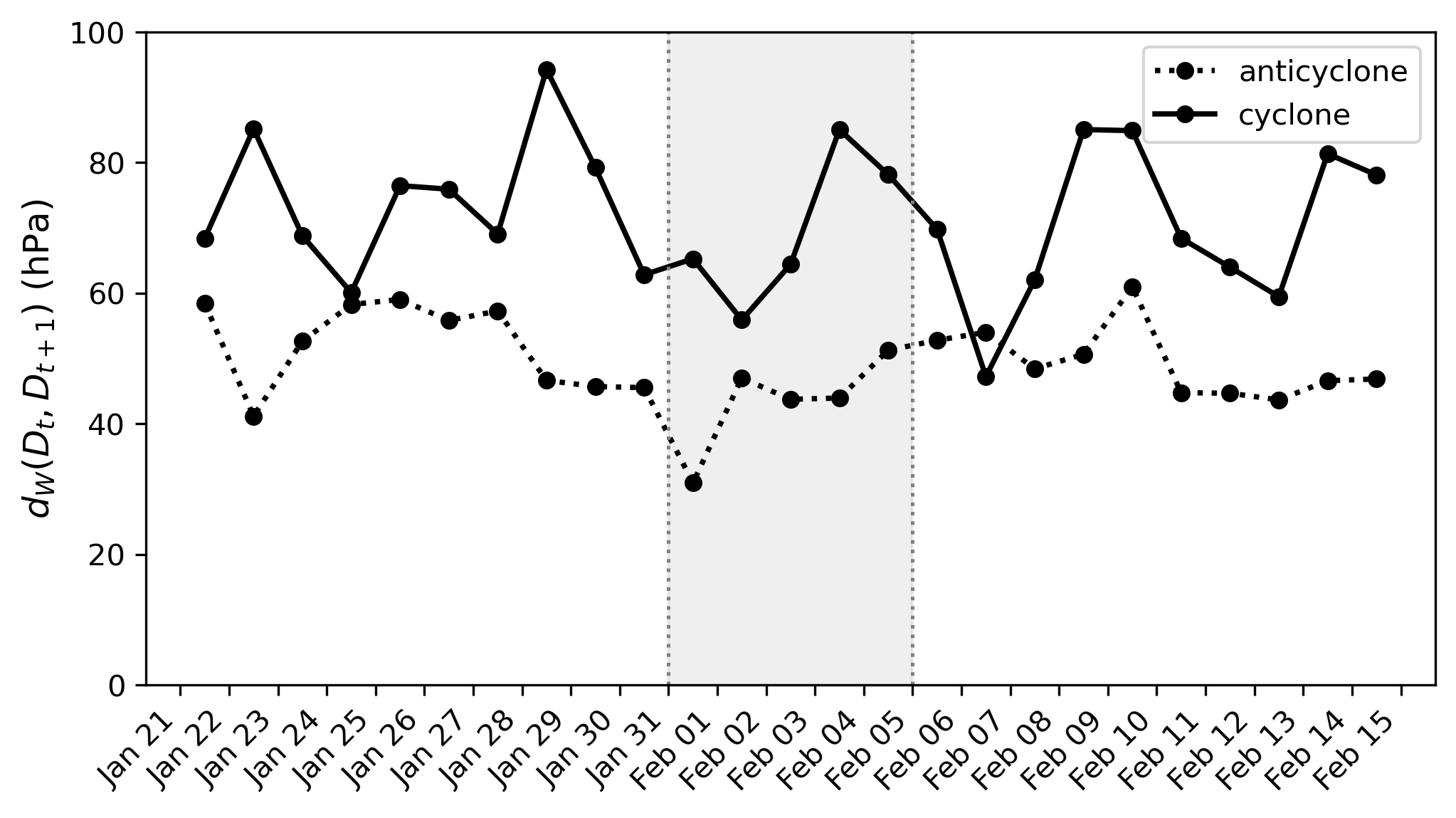}
    }

    \caption{
    Comparison of persistence-based descriptors during the July--August 2003 and February 2012 blocking events. The first row shows the evolution of the first-order total persistence ($\mathrm{TP}_1$), the second row shows the second-order total persistence ($\mathrm{TP}_2$), and the third row shows the Wasserstein distance between persistence diagrams of consecutive days, $d_W(D_t,D_{t+1})$. The left column corresponds to the July--August 2003 blocking event, while the right column corresponds to the February 2012 blocking event.
    }
    \label{fig:blocking_tp_wasserstein}
\end{figure}

Figure~\ref{fig:blocking_tp_wasserstein} provides a deeper look into the global organization and dynamical stability of these events through persistence-based metrics. The first-order total persistence ($\mathrm{TP}_1$, top row) reflects the overall topological activity. In both events, 1-cyclones generally contribute more to the global signal than 1-anticyclones; however, the February 2012 episode exhibits higher baseline values (averaging around 150~hPa) compared to the July--August 2003 event (averaging around 110~hPa), suggesting a more active and complex synoptic background in the 2012 case.

The second-order total persistence ($\mathrm{TP}_2$, middle row) emphasizes large-scale, robust structures by penalizing short-lived topological noise. A key distinction emerges here: in the 2003 event, the $\mathrm{TP}_2$ of the 1-anticyclone dominates for a sustained period during the blocking phase, even exceeding the cyclonic signal. In contrast, while the 2012 event reaches higher absolute $\mathrm{TP}_2$ values (peaking near 3500~hPa$^2$), it is characterized by much stronger fluctuations. This indicates that while the 2012 blocking was topologically more ``intense'', it lacked the steady dominance observed in the 2003 heatwave.

Finally, the Wasserstein distance $d_W(D_t, D_{t+1})$ (bottom row) measures the daily rate of topological reorganization for both cyclonic (solid line) and anticyclonic (dotted line) structures. In both events, the Wasserstein distance for 1-cyclones is consistently higher than for 1-anticyclones. This indicates that the cyclonic activity is inherently more volatile and undergoes more significant day-to-day topological changes. However, the comparison between the two episodes reveals key dynamical differences:
\begin{itemize}
    \item In the July--August 2003 event (panel e), the distance for the 1-anticyclone remains remarkably low and stable (around 35--40~hPa) throughout the blocking phase. This confirms that the anticyclonic core of the heatwave was topologically stationary, changing very little from one day to the next.
    \item In the February 2012 event (panel f), the values for both features are significantly higher. The 1-cyclone distance shows extreme volatility with multiple spikes exceeding 80--90~hPa. Notably, the 1-anticyclone distance is also less stable than in 2003, reflecting a blocking system that was undergoing constant internal reorganization.
\end{itemize}


\subsection{Methodological Benchmarking: Comparison with the Murray--Simmonds algorithm}\label{sec:MSalg}

The Murray--Simmonds (M\&S) algorithm represents a classical geometric approach to cyclone identification, relying on the detection of local minima in the sea level pressure (SLP) field. While robust for tracking intense extratropical storms, it treats cyclones as zero-dimensional points, effectively reducing complex atmospheric circulations to a single coordinate of minimum pressure. In contrast, the Topological Data Analysis (TDA) framework proposed in this work shifts the focus from local extrema to the global structural organization of the SLP anomaly field. By identifying persistent 1-cycles, the TDA approach captures the holistic ``loop-like'' nature of cyclonic circulation, offering a more nuanced representation of atmospheric dynamics.

\subsubsection{Qualitative Methodological Differences}

Table~\ref{tab:algo_comparison} summarizes the fundamental differences between the two paradigms. A key distinction lies in the significance filter: while M\&S uses a local depth criterion relative to a $5^{\circ}$ radius, TDA employs topological depth, which measures the ``lifetime'' of a feature across all pressure thresholds in a filtration. This allows TDA to filter out small-scale noise more effectively while preserving coherent structures that might lack a deep local minimum but possess a robust and organized circulation pattern.

\begin{table}[ht]
\centering 
\caption{Qualitative comparison of the TDA and Murray--Simmonds (M\&S) cyclone-tracking algorithms.} \label{tab:algo_comparison} \footnotesize \renewcommand{\arraystretch}{1.3} 
\begin{tabular}{|p{3.2cm}|p{4.5cm}|p{4.5cm}|} \hline 
\textbf{Aspect} & \textbf{M\&S (1991)} & \textbf{TDA (this work)} \\ \hline 
\textbf{Detection basis} & Local SLP minima ($3\times3$ search) & Topological depth of SLP anomalies \\ \hline 
\textbf{Significance filter} & Depth relative to a $5^{\circ}$ ring & Birth--Death interval (Persistence) \\ \hline 
\textbf{Representation} & Zero-dimensional point (center) & 1-cycle (spatial extent, shape, and area) \\ \hline 
\textbf{Track linking} & Nearest-neighbor (Euclidean distance) & Wasserstein distance (Topological stability) \\ \hline 
\textbf{Computational cost} & Low (Local extrema search) & High (Filtration and diagram matching) \\ \hline 
\textbf{Geographical bias} & Favors deep extratropical systems & Latitude-independent (anomaly-based) \\ \hline 
\end{tabular} 
\end{table}

\subsubsection{Quantitative Analysis and Latitudinal Distribution}

The quantitative statistics in Table~\ref{tab:quant_comparison} reveal that the TDA tracker is significantly more inclusive than M\&S, particularly at higher intensity thresholds. For instance, at 20~hPa, M\&S identifies only 6 tracks in 75 years, whereas TDA detects over 30,000. This disparity highlights a major limitation of point-based methods: as the threshold increases, local gradients often vanish or become too sharp for point-detection, whereas the topological signature of a large-scale anomaly remains robustly identifiable in the persistence diagram.

\begin{table}[ht] \centering \caption{Track statistics for M\&S and TDA (1948--2023, NCEP daily SLP, $0$--$90^{\circ}$N). $N$ = number of tracks; Mean, Max = lifetime (days); Lat = genesis latitude; Displ. = total displacement (km).} \label{tab:quant_comparison} \footnotesize \renewcommand{\arraystretch}{1.3} 
\begin{tabular}{|l|l|r|r|r|r|r|} \hline 
\textbf{Thresh.} & \textbf{Algo.} & \textbf{\textit{N}} & \textbf{Mean} & \textbf{Max} & \textbf{Lat} & \textbf{Displ.} \\ \hline 
5~hPa & M\&S & 147,050 & 3.3 & 146 & 73.9 & 386 \\ 
5~hPa & TDA & 178,515 & 1.2 & 11 & 37.2 & 106 \\ \hline 
10~hPa & M\&S & 9,938 & 2.4 & 34 & 66.5 & 345 \\ 
10~hPa & TDA & 94,169 & 1.2 & 11 & 35.7 & 132 \\ \hline 
20~hPa & M\&S & 6 & 2.7 & 5 & 41.7 & 268 \\ 
20~hPa & TDA & 30,084 & 1.3 & 11 & 34.4 & 182 \\ \hline 
\end{tabular} 
\end{table}

The difference in mean genesis latitude is perhaps the most striking result. M\&S shows a strong bias toward high latitudes ($>66^{\circ}$N), identifying only the deepest extratropical cyclones. TDA, by contrast, identifies systems at an average latitude of $\sim35^{\circ}$N. This indicates that TDA captures a wider variety of cyclonic phenomena, including Mediterranean cyclones and subtropical systems, which lack the extreme pressure depths required by M\&S but exhibit clear topological organization.

\subsubsection{Computational Trade-offs and Topological Rigor}

Despite its advantages, the TDA framework involves certain trade-offs. Computationally, the M\&S algorithm is more efficient due to its simpler local extrema search. TDA entails a higher overhead for constructing persistent filtrations and calculating Wasserstein distances. 

However, this cost is justified by the ``topological rigor'' of the framework. Unlike M\&S, which can bridge physically distinct systems via nearest-neighbor proximity (leading to unrealistic lifetimes of 146 days), TDA trajectories are strictly limited by topological continuity. A track is terminated immediately upon any merging or splitting event. This explains the shorter mean lifetimes in TDA results, ensuring that trajectories represent intervals of genuine structural stability rather than mere geometric proximity.

\section{Discussion}\label{sec:disc}

\subsection{Climatological Contrast between 1-Cyclones and 1-Anticyclones}
In this study, we have introduced a comprehensive framework based on Topological Data Analysis (TDA) to identify and track cyclonic and anticyclonic structures in Northern Hemisphere sea-level pressure (SLP) anomaly fields. By analyzing a historical record spanning seven decades (1948--2023), we have demonstrated that persistent homology provides a robust and mathematically grounded proxy for atmospheric circulation dynamics.
Our results reveal a clear statistical asymmetry between cyclonic and anticyclonic topological features in the Northern Hemisphere. 1-cyclone trajectories are roughly $2.5\times$ more numerous than 1-anticyclone trajectories (94{,}169 versus 38{,}221). Despite their higher total count, individual 1-cyclone trajectories are shorter on average (1.25 versus 1.47 days) and less spatially mobile, travelling on average 132.42~km per trajectory compared to 225.58~km for 1-anticyclones, with mean daily displacements of 99.73~km/day and 138.92~km/day, respectively. At the same time, 1-cyclones exhibit larger mean topological depth (18.32 versus 16.59~hPa, approximately 10\% higher), consistent with the stronger pressure gradients typically associated with cyclonic systems.

A key result of this study is that 1-anticyclones are systematically larger than 1-cyclones throughout the annual cycle (see Figure ~\ref{fig:traj_sup_sub_level}). This difference in spatial extent provides a natural explanation for the observed frequency asymmetry. Because anticyclonic structures occupy substantially larger areas, a relatively small number of them can cover a significant fraction of the atmospheric domain at any given time. In contrast, cyclonic activity is distributed among a larger number of smaller-scale features. The higher frequency of 1-cyclones is therefore consistent with their smaller characteristic area, whereas the lower frequency of 1-anticyclones reflects their broader spatial organization.

This structural asymmetry is also consistent with fundamental constraints of atmospheric dynamics. Although 1-anticyclones occupy larger areas, they exhibit lower mean topological depth than 1-cyclones. In the mid-latitudes, the intensity and compactness of an anticyclone are constrained by the requirement to avoid a local reversal of the absolute meridional vorticity gradient, which would favor barotropic instability. Consequently, while cyclones can develop into relatively intense and compact vortices with large topological depth, anticyclones tend to be broader structures with lower relative intensity. The joint analysis of topological depth and geodetic area therefore captures a physically meaningful distinction between the organization of cyclonic and anticyclonic circulation.

Additional differences emerge in the duration statistics. The distribution of 1-anticyclone lifetimes exhibits a substantially longer tail, with 46 trajectories exceeding 10 days and a maximum duration of 17 days. In contrast, only two 1-cyclone trajectories exceed 10 days, with a maximum duration of 11 days. This behavior suggests that anticyclonic structures, once established, are capable of maintaining their topological identity for longer continuous periods, consistent with the well-known persistence of blocking highs \cite{rex1950blocking,kautz2021atmospheric}.

Despite these marked contrasts in frequency, area, intensity, and duration, both filtrations display a similar seasonal cycle. The frequency of identified features and their associated topological depth increase during winter and decrease during summer, reaching a minimum around mid-summer. This common seasonal behavior suggests that cyclonic and anticyclonic structures remain coupled to the same large-scale atmospheric forcing, even though their statistical properties differ substantially.

\subsection{Synoptic Interpretation and Physical Drivers}

The long 1-cyclone trajectories identified in winter are consistent with the seasonal deepening of the Icelandic Low and its role as a quasi-permanent cyclonic structure \cite{hurrell1995decadal,dong2025key}. The enhanced topological depth during winter also aligns with the higher baroclinicity and stronger jet stream, which sustain large, robust cyclonic systems \cite{woollings2010variability}. 

However, high baroclinicity also favors rapid, small-scale secondary cyclogenesis. As noted by Stanković et al. (2024) \cite{stankovic2024large}, secondary lows often form at the southern edge of primary extratropical cyclones, deepening to replace or interact with pre-existing systems. Due to the topological rigor of our tracking algorithm, which terminates a trajectory at any merging or splitting event, these secondary developments are typically identified as distinct, shorter-lived features rather than as part of a single, long-duration event. While this may be viewed as a limitation for tracking the continuous envelope of a storm's center, it remains a significant strength for identifying intervals of genuine structural and topological stability.

For 1-anticyclones, the features capture the persistence of high-pressure systems, including the Azores High in the subtropics and blocking highs over the eastern Atlantic and Europe. The shorter mean lifetimes compared to the longest cyclonic events, but with a heavier long-duration tail (maximum 17 days), are consistent with the transient nature of many anticyclones alongside the occurrence of multi-week blocking episodes that exert a strong influence on European climate \cite{rex1950blocking,kautz2021atmospheric}. The greater mean spatial mobility of 1-anticyclone trajectories (225.58~km vs.\ 132.42~km) reflects the tendency of anticyclonic systems to migrate across larger distances, consistent with the known eastward propagation of blocking highs over the Euro-Atlantic sector \cite{kautz2021atmospheric}. The seasonal frequency pattern we observe in 1-anticyclone trajectories, with significantly higher activity during the cold season, is consistent with the documented climatology of blocking over the Euro-Atlantic sector \cite{kautz2021atmospheric}, which typically peaks during winter and early spring.

\subsection{Dynamical Stability and Volatility in Blocking Events}
The case studies of the July--August 2003 and February 2012 blocking events demonstrate that TDA-based metrics are capable of distinguishing between markedly different dynamical regimes. Although both events manifested as large-scale anticyclonic blocking structures, their topological evolution revealed a fundamental contrast. The 2003 heatwave was characterized by a remarkably stable topological skeleton, reflected in persistently low Wasserstein distances and sustained high values of $\mathrm{TP}_2$. Such behavior is consistent with a long-lived, quasi-stationary blocking configuration. In contrast, the 2012 event exhibited pronounced topological variability, with recurrent spikes in the Wasserstein distance accompanied by substantial fluctuations in topological area and depth. These signatures indicate a more dynamic and continuously reorganizing structure.

Taken together, these results suggest that persistence-diagram-based metrics provide a quantitative characterization of the nature of a blocking event, capturing its degree of stationarity, structural coherence, and topological stability. Such information is difficult to extract from conventional local pressure-based indices alone, highlighting the added value of a topological perspective for the analysis of atmospheric blocking.

\subsection{Methodological Advancements over Geometric Tracking}

Comparing our framework with the classical Murray--Simmonds (M\&S) algorithm highlights the advantages of prioritizing topological organization over local extrema. While M\&S is computationally efficient, it is intrinsically biased toward the deep pressure minima of high-latitude storm tracks. In contrast, TDA identifies a more diverse census of cyclonic structures, such as Mediterranean or subtropical systems, which may lack extreme pressure depths but exhibit clear structural organization.

Furthermore, a fundamental divergence concerns track lifetimes. The M\&S algorithm produces extremely long trajectories (up to 146 days) because its nearest-neighbor linking can bridge separate systems as long as a local minimum exists. In contrast, a TDA track terminates immediately upon any physical bifurcations such as merging or splitting events. This ``topological rigor'' explains the shorter mean lifetimes ($\sim1.3$ days) compared to M\&S.  While M\&S is optimized for following the long-range trajectory of a storm's center, TDA is designed to identify intervals of genuine structural stability, providing a more physically consistent description of the life cycle of synoptic weather systems.

\subsection{Sensitivity to the Analysis Domain}

An important methodological consideration concerns the dependence of the topological description on the spatial domain under investigation. Because persistent homology characterizes the topology of the observed pressure field, enlarging the geographical domain naturally increases the number of pressure systems available to generate topological loops. Consequently, the absolute number of identified 1-cyclones and 1-anticyclones is expected to depend on the size and location of the selected region.

The Northern Hemisphere analysis presented in this study provides an opportunity to assess the behavior of the method on a large and dynamically diverse atmospheric domain. As expected, a larger domain contains a greater number of circulation systems and therefore generates a larger number of topological features. However, the principal statistical relationships remain qualitatively robust. In particular, the contrast between cyclonic and anticyclonic structures, the distributions of topological depth, and the relationships between area, duration, and frequency persist across the hemispheric circulation.

These results suggest that the TDA framework should be regarded as a domain-dependent description of atmospheric circulation, in the sense that the number and arrangement of topological features depend on the region under consideration. At the same time, the dynamical signatures captured by the method appear robust to substantial changes in spatial scale. This robustness is particularly encouraging for future applications to other ocean basins, regional circulation systems, or larger-scale atmospheric domains.

For future global applications, an additional technical consideration would be the implementation of periodic boundary conditions at the $0^\circ/360^\circ$ longitude interface, ensuring topological continuity across the full zonal domain.

\subsection{Limitations}\label{sec:limitations}
Several limitations of the proposed framework should be acknowledged to guide its future use. First, regarding the treatment of the domain boundaries, a 1-hole can only be identified once a closed loop is fully contained within the cubical complex. Systems entering through a boundary, or quasi-stationary features anchored near it such as the subtropical Azores High, therefore generate a topological feature only once they are sufficiently inside the domain to close a cycle, and their trajectories terminate naturally as they approach the edge and can no longer support a closed loop. Because the hemispheric domain covers all longitudes, only the equatorial ($0^\circ$N) boundary and the polar cap require special consideration, and the 10~hPa topological-depth threshold further limits the influence of spurious boundary artifacts. Second, the method is intrinsically biased toward robust, well-organized structures: because detection relies on topological depth, highly transient, rapidly evolving, or shallow systems may not persist across enough filtration thresholds to be retained. This bias, combined with the topological rigor of the tracking (which terminates a trajectory at any merging or splitting event), explains the short mean track lifetimes ($\sim$1.3~days) and the predominance of single-day trajectories; the framework is consequently better suited to characterizing intervals of genuine structural stability than to following the continuous envelope of an individual storm centre, including the rapidly developing secondary cyclones discussed above. Finally, the construction of persistent filtrations and the computation of Wasserstein distances entail a higher computational cost than local-extrema searches, and the resulting statistics retain some sensitivity to the chosen topological-depth threshold and to the extent of the spatial domain.

\section{Conclusion} \label{sec:conc}

In this study, we have introduced a comprehensive Topological Data Analysis (TDA) framework for identifying and tracking cyclonic and anticyclonic structures in Northern Hemisphere sea-level pressure (SLP) anomaly fields.
 By analyzing a historical record spanning seven decades (1948--2023), we have demonstrated that persistent homology provides a robust and mathematically grounded proxy for atmospheric circulation dynamics. The method quantifies the size and duration of pressure anomalies in a way that standard storm-track metrics do not capture. 
Our results show that the topological organization of the Northern Hemisphere exhibits a pronounced seasonal cycle, with larger and more robust structures during boreal winter. This seasonality reflects the intensification of the Icelandic Low and the Azores High, as well as the enhanced baroclinicity of the cold season, while summer fields display a more fragmented and transient topological configuration.

Interpreting sea-level pressure anomalies as topological surfaces provided a geometric view of cyclones and anticyclones and their spatial organization. The method avoids feature-tracking choices and produces quantitative descriptors of circulation regimes. 

 A key contribution of this work is the characterization of atmospheric blocking through the lens of topological stability. By contrasting the 2003 European heatwave with the 2012 cold spell, we showed that TDA metrics, specifically second-order total persistence ($\mathrm{TP}_2$) and the Wasserstein distance, can differentiate between distinct dynamical regimes. While both events manifest as dominant anticyclonic structures, the 2003 episode was characterized by a ``frozen'' and quasi-stationary topological skeleton, whereas the 2012 event exhibited high volatility and frequent structural reorganizations. These findings suggest that TDA offers a unique diagnostic tool for assessing the stability and coherence of extreme weather events, providing insights that go beyond traditional local pressure indices. TDA offers a deeper understanding of the atmospheric ``skeleton'' and provides new tools to assess the evolution and impact of large-scale circulation anomalies in a changing climate.

Furthermore, our comparative assessment with the classical Murray--Simmonds (M\&S) algorithm underscores a fundamental shift from local geometric detection to global topological characterization. While traditional point-based methods are computationally efficient and well-suited for tracking intense polar centers, they often exhibit significant latitudinal biases and can produce tracking artifacts by bridging physically distinct systems. In contrast, the TDA framework offers a more physically consistent description of the atmospheric circulation. By prioritizing the stability of the topological ``skeleton'', it succeeds in identifying a broader and more diverse census of cyclonic structures regardless of their geographical location or absolute pressure depth. TDA provides a much broader census of the cyclonic structures that drive global weather patterns, albeit at a higher computational cost and with a stricter definition of temporal continuity. Although the topological rigor of TDA results in shorter track lifetimes due to its inherent sensitivity to merging and splitting events, it ensures that the identified trajectories correspond to intervals of genuine structural stability.

Ultimately, this work demonstrates that TDA can recover and enrich the well-established climatology of cyclones, anticyclones, and blocking from first principles, without the need for arbitrary filtering or vortex-tracking parameters. The noise-robust and multiscale nature of persistent homology makes it a powerful complementary tool for climate science. The successful application of the framework to seven decades of Northern Hemisphere circulation further demonstrates its ability to characterize atmospheric organization across large spatial scales. Future research could extend this framework to other atmospheric variables, fully global domains, or three-dimensional fields, potentially uncovering new topological signatures of climate variability and change in the global circulation.

\appendix
\section {Cubical complexes} \label{app:cubical-complexes}

Before establishing a definition of the term {\textit{cubical complex}}\cite{kaczynski2006computational}, it is necessary to define certain concepts.

\begin{definition}
    An \textit{elementary cube} $\tau\subset\R^n$ is defined as a finite product of \textit{elementary intervals}:
    
    $$\prod_{i=1}^n [m_i, m_i + \epsilon_i],$$
    where $m_i\in\Z$, and $\epsilon_i=\{0,1\}$. 

\end{definition}
There are two types of elementary intervals: the \textit{degenerate intervals} $[m_i, m_i]$ (when $\epsilon_i = 0$), which represent vertices, and the \textit{non-degenerate intervals} $[m_i, m_i + 1]$ (when $\epsilon_i = 1$), which represent segments.

\begin{definition}
Let $\tau \subset \mathbb{R}^n$ be an elementary cube. The \textit{dimension} of $\tau$ is defined as the number $l$ of non-degenerate intervals in its product representation, and $\tau$ is called an \textit{$l$-cube}.

\end{definition}

If an elementary cube is contained in another, it is called a \textit{face} of the larger cube.

\begin{definition}
 A cubical complex $K\subseteq \mathbb{R}^n$ is a collection of cubes such that every non-empty intersection of two cubes is a face of each, and every face of a cube in $K$ is also contained in $K$.   
\end{definition}

\begin{example}
The tessellation in \cref{subfig:tesl} is an example of a cubical complex, which contains 2-cubes as unit squares, non-degenerate elementary intervals as the edges of the squares, and degenerate elementary intervals as their vertices. One can easily verify that the tessellation in \cref{subfig:tesl} satisfies the two defining properties of a cubical complex.
\end{example}
\section{Homology groups} \label{app:hom_groupcomplex}
For a cubical complex $K$, a chain group $C_{l}$ ($1\leq l\leq n$) represents how the $l$-cubes are assembled in the complex. The present paper will deal with the notion of homology in the context of coefficients of the multiplicative group $Z_2$.
The chain group $C_{l}$ is an abelian group given by
$$C_{l}(K;\Z_{2}) \coloneqq
\bigg\{\sum_{i=1}^{q}\tau_{i}
:\tau_{i}\in\C
,q\in\N
\text{ , and }
\tau_{i}
\text{ is }
l\text{-cube}\bigg\}$$

\begin{definition}
 A \textit{boundary} $\partial$ of an elementary interval $[a,b]$, where $a,b\in\Z$ is defined as $$\partial[a,b] = a+b.$$  
\end{definition}

If $[a,b]$ is a degenerate interval ($a=b$), $\partial[a,a] = a+a = 2a = 0$ as $2=0$ in $\Z_2$. Similarly, let $\tau$ be an $l$-cube, $\tau = I_{1}\times\cdots\times I_h\times\cdots\times I_l$, where $I_h$ is an elementary interval $1\leq h\leq l\leq n$. The boundary of  $\tau$ boundary is defined as:
$$ \partial\tau = (\partial I_{1}\times\cdots\times I_h\times\cdots\times I_l) +$$
$$\cdots +
(I_{1}\times\cdots\times \partial I_h\times\cdots\times I_l) +
\cdots + (I_{1}\times\cdots\times I_h\times\cdots\times \partial I_l).$$
Using this definition for boundary of a cube we extend it linearly to define boundary map between chain groups $\partial_{l+1}:C_{l+1}(K;\Z_{2})\to C_{l}(K;\Z_{2})$ . For simplicity we will drop $\Z_2$ from notation of chain complex and chain group.
\begin{definition}
 A \textit{chain complex} $C_{*}(K;\Z_{2})$ consists of a sequence of chain groups connected by boundary maps. The chain complex $C_{*}\subseteq\R^n$, can be represented as:
$$0\overset{\partial_{n+1}}{\to} C_{n}(K)
\overset{\partial_{n}}{\to}C_{n-1}(K)
\overset{\partial_{n-1}}{\to}
\cdots
\overset{\partial_{2}}{\to}C_{1}(K)
\overset{\partial_{1}}{\to}C_{0}(K)
\overset{\partial_{0}}{\to}0.$$

Each map $\partial_n:C_n\to C_{n-1}$ satisfies: $\partial_n \circ \partial_{n-1}=0$, $\forall n.$

\end{definition}

Chain complexes form the basis for homological algebra and are used to define homology groups.

\begin{definition}
    The quotient group $H_{l}(K)\coloneqq\Ker\partial_{l}/\Ima\partial_{l+1}$ is called the \textit{$l$-th homology group} or \textit{homology in degree $l$} of the cubical complex $K$. 

\end{definition}

The elements of the kernel of $\partial_{l}$ ($\Ker\partial_{l}$) are called \textit{cycles} and the elements of the image of $\partial_{l}$ ( $\Ima\partial_{l+1}$) are called \textit{boundary cycles}.
Thus, homology captures the cycles that are not the boundary of any higher dimensional cube: $H_{0}(K)$ measures the connected components of $K$, $H_{1}(K)$ the non-equivalent 1-holes, $H_{2}(K)$ the voids, and $H_{n}(K)$ the n-dimensional voids $\forall n \ge 3$.

\section{Supplementary Information}

All the codes required to reproduce the numerical results presented in this paper are publicly available at the following repository: \url{https://github.com/Himanshu484/topological_interaction_cyclone_and_anticyclone}

\section*{Data availability}

The NCEP-NCAR Reanalysis 1 dataset used in this study is publicly available from the NOAA Physical Sciences Laboratory (PSL) (\url{https://psl.noaa.gov/data/gridded/data.ncep.reanalysis.html}). The data and code used in this study are openly accessible at the GitHub repository linked above.

\section*{Acknowledgments}
This work was supported by the COST Action CA22162 FutureMed (European Cooperation in Science and Technology), the ANR project PowDev (Strategic Development of the power grids of the future, ANR reference: ANR-22-PETA-0016), and the ANR project Templex (ANR-23-CE56-0002). We acknowledge useful discussions in the Mathematical Research Community of the American Mathematical Society. 
\section*Figure 7: Sea level pressure (SLP) anomalies over the Northern Hemisphere on 31 Decem-{Declaration of generative AI and AI-assisted technologies in the manuscript preparation process}
During the preparation of this work, the authors used OpenAI tools to improve the readability and language of the manuscript. After using these tools, the authors reviewed and edited the content as needed and take full responsibility for the content of the published article.

\clearpage
\bibliographystyle{elsarticle-num}
\bibliography{TICAC.bib}

\end{document}